\documentclass[acmsmall]{acmart}

\usepackage{booktabs} % For formal tables
\usepackage{subcaption}
\usepackage{float}
\usepackage{url}
\usepackage{epstopdf}
\usepackage{graphicx}
\usepackage{array}
\usepackage{pifont}
\usepackage{multirow}
\usepackage{amssymb}
\usepackage{amsmath}
\usepackage{caption}
\usepackage{algorithm, algpseudocode}

\usepackage{paralist, tabularx}

\renewcommand\footnotetextcopyrightpermission[1]{} 

\makeatletter
\def\@copyrightspace{\relax}
\makeatother

\begin{document}

\title{Summarizing User-generated Textual Content: Motivation and Methods for Fairness in Algorithmic Summaries}
\thanks{\textcolor{red}{This work has been accepted for presentation at the ACM Conference on Computer-Supported Cooperative Work and Social Computing (CSCW) 2019. Please cite the CSCW proceedings version.}\\ 
We acknowledge the human annotators who developed the gold standard summaries for the datasets used in this study. This research was supported in part by a European Research Council (ERC) Advanced Grant for the project ``Foundations for Fair Social Computing", funded under the European Union's Horizon 2020 Framework Programme (grant agreement no. 789373). 
A. Dash is supported by a Fellowship from Tata Consultancy Services.}

\author{Abhisek Dash}
%\email{dash.abhi93@iitkgp.ac.in}
\affiliation{
    \institution{Indian Institute of Technology Kharagpur, India}
    %\country{India}
}

\author{Anurag Shandilya}
%\email{anurag.k.shandilya@gmail.com}
\affiliation{
    \institution{Indian Institute of Technology Kharagpur, India}
    %\city{Kharagpur}
    %\country{India}
}

\author{Arindam Biswas}
%\email{aribis369@gmail.com}
\affiliation{
    \institution{Indian Institute of Technology Kharagpur, India}
    %\city{Kharagpur}
    %\country{India}
}

\author{Kripabandhu Ghosh}
%\email{kripa.ghosh@gmail.com}
\affiliation{
    \institution{Tata Research Development and Design Centre, India}
    %\city{Pune}
    %\country{India}
}

\author{Saptarshi Ghosh}
%\email{saptarshi@cse.iitkgp.ac.in}
\affiliation{
    \institution{Indian Institute of Technology Kharagpur, India}
    %\city{Kharagpur}
    %\country{India}
}

\author{Abhijnan Chakraborty}
%\email{achakrab@mpi-sws.org}
\affiliation{
    \institution{Max Planck Institute for Software Systems, Germany}
    %\city{Saarbruecken-Kaiserslautern}
    %\country{Germany}
}

\begin{abstract}
As the amount of user-generated textual content grows rapidly, text summarization algorithms are increasingly being used to provide users a quick overview of the information content. Traditionally, summarization algorithms have been evaluated only based on how well they match human-written summaries (e.g. as measured by ROUGE scores). 
In this work, we propose to evaluate summarization algorithms from a completely new perspective that is important when the user-generated data to be summarized comes from different socially salient user groups, e.g. men or women, Caucasians or African-Americans, or different political groups (Republicans or Democrats). 
In such cases, we check whether the generated summaries {\it fairly represent} these different social groups. 
Specifically, considering that an {\it extractive} summarization algorithm selects a subset of the textual units (e.g. microblogs) in the original data for inclusion in the summary, we investigate whether this selection is {\it fair} or not. 
Our experiments over real-world microblog datasets show that existing summarization algorithms often represent the socially salient user-groups very differently compared to their distributions in the original data. 
More importantly, some groups are frequently {\it under-represented} in the generated summaries, and hence get far less exposure than what they would have obtained in the original data.
To reduce such adverse impacts, we propose novel fairness-preserving summarization algorithms which produce high-quality summaries while ensuring fairness among various groups. 
To our knowledge, this is the first attempt to produce fair text summarization, and is likely to open up an interesting research direction.
\end{abstract}




\if 0
\setcopyright{acmlicensed}
\acmJournal{PACMHCI}
\acmYear{2019} \acmVolume{3} \acmNumber{CSCW} \acmArticle{172} \acmMonth{11}
\acmPrice{15.00}\acmDOI{10.1145/3359274}
\fi 

\maketitle

%\vspace{-4.5mm}

\section{Introduction} 
\label{sec:intro}
\noindent 
Recently, there has been an explosion in the amount of user-generated information on the Web. To help Web users deal with the information overload, text summarization algorithms are commonly used to get a quick overview of the textual information. Recognizing the business opportunities, many startups have mushroomed recently to offer content summarization services. 
%, and many companies are providing interfaces for summarisation, 
For example, Agolo ({\tt agolo.com/splash}) provides a summarization platform to get the most relevant information from both public and private documents. 
Aylien ({\tt aylien.com/text-api/summarization}) or 
Resoomer ({\tt resoomer.com}) present relevant points and topics from a piece of text. 
%A list of startups: https://angel.co/content-summarization
%Moreover, some news websites, e.g., Harvard Business Review\footnote{\url{https://www.hbr.org}}, have started providing an `executive summary' of the news articles, to help the readers lacking time to go through the full article.  
%E.g., see https://hbr.org/2013/05/why-the-lean-start-up-changes-everything and click the “summary” on the menu.
Multiple smartphone apps (e.g. News360, InShorts) have also been launched to provide short summaries of news stories.

A large number of text summarization algorithms have been devised, 
%the research community,
including algorithms to  summarize a single large document, as well as 
%multi-document summarization algorithms for 
for summarizing a set of documents (e.g. a set of microblogs or tweets); interested readers can check~\cite{text-summarization-survey} for a survey on summarization algorithms. 
Most of these summarization algorithms are {\it extractive} in nature, i.e.
they form the summary by extracting some of the textual units in the input~\cite{extractive-summarization-survey-gupta} (e.g. individual sentences in a document, or individual %microblogs 
tweets in a set of tweets). 
Additionally, some {\it abstractive} algorithms have also been devised, that attempt to generate natural language summaries~\cite{text-summarization-survey}. In this paper, we restrict our focus to the more prevalent extractive summarization. 

Extractive summarization algorithms essentially perform a selection of a (small) subset of the textual units in the input, for inclusion in the summary, based on some measure of the relative quality or importance of the textual units. 
Traditionally, these algorithms are judged based on how closely the algorithmic summary matches gold standard summaries that are usually written by human annotators. 
To this end, measures such as ROUGE scores are used to evaluate the goodness of algorithmic summaries~\cite{lin-rouge}. The underlying assumption behind this traditional evaluation criteria is that {\it the data to be summarized is homogeneous, and the sole focus of summarization algorithms should be to identify summary-worthy information}.

%Information on the Web today is often an amalgamation of information of different classes, that come from multiple sources or groups, and often cover different perspectives. 
However, user-generated content constitutes a large chunk of information generated on the Web today, and such content is often heterogeneous, coming from users belonging to different social groups. 
For example, on social media, different
user groups (e.g. men and women, Republicans and Democrats) discuss socio-political issues, and it has been observed that 
different social groups often express very different opinions on the same topic or event~\cite{chakraborty2017makes}. 
%Similarly, news media sources having different ideological leanings publish different articles on the same topic or event, covering different political parties, different gender issues, etc. 
Hence, while summarizing such heterogeneous %crowdsourced 
user-generated data, one needs to check whether the %generated 
summaries are properly representing the opinions of these different social groups.
Since the textual units (e.g. tweets) that are included in the summary get much more exposure than the rest of the information (similar to how top-ranked search results get much more exposure than other documents~\cite{Zehlike2017,Biega-fair-rank-sigir18}), if a particular group is under-represented in the summary, their opinion will get much less exposure than %as compared to 
the opinion of other groups.

Therefore, in this paper, we propose to look at %extractive 
summarization algorithms from a completely new perspective, and 
%We propose to 
investigate whether the selection of the textual units in the summary is fair, i.e. {\it whether the generated summary fairly represents every social group in the input data}.
%In this paper, 
We experiment with three %microblog 
datasets of %that contain %information crowdsourced from 
tweets generated by different user groups (men and women, pro-republican and pro-democratic users).  
We find that %many 
most existing summarization algorithms do {\it not} fairly represent %the various social 
different groups in the generated summaries, even though the tweets written by %different social 
these groups are of comparable textual quality. More worryingly, some groups are found to be systemically under-represented in the process. Note that we, by no means, claim such under-representation to be  intentionally caused by the existing algorithms. Rather it is most likely an inadvertent perpetuation of the metrics that the algorithms are trying to optimize. %\new{Summarization algorithms are increasingly important with the rapid growth of user generated content on the Web; since such, 
Since the applications of summarization algorithms may extend from product reviews to citizen journalism, the question of whether existing algorithms are fair and how we can potentially improve them become even more important.

Having observed that existing summarization algorithms do not give fair summaries in most cases, we next attempt to develop algorithms for fair summarization. 
Recently, there have been multiple research works attempting to incorporate fairness in machine learning algorithms~\cite{hajian2016algorithmic, dwork2012fairness, kleinberg2018discrimination}. Primarily, there are three ways in which these research works make fairness interventions in an existing system -- {\it  pre-processing, in-processing} and {\it post-processing}, depending on whether the interventions are applied at the input, algorithm or the output stage~\cite{friedler2019comparative}.
Following this line of work, in this paper, %To reduce the above mentioned unfairness, 
we develop three novel fairness-preserving summarization algorithms which select highly relevant textual units in the summary while maintaining fairness in the process. Our proposed in-processing algorithm is based on constrained sub-modular optimization (where the fairness criteria are applied as constraints).  The post-processing algorithm is based on fair ranking of textual units based on some goodness measure, and the pre-processing approach groups the tweets on the basis of their association to different classes, and then summarizes each group separately to generate fair summaries.
Extensive evaluations show that our proposed algorithms are able to generate %fair 
summaries having quality comparable to %what is generated by several well-known 
state-of-the-art summarization algorithms (which often do not generate fair summaries), while being fair to different user groups.

In summary, we make the following contributions in this paper: 
(1)~ours is one of the first attempts to consider the notion of fairness in summarization, and the first work on fair summarization of textual information;  
%(see Section~\ref{sec:related} for a literature survey).
%\footnote{Fairness in summarization was briefly introduced in our prior work~\cite{shandilya2018fairness}.},  
(2)~we show that, while summarizing content generated by different user groups, existing summarization algorithms often do not  represent the user groups fairly; and 
(3)~we propose summarization algorithms that produce summaries that are of good quality as well as fair according to different fairness notions, including equal representation, proportional representation, and so on.
We have made the implementation of our fair summarization algorithms and our datasets publicly available at {\tt \url{https://github.com/ad93/FairSumm}}. 

We believe that this work will be an important addition to the growing literature on incorporating fairness in algorithmic systems. 
Generation of fair summaries would not only benefit the end users of the summaries, but also many downstream applications that use the summaries of crowdsourced information, e.g., summary-based opinion classification and rating inference systems~\cite{lloret2010experiments}. %(e.g., Lloret {\it et al.}~\cite{lloret2010experiments} proposed a summary-based opinion classification and rating inference mechanism).
%Such applications would also benefit from a fair summary, e.g., one which fairly represents the positive and negative opinions in the input data. 

The rest of the paper is structured as follows. Section~\ref{sec:related} gives a background on summarization and discusses related works. %and background information, that are relevant to our work, succeeded 
Section~\ref{sub:datasets} describes the datasets we use throughout the paper. 
Thereafter, we motivate the need for fair  summarization in Section~\ref{sec:background}. Section \ref{sec: notions} introduces some possible notions of fairness in summarization, and %the subsequent 
Section~\ref{sec:existing} shows how existing text summarization algorithms do not adhere to these fairness notions. 
In Section~\ref{sec: Framework}, we discuss a principled framework %to introduce the different approaches of
for achieving fairness in summarization, followed by details of three fair summarization algorithms in Sections~\ref{sec:fairsumm} and ~\ref{sec:otherMeth}. 
We evaluate the %goodness of the summaries produced by
performance of our proposed algorithms in Section~\ref{sec:expts}. Finally, we conclude the paper, discussing some limitations of the proposed algorithms and possible future directions.
%\vspace{-2mm}
\section{Background and Related Work} \label{sec:related}

In this section, we discuss two strands of prior works that are 
relevant to our paper. 
First, we focus on text summarization. Then, 
we relate this paper to prior works on bias and fairness in information systems.

\subsection{Text Summarization}
Text summarization is a well-studied problem in Natural Language Processing, where the task is to produce a fluent and informative summary given a piece of text or a collection of text documents. 
A large number of text summarization algorithms have been proposed in literature; the reader can refer to~\cite{extractive-summarization-survey-gupta,text-summarization-survey} for  surveys.
As discussed in the introduction, there are two variants of summarization algorithms -- extractive and abstractive summarization algorithms.
While most classical summarization algorithms were unsupervised, the recent years have seen the proliferation of many supervised neural network-based models for summarization; the reader can refer to~\cite{dong2018survey} for a survey on neural summarization models. To contextualise our work, next we discuss different types of extractive text summarization algorithms in the literature.

\vspace{1 mm}
\noindent
\textbf{Single-Document Summarization:}
Traditional single document extractive summarization deals with extraction of useful information from a single document.
A series of single-document summarization algorithms have been proposed~\cite{GargFRH09, He-dsdr, Erkan:2004, Gong:2001, Luhn:1958, Nenkova05theimpact, nallapati2017summarunner}.
We will describe some of these algorithms in Section~\ref{sec:existing}.
One of the most commonly used class of summarization algorithms is centered around the popular TF-IDF model~\cite{salton1989automatic}. 
%introduced by Salton in 1989. 
Different works have used TF-IDF based similarities for summarization~\cite{Radev:2002,alguliev2011mcmr}. 
Additionally, there has been a series of works where summarization has been treated as a sub-modular optimization problem~\cite{Lin2011,badanidiyuru2014streaming}. 
One of the fair summarization algorithms proposed in this work, is also based on a sub-modular constrained optimization framework, and uses the notion of TF-IDF similarity.

\vspace{1 mm}
\noindent
\textbf{Multi-Document Summarization:}
Multi-document extractive summarization deals with extraction of information from multiple documents (pieces of text) written about the same topic. %Resulting summary allows individual users to quickly familiarize themselves with information contained in a large cluster of documents. 
%Multi-document summarization creates summaries that are both concise and comprehensive with different perspectives being put together for the consumption of the end user. 
For instance, NeATS~\cite{lin2002single} is a multi-document summarization system that, given a collection of newspaper articles as input, generates a summary in three stages -- content selection, filtering, and presentation. Hub/Authority~\cite{zhang2005cue} is another multi-document summarization system which
%firstly detect the sub-topics (in the input document collection) by sentence clustering, and then extract the feature words (or phrase) of different sub-topics. This algorithm
uses the Markov Model to order the sub-topics that the final summary should contain, and then outputs the summary according to the sentence ranking score of all sentences within one sub-topic. 
Generic Relation Extraction (GRE)~\cite{hachey2009multi} is another multi-document text summarization approach, which aims to build systems for relation identification and characterization that can be transferred across domains and tasks without modification of model parameters. 
Celikyilmaz \textit{et al.}~\cite{celikyilmaz2010hybrid} described multi-document summarization as a prediction problem based on a two-phase hybrid model and proposed a hierarchical topic model to discover the topic structures of all sentences. 
Wong \textit{et al.}~\cite{wong2008extractive} proposed
a semi-supervised method for extractive summarization, by
co-training two classifiers iteratively. %to exploit unlabeled data. 
In each iteration, the unlabeled training sentences with top scores are included in the labeled
training set, and the classifiers are trained on the new training data.

\if 0
%The increasing availability of textual information has led to exhaustive research on automatic text summarization. 
%A large number of text summarization algorithms have been proposed in literature; the reader can refer to~\cite{extractive-summarization-survey-gupta,text-summarization-survey} for  surveys.
%One popular track of work considers documents as a network in which sentences are analogous to nodes and their similarity scores can be treated as a notion of weighted edges~\cite{mani1997multi,erkan-lexrank,mihalcea-texrank,wan2007towards,giannakopoulos2008testing,morales2008concept}.
%While most classical summarization algorithms were unsupervised, the recent years have seen the proliferation of many supervised neural network-based models for summarization; the reader can refer to~\cite{dong2018survey} for a survey on neural summarization models. 
%One of the most commonly used class of summarization algorithms is centered around the popular TF-IDF model~\cite{salton1989automatic}. 
%introduced by Salton in 1989. 
%Different works have used TF-IDF based similarities for summarization~\cite{Radev:2002,alguliev2011mcmr}. 
%Additionally, there has been a series of works where summarization has been treated as a sub-modular optimization problem~\cite{Lin2011,badanidiyuru2014streaming}. 
%Algorithm I, proposed in this work, is also based on a sub-modular constrained optimization framework, and uses the notion of TF-IDF similarity . %and solve it using a greedy algorithm proposed by Du et al.~\cite{Du2013}. 
\fi

\vspace{1 mm}
\noindent
\textbf{Summarization of User Generated Text on Social Media:}
With the proliferation of user generated textual content on social media (e.g., Twitter, Facebook), a number of summarization algorithms have been developed specifically for such content. 
For instance, Carenini {\it et al.}~\cite{carenini2007summarizing} proposed a novel summarization algorithm that summarizes e-mail conversations using fragment quotation graph and clue words. 
Nichols {\it et al.}~\cite{nichols2012summarizing} described an algorithm that generates a journalistic summary
of an event using only status updates from Twitter as information source. They used temporal cues to find important moments within an event and a sentence ranking method to extract the most relevant sentences describing the event. 
Rudra \textit{et al.}~\cite{rudra2015extracting} proposed a summarization algorithm for tweets posted during disaster events.
Kim \textit{et al.}~\cite{kim2016storia} used narrative theory as a framework for identifying the links between social media content and designed crowdsourcing tasks to generate summaries of events based on commonly used narrative templates.
Zhang {\it et al.}~\cite{zhang2017wikum} proposed a recursive summarization workflow where they design a summary tree that enables readers to digest the entire abundance of posts. 
Zhang \textit{et al.}~\cite{zhang2018making} developed Tilda, %a prototype system built for Slack that 
which allows participants of a discussion to collectively tag, group, link, and summarize chat messages in a variety of ways, such as by adding emoji reactions to messages or leaving written notes.

%%%%%%%%%%%%%%%%%%%%%%%%%%%%%%%
\subsection{Bias and Fairness in Information Filtering Algorithms}

\vspace{2mm}
\noindent {\bf Bias in applications on user-generated content:} 
Powerful computational resources along with the enormous amount of data from social media sites has driven a growing school of works that uses a combination of machine learning, natural language processing, statistics and network science for decision making. %However, analysis suggest that such forecasts and predictions often misrepresent the entire real world population \cite{tufekci2014big}.
In~\cite{baeza2018bias}, Baeza-Yates has discussed how human perceptions and societal biases creep into social media, and how different algorithms fortify them. 
These observations raise questions of bias in the decisions derived from such analyses. 
Friedman \textit{et al.}~\cite{friedman1996bias} broadly categorized these biases into 3 different classes, and essentially were the first to propose a framework for comprehensive understanding of the biases. 
Several recent works have investigated different types of biases (demographic, ranking, position biases etc.) and their effects on online social media~\cite{bonchi2017exposing,chakraborty2017makes,chakraborty2016dissemination}. %, dupret2008user}.
Our observations in this work show that summaries generated by existing algorithms (which do not consider fairness) can lead to biases towards/against socially salient demographic groups.

\vspace{2mm}
\noindent 
\textbf{Rooney Rule:}
The notion of implicit bias has been an important component in understanding discrimination in activities such as hiring, promotion, and school admissions. Research on implicit bias hypothesizes that when people evaluate others -- e.g., while hiring for a job -- their unconscious biases about membership in particular groups can have an effect on their decision-making, even when they have no deliberate intention to discriminate
against members of these groups. 
To this end, the \textit{Rooney Rule} was proposed hoping to reduce the adverse effects of such implicit biases.
The Rooney Rule is a National Football League policy in the USA, that requires league teams to interview ethnic-minority candidates for head coaching and senior football operation jobs. 
Roughly speaking, it requires that while recruiting for a job opening, one of the candidates interviewed must come from an underrepresented group.
As per~\cite{collins2007tackling}, there are two variants of the Rooney rule. The `soft' affirmative action programs encompass outreach attempts like minority recruitment and counseling etc., while the `hard' affirmative action programs usually include explicit preferences or quotas that reserve a specific number of openings exclusively for members of the preferred group.

In the context of summarization, any summarization algorithm will adhere to the `soft' variant of Rooney Rule, since all the textual units (be it from majority or minority groups) have candidature to enter the summary. However, existing summarization algorithms are {\it not} guaranteed to adhere to the `hard' variant of the Rooney rule. 
The algorithms proposed in this paper (detailed in Sections~\ref{sec:fairsumm} and~\ref{sec:otherMeth}) are guaranteed to also cohere to the `hard' variant of the Rooney Rule since they maintain a specific level of representation of various social groups in the final summary.

\vspace{2mm}
\noindent {\bf Fairness in information filtering algorithms:} Given that information filtering algorithms (search, recommendation, summarization algorithms) have far-reaching social and economic consequences in today's world, fairness and anti-discrimination have been recent inclusions in the algorithm design perspective~\cite{hajian2016algorithmic, dwork2012fairness, kleinberg2018discrimination}. 
There have been several recent works on defining and achieving different notions of fairness~\cite{hardt2016equality, zafar2017fairness,zemel2013learning, kleinberg2019simplicity}
%angwin2016machine,
as well as on removing the existing unfairness from different  methodologies~\cite{hajian2014generalization,zemel2013learning, kleinberg2018selection}. 
Different fairness-aware algorithms have been proposed to achieve group and/or individual fairness for tasks such as clustering~\cite{chierichetti2017fair}, classification~\cite{zafar2017fairness}, ranking~\cite{Zehlike2017}, matching~\cite{suhr2019two},  recommendation~\cite{chakraborty2018equality} and sampling~\cite{celis2016fair}. 
%However, to the best of our knowledge, there has not been prior exploration of fairness in the context of summarization. 
%(which we do in the present work). 
%In this work, we have tried to show the existing discrimination in the already existing summarization algorithms and tried to produce summaries which will satisfy an underlying notion of fairness as per the requirements of a given application.

\vspace{2 mm}
\noindent
To our knowledge, only two prior works have looked into fairness in summarization.
Celis {\it et al.} proposed a methodology to obtain fair and diverse summaries~\cite{fair-diverse-summarization}. 
%There are some points of similarity between this prior work and the present work -- both~\cite{fair-diverse-summarization} and the present work have the same objective, both rewards diversity among the textual units, and both can accommodate various notions of fairness. 
%e.g., a similar goal of producing fair summaries, and our algorithm also rewards diversity among the textual units selected for inclusion in the summary.
%Additionally, both the present work and~\cite{fair-diverse-summarization} can accommodate various notions of fairness.
%However, the methodologies in the two works are very different -- while Celis {\it et al.} employs sampling from determinantal point processes (DPPs)~\cite{fair-diverse-summarization}, we formulate fair summarization as a sub-modular constrained optimization problem.
%Additionally, Celis {\it et al.} performed experiments 
They applied their determinantal point process based algorithm on an image dataset and a categorical dataset (having several attributes), and not on textual data.
The problem of fair {\it text} summarization was first introduced in our prior work~\cite{fair-summ-www18-poster}, which showed that many existing text summarization algorithms do not generate fair summaries; however, no algorithm for fair summarization was proposed in~\cite{fair-summ-www18-poster}.
To our knowledge, ours is the first work to propose algorithms for fair summarization of textual data.

%\vspace{-2 mm}
\section{Datasets Used}
\label{sub:datasets}

\noindent Since our focus in this paper is to understand the need for fairness while summarizing user-generated content, we consider  
datasets containing %textual units  
tweets posted by different groups of users, e.g. different gender groups, or groups of users with different political leanings. 
Specifically, we use the following three  datasets throughout this paper.

\vspace{2mm}
\noindent \underline{(1) {\bf Claritin dataset}}: 
Patients undergoing medication often post the consequences of using different drugs on social media, especially highlighting the side-effects they endure~\cite{o2014pharmacovigilance}. 
Claritin (loratadine) is an %antihistamine
anti-allergic drug that reduces the effects of natural chemical histamine in the body, which can produce symptoms of sneezing, itching, watery eyes and runny nose. However, this drug may also have some adverse effects on the patients. %Given the consequences of the usage of drugs, many drugs are withdrawn from the market due to safety concerns as a result of serious adverse events. 

To understand the sentiments of people towards Claritin and different side-effects caused by it, tweets posted by users about Claritin were collected, analyzed and later publicly released by `Figure Eight' (erstwhile CrowdFlower). This dataset contains tweets in English about the effects of the drug. Each tweet is annotated with the gender of the user (male/female/unknown) posting it~\cite{claritin-dataset}. %\footnote{The dataset is described in detail at {\it https://www.figure-eight.com/discovering-drug-side-effects-with-crowdsourcing}.}.  
%In this dataset, we consider that the sensitive attribute is the gender of the user who posted a tweet.
Initial analyses on these tweets reveal that women mentioned some serious side effects of the drug (e.g. heart palpitations, shortness of breathe, headaches) while men did not~\cite{claritin-dataset}. 
%This can be due to how different genders react to the doses of claritin. 
From this dataset, we ignored those tweets for which the gender of the user is unknown. 
We also removed exact duplicate tweets, since they do not have any meaningful role in summarization. 
Finally, we have $4,037$ tweets in total, of which $1,532$ ($37.95\%$) are written by men, and $2,505$ ($62.05\%$) by women.

\vspace{2mm}
\noindent \underline{(2) {\bf US-Election dataset}:}
This dataset consists of tweets related to the 2016 US Presidential Election collected by the website TweetElect (\url{https://badrit.com/work/Tweetelect}) during the period from  September 1, 2016 to
November 8, 2016 (the election day)~\cite{us-election-dataset}.  
%TweetElect is a public website that aggregated tweets relevant to the 2016 US presidential election.
%The site uses state-of-the-art adaptive filtering methods for detecting relevant tweets on broad and dynamic topics, such as politics and elections~\cite{magdy2014adaptive, magdy2016unsupervised}. 
TweetElect used an initial set of 38 keywords related to the election (including all candidate names and common hashtags about the participating parties) for filtering relevant tweets. %Consequently,
Subsequently, state-of-the-art adaptive filtering methods 
%filtering continuously enriches
were used to expand the set of keywords with additional terms that emerged over time~\cite{magdy2016unsupervised}, and their related tweets were added to the collection.
%The 38 seeding keywords included .
%This dataset, provided by Darwish {\it et al.}, contains 
%English tweets posted during the 2016 US Presidential election. 

In this dataset released by Darwish {\it et al.}~\cite{us-election-dataset}, each tweet is annotated as supporting or attacking one of the presidential candidates (Donald Trump 
and Hillary Clinton) or neutral or attacking both. 
For simplicity, we grouped the tweets into three classes: 
(i)~{\it Pro-Republican}: tweets which support Trump and / or attack Clinton, 
(ii)~{\it Pro-Democratic}: tweets which support Clinton and / or attack Trump, and
(iii)~{\it Neutral}: tweets which are neutral or attack both candidates.
%The number of tweets in the different classes is reported in Table~\ref{tab:summ-results-uselection} (first row).
After removing duplicates, we have $2,120$ tweets, out of which $1,309$ ($61.74\%$) are Pro-Republican, $658$ ($31.04\%$) tweets are Pro-Democratic, and remaining $153$ ($7.22\%$) are Neutral tweets.
%After removing duplicates, we have $2,120$ tweets, out of which $1,309$ are Pro-Republican, $658$ are Pro-Democratic, and $153$ are Neutral tweets.
%\new{Note that prior works have shown that the political leaning of a tweet often corresponds to the political leaning of the user who posted the tweet~\cite{search-bias-cscw17}, hence the tweets in this dataset can be thought as representing the opinions of the three social groups -- pro-Republican, pro-Democratic and neutral.}\todo{do we include the previous sentence?}

\vspace{2mm}
\noindent \underline{(3) {\bf MeToo dataset}:}
%\textcolor{red}{AC: Mention when the tweets were collected. Put reference to Search API.}
We collected a set of tweets related to the %currently ongoing 
\#MeToo movement in October 2018. %, a topic where the opinions of male users can potentially be very different from that of female users.  
We initially collected $10,000$ English tweets containing the hashtag \#MeToo using the Twitter Search API~\cite{SearchAPI}. 
After removing duplicates, we were left with $3,982$ distinct tweets. 
We asked three human annotators to examine the name and bio of the Twitter accounts who posted the tweets. 
The annotators observed %that there are 
three classes of tweets based on who posted the tweets -- tweets posted by male users, tweets posted by female users, and tweets posted by organizations (mainly news media agencies). Also, there were many tweets for which the annotators could not understand the type/gender of the user posting the tweet.
For purpose of this study, we decided to focus only on those tweets for which all the annotators were certain that they were written by men or women. 
In total, we had $488$ such tweets, out of which $213$ are written by men and $275$ are written by women.

\vspace{1 mm}
\noindent In summary, two of our datasets contain tweets posted by two social groups (men and women) which the other dataset contains three categories of tweets (pro-democratic, pro-republican and neutral tweets, presumably written by users having the corresponding political leanings).

\vspace{2mm}
\noindent \underline{\bf Human-generated summaries for evaluation:} The traditional way of evaluating the `goodness' of a summary is to match it with one or more human-generated summaries (gold standard), and then compute ROUGE scores~\cite{lin-rouge}.
ROUGE scores are between $[0,1]$, where a higher ROUGE score means a better algorithmic summary that has higher levels of `similarity' with the gold standard summaries. 
Specifically, the similarity is computed in terms of common unigrams (in case of ROUGE-1) or common bigrams (in case of ROUGE-2) between the algorithmic summary and the human-generated summaries. 
For creating the gold standard summaries, we asked three human annotators to summarize the datasets. Each annotator is well-versed with the use of social media like Twitter, is fluent in English, and none is an author of this paper. 
The annotators were asked to generate extractive summaries %of length 100 tweets, 
independently, i.e., without consulting one another. We use these three human-generated summaries for the evaluation of algorithmically-generated summaries, by computing the average ROUGE-1 and ROUGE-2 Recall and $F_1$ scores~\cite{lin-rouge}.

\section{Why do we need fair summaries?}
\label{sec:background}

%\noindent 
Traditionally, summarization algorithms have only considered including (in the summary) those textual units (tweets, in our case) whose contents are most `summary-worthy'.
In contrast, in this paper, we argue for giving a fair chance to textual units written by different social groups to appear in the summary. 
Before making this argument, two questions need to be investigated --\\
(1)~Are the tweets written by different social groups of comparable textual quality? If not, someone may argue for discarding lower quality tweets generated by a specific user group. \\
(2)~Do the tweets written by different social groups actually reflect different opinions? This question is important since, if the opinions of the different groups are not different, then it can be argued that selecting tweets of any group (for inclusion in the summary) is %equivalent.\\
sufficient.\\ 
We attempt to answer these two questions in this section.

\subsection{Are tweets written by different social groups of comparable quality?}

We use three measures for estimating the textual quality of individual tweets.
(i)~First, the NAVA words (nouns, adjectives, verbs, adverbs) are known to be the most informative words in an English text~\cite{wordnet-miller}.
Hence we consider the count of NAVA words in a tweet as a measure of its textual quality. 
We consider two other measures of textual quality that are specific to the application of text summarization -- (ii)~ROUGE-1 precision and (iii)~ROUGE-2 precision scores. Put simply, the ROUGE-1 (ROUGE-2) precision score of a tweet measures what fraction of the unigrams (bigrams) in the tweet appears in the gold standard summaries for the corresponding dataset (as described in Section~\ref{sub:datasets}). 
Thus, these scores specifically measure the utility of selecting a particular tweet for inclusion in the summary.

\begin{figure}[tb]
	\centering
	\begin{subfigure}{.48\columnwidth}
		\centering
		\includegraphics[width=\textwidth, height=2.5cm]{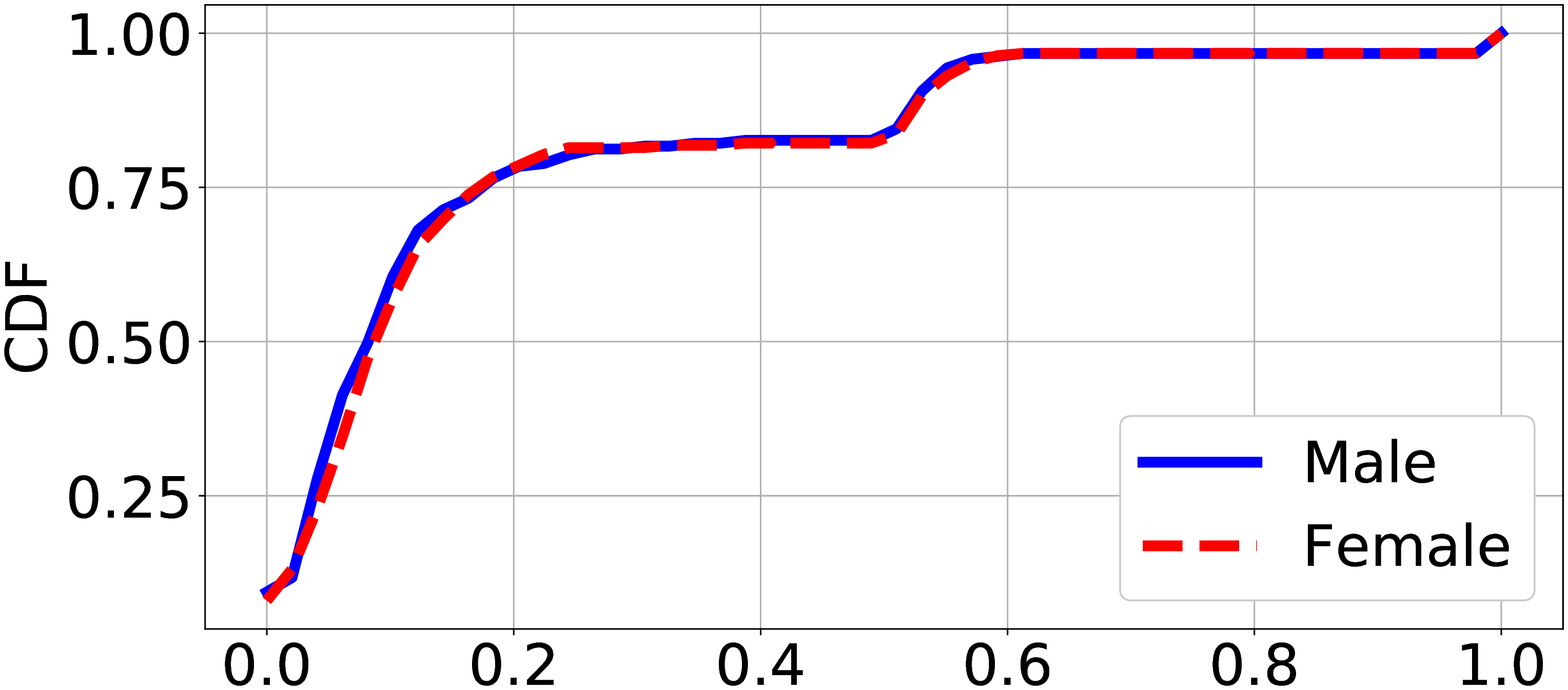}
		\caption{ROUGE-2 Precision}
		\label{fig:R2_Metoo}
	\end{subfigure}%
	\hfill
	~\begin{subfigure}{.48\columnwidth}
		\centering
		\includegraphics[width=\textwidth, height=2.5cm]{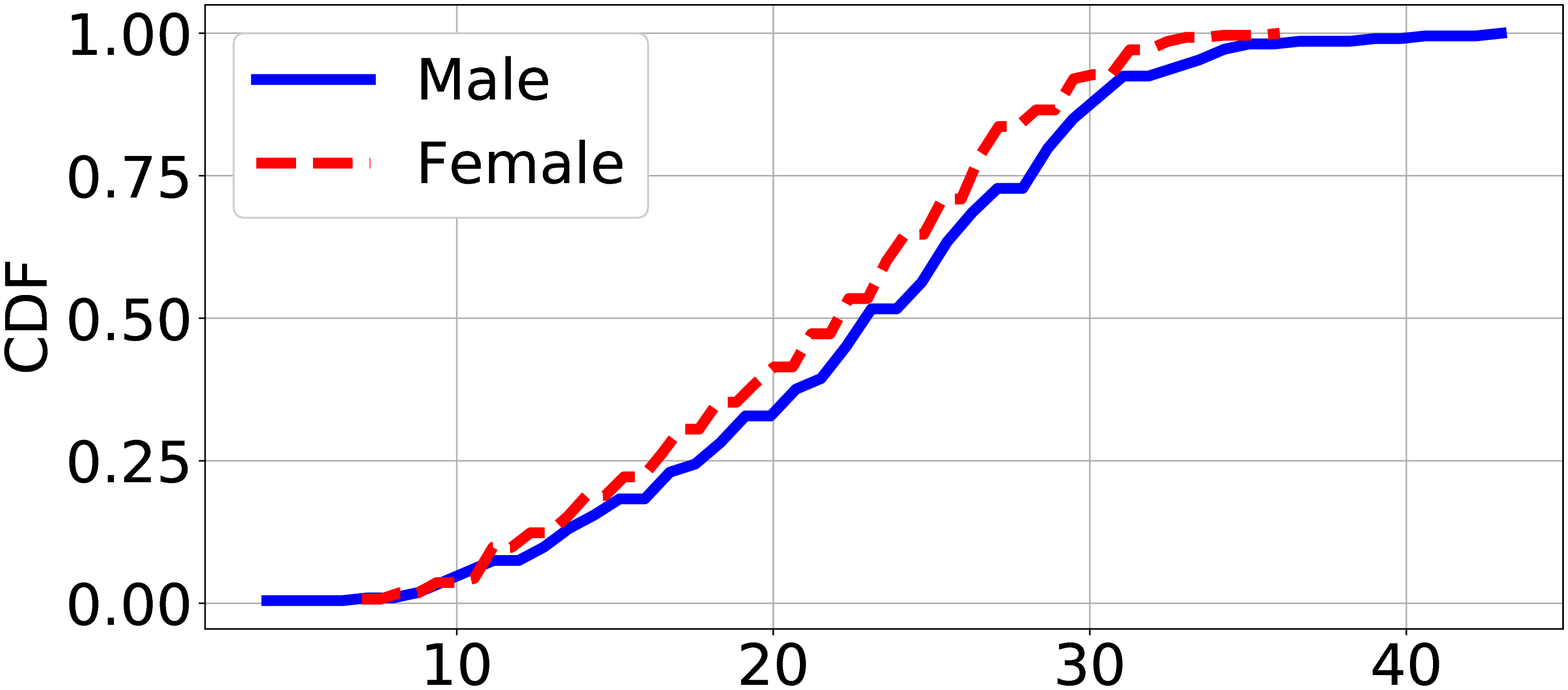}
		\caption{NAVA word count}
		\label{fig:NAVA_Metoo}
	\end{subfigure}%
	\vspace{-3mm}
	\caption{\textbf{Comparing textual quality of individual tweets of the two user groups in MeToo dataset -- distributions of (a)~ROUGE-2 Precision scores and (b)~Count of NAVA words, of individual tweets.}}
	\label{fig:similarity_metoo}
	\vspace*{-2mm}
\end{figure}

\begin{figure}[tb]
	\centering
	\begin{subfigure}{.48\columnwidth}
		\centering		\includegraphics[width=\textwidth, height=2.5cm]{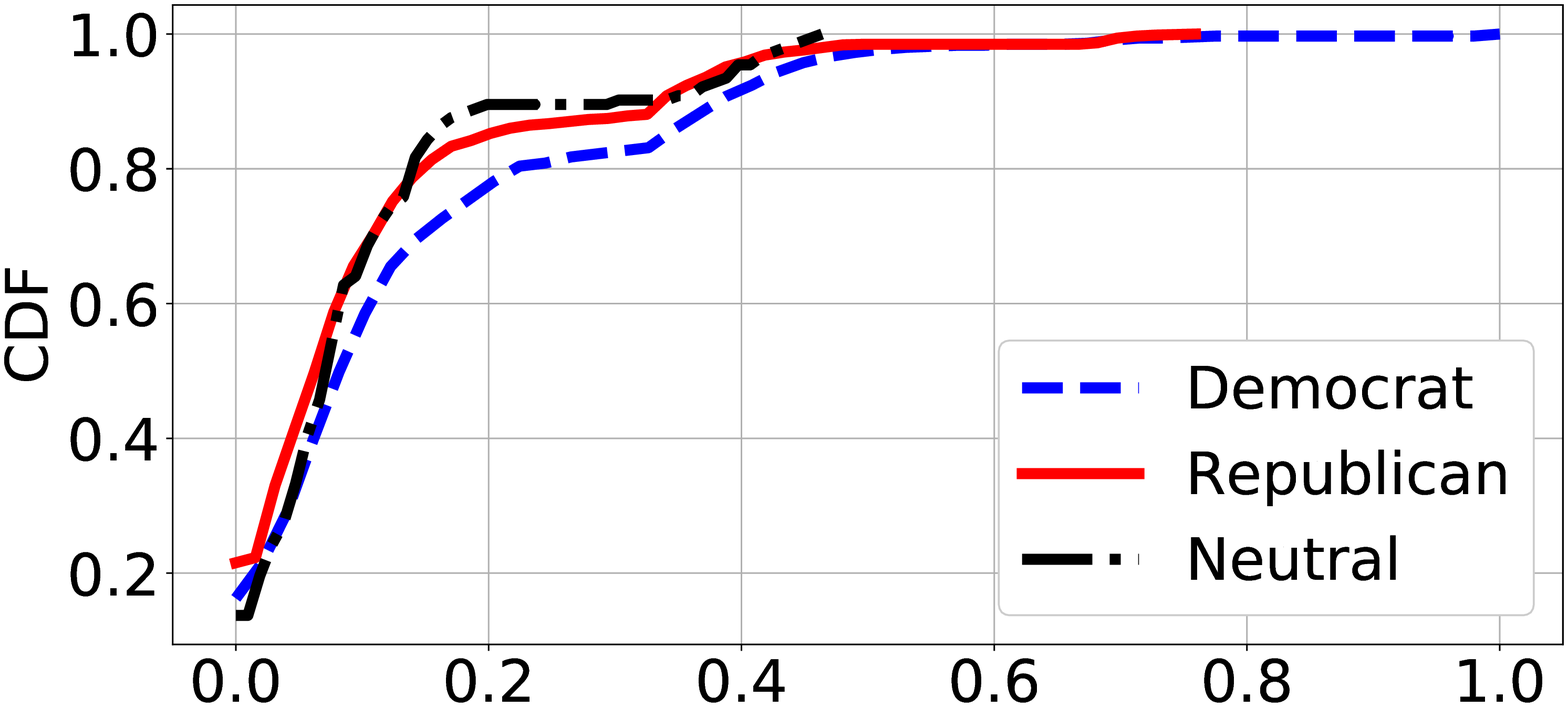}
		\caption{ROUGE-2 Precision}
		\label{fig:R2_USE}
	\end{subfigure}%
	\hfill
	~\begin{subfigure}{.48\columnwidth}
		\centering
	\includegraphics[width=\textwidth, height=2.5cm]{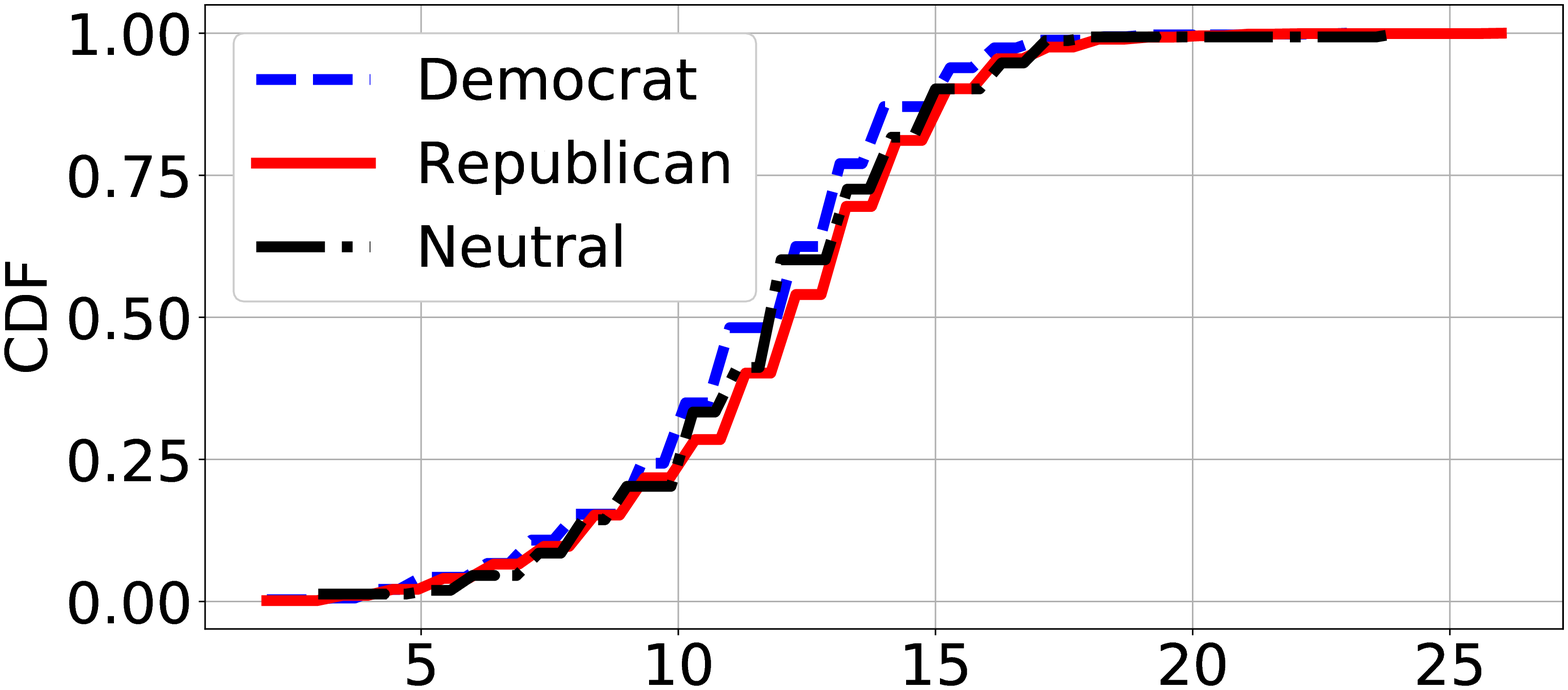}
		\caption{NAVA word count}
		\label{fig:NAVA_USE}
	\end{subfigure}%
	\vspace{-3mm}
	\caption{\textbf{Comparing textual quality of individual tweets of the three groups in US-Election dataset --  distributions of (a)~ROUGE-2 Precision scores and (b)~Count of NAVA words.}}
	\label{fig:similarity_use}
	\vspace*{-2mm}
\end{figure}

For a particular dataset, we compare the distributions of the three scores -- ROUGE-1 precision score, ROUGE-2 precision score, and count of NAVA words -- for the subsets of tweets written by different user groups.
For all cases, we found that ROUGE-1 precision scores and ROUGE-2 precision scores show similar trends; hence we report only the ROUGE-2 precision scores. 
%Figure~\ref{fig:similarity_claritin}(a) and Figure~\ref{fig:similarity_claritin}(b) respectively compare the distributions of ROUGE-2 precision scores and NAVA word counts among the tweets written by male and female users in the Claritin dataset. We find that the distributions are very close to each other, thus implying that the tweets written by both groups are of comparable textual quality.
Figure~\ref{fig:similarity_metoo}(a) and Figure~\ref{fig:similarity_metoo}(b) respectively compare the distributions of ROUGE-2 precision scores and NAVA word counts among the tweets written by male and female users in the MeToo dataset. We find that the distributions are very close to each other, thus implying that the tweets written by both groups are of comparable textual quality. 
%The mean ROUGE-1 precision scores are $0.55$ and $0.56$ for the tweets written by male and female users respectively, while the mean ROUGE-2 precision scores are $0.17$ (for male) and $0.18$ (for female).
Similarly, Figure~\ref{fig:similarity_use} shows that, in the US-Election dataset, the pro-democratic, pro-republican and neutral tweets are of comparable textual quality.
The textual quality of the tweets written by male and female users in the Claritin dataset are also very similar -- the mean number of NAVA words are $8.19$ and $8.61$ respectively for tweets written by male and female users, while the mean ROUGE-2 Precision scores are $0.22$ for male and $0.20$ for female (detailed results omitted for brevity). 
All these values show that the textual quality is very similar for the different groups of tweets, across all the three datasets.

%\new{Table~\ref{tab:QualityAnalysis} shows the mean and median values for the three textual quality measures (number of NAVA words, ROUGE-1 precision score, and ROUGE-2 precision score), for the tweets of the different groups in all three datasets. For all datasets, the values for the mean and median are very similar, thus showing that the tweets written by different groups of users are of comparable textual quality.}

%%%%%%%% FIGURE COMMENTED OUT %%%%%%
\if 0
\begin{figure}[tb]
	\centering
	\begin{subfigure}{.48\columnwidth}
		\centering
		\includegraphics[width=\textwidth, height=3cm]{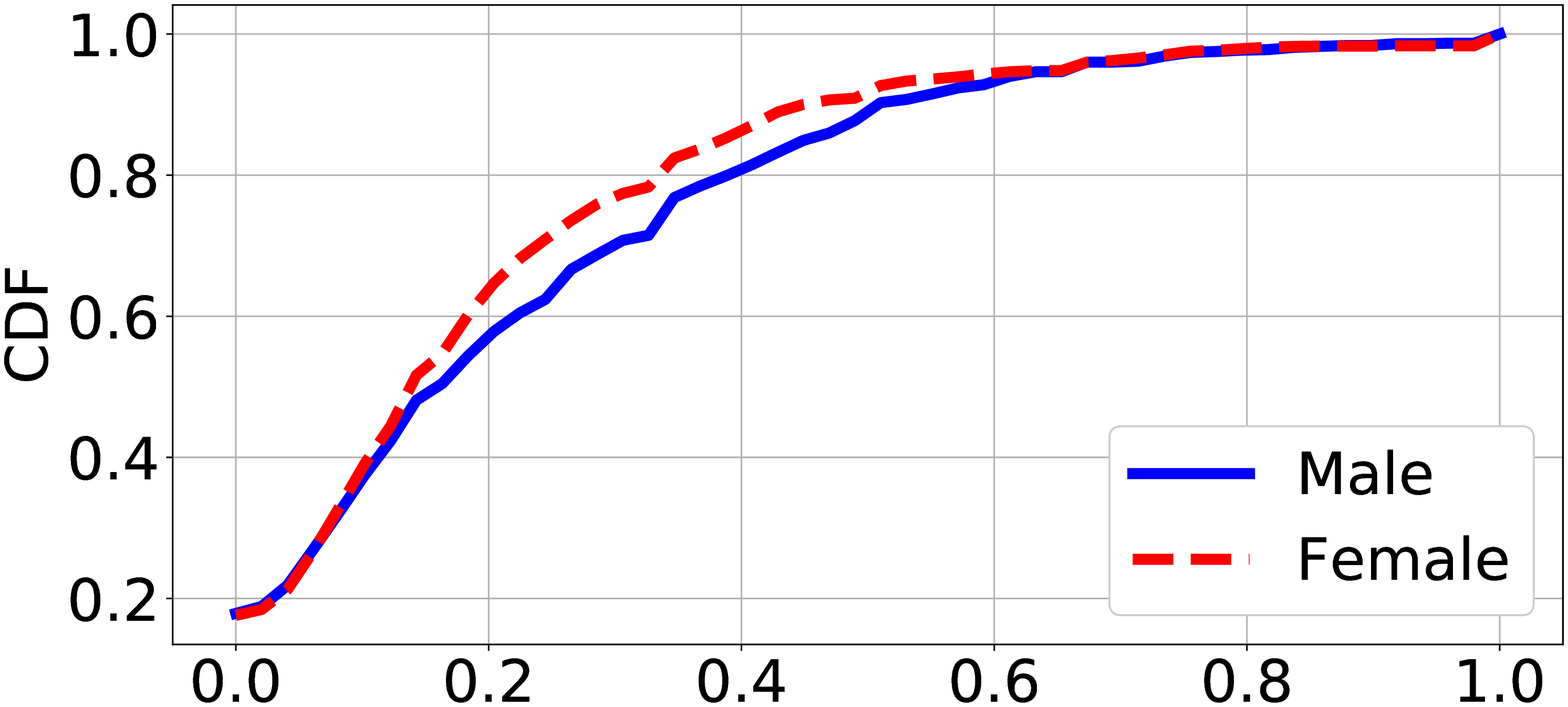}
		\caption{ROUGE-2 Precision of individual tweets}
		\label{fig:R2_Claritin}
	\end{subfigure}%
	\hfill
	~\begin{subfigure}{.48\columnwidth}
		\centering
		\includegraphics[width=\textwidth, height=3cm]{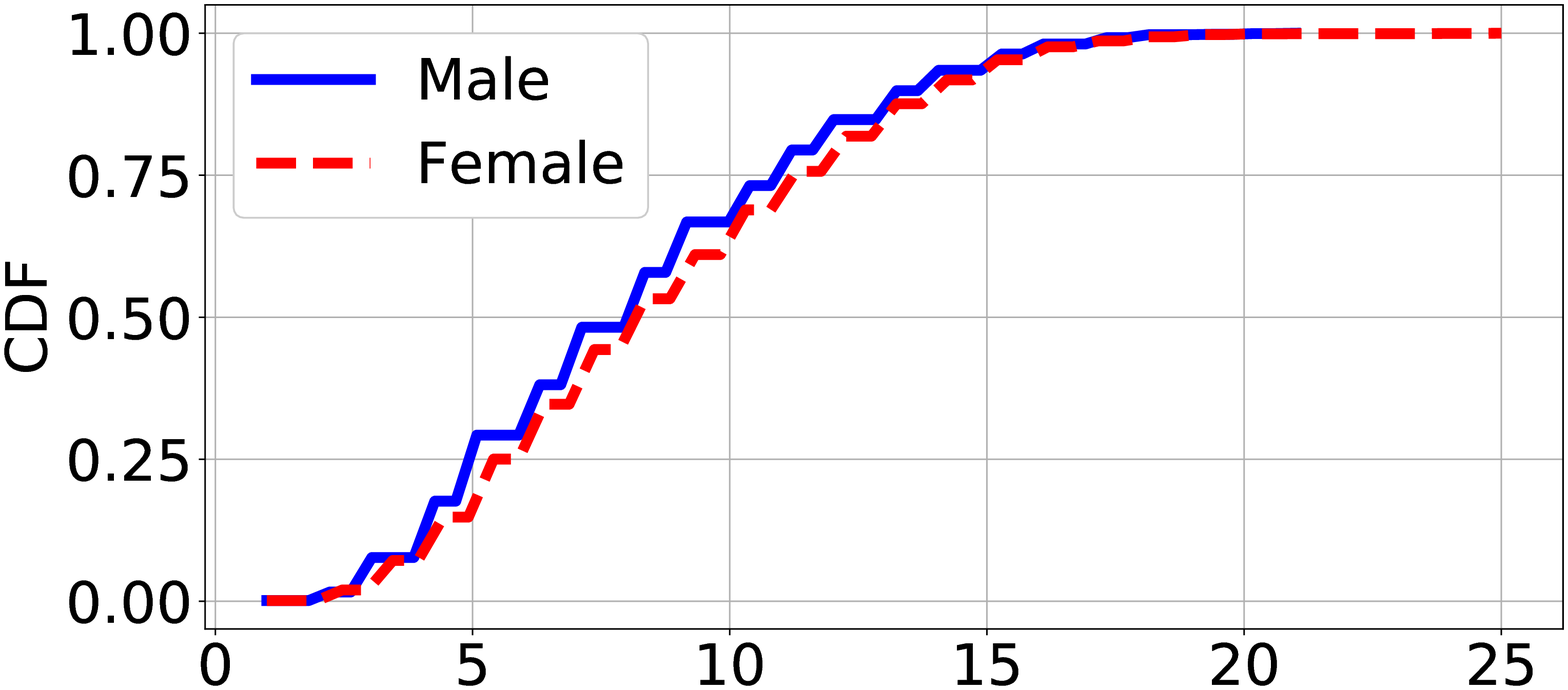}
		\caption{NAVA word count of individual tweets}
		\label{fig:NAVA_Claritin}
	\end{subfigure}%
	\vspace{-3mm}
	\caption{\textbf{Comparing textual quality of the tweets of the two groups in Claritin dataset -- comparing distributions of (a)~ROUGE-2 Precision scores and (b)~Count of NAVA words, of individual tweets.}}
	\label{fig:similarity_claritin}
	\vspace*{-5mm}
\end{figure}
\fi
%%%%%%%% FIGURE COMMENTED OUT %%%%%%

%%%%%%%%% TABLE COMMENTED OUT %%%%%%%%%

\if 0

\begin{table}
\small
%\footnotesize
%\tiny
\centering
\begin{tabular}{|>{\centering}p{1.1 cm}||p{0.6 cm}|p{0.7 cm}||p{0.6 cm}|p{0.7 cm}||p{0.6 cm}|p{0.7 cm}|}
\hline
Group & \multicolumn{2}{c||}{\#NAVA Words}  & \multicolumn{2}{c|}{ROUGE-1 P} & \multicolumn{2}{c|}{ROUGE-2 P}
\\ \cline{2-7}
 & Mean & Median & Mean & Median & Mean & Median
\\ \hline \hline
\multicolumn{7}{|c|}{{\bf Claritin Dataset}} 
\\ \hline
Male & 8.19  & 8  & 0.60  &  0.61    &   0.22 & 0.15
\\ \hline
Female & 8.61  & 8  & 0.59  &  0.58    &   0.20 & 0.14
\\ \hline
\multicolumn{7}{|c|}{{\bf MeToo Dataset}} 
\\ \hline
Male & 22.54  & 23  &  0.55 &  0.53    &  0.17    & 0.08
\\ \hline
Female & 21.14  & 22  & 0.56  & 0.55     &  0.18  & 0.09
\\ \hline
\multicolumn{7}{|c|}{{\bf US-Election Dataset}} 
\\ \hline
Republican & 11.85  & 12  & 0.50  & 0.51 &  0.10 & 0.06
\\ \hline
Democrat & 11.37 & 12  & 0.56  &  0.57 &  0.10 & 0.08
\\ \hline 
Neutral & 11.79 & 12 & 0.50 & 0.51 & 0.13 & 0.07
\\ \hline
\end{tabular}
\caption{Comparison of mean and median values of the textual quality measures of tweets of different groups, in the three datasets. The textual quality of tweets of the different groups are very similar in all datasets.
}
\label{tab:QualityAnalysis}
\vspace*{-5mm}
\end{table}

\fi
%%%%%%%%% TABLE COMMENTED OUT %%%%%%%%%

%\input{quality}
\begin{table*}[tb]
	\small
	\centering
%	\begin{tabular}{ |>{\centering\arraybackslash}p{6.5 cm}| >{\centering\arraybackslash}p{6.5 cm}| }
	\begin{tabular}{ |p{6.5 cm}|p{6.5 cm}| }
		\hline
		\multicolumn{1}{|c|}{\textbf{Tweets on \#MeToo from male users}} & \multicolumn{1}{|c|}{\textbf{Tweets on \#MeToo from female users}} \\
		\hline\hline
		If a woman shares a \textbf{\#metoo} without evidence, it`s taken to be true coz it`s a women`s testimony, a man coming out with \textbf{\#HeToo} story, people would be doubtful,  \& question the evidences, the intent \& will never except the man as victim. \textbf{\#misandry} must be understood. \textbf{\#SpeakUpMan} & If a woman is unveiled it gives a man the right 2 demand sexual favors.When it comes 2 sexual harassment in Islamic Republic it is always your fault if U dont wear hijab. Women using camera to expose sexual harassment. \textbf{\#MyCameraIsMyWeapon} is like  \textbf{\#MeToo} movement in Iran \\
	%	\hline
	%	Woman assaulted security guard \& put the blame on him. Policeman visited the lobby of the housing society. The women went out of control and stripped in front on policemen and security guard to run away from the situation. \#MeghaSharma \textbf{\#HimToo \#MeToo} & Since my \textbf{\#metoo} moment:  * Someone tried to unsuccessfully shut down my Shopify stores. * Someone unsuccessfully reported my herbal business to Health Canada. *One abuser had a lawyer unsuccessfully send me a threatening cease \& desist letter. * Someone hacked my FB \& messages. \\
		\hline
		Instead of arresting this women @CPMumbaiPolice taking common man coz its \textbf{\#MeToo \#MeTooIndia  \#MeToo4Publicity} This is why \textbf{\#FeminismIsCancer  \#feminismIsMisandry \#CrimeByWomen} & Whatever happens to you in your life, you always have the choice to rise above your challenges. Choose NOT to be a victim. \textbf{\#feminism \#metoo} \\
		\hline
		Pain knows no gender. When it hurts, it hurts equally, whether its a man or woman. Why there is discrimination on Gender. Every person deserves dignified treatment and happy life. \textbf{\#MeToo  \#MeToo4Publicity} & ONLY 40 charges and thousands of cries for help. Too many are victim to \textbf{\#UberRape} and their voices aren`t being heard. \textbf{\#TimesUp \#Metoo} \\
%		\hline
%	Don't get fooled by \textbf{\#MeToo gang} . \textbf{\#SpeakUpMan} to Stop  \#MeToo4Publicity Don't get fooled by \textbf{\#MeToo gang} . \textbf{\#SpeakUpMan} to Stop abuse of Men and Demand Gender Neutral law. If Women do not do such crime , why they afraid for Gender neutral Law? &\textbf{\#MeToo} Pls empower the children in your houses to speak up. Let us discuss the evil in society if we want to tackle them! Increase collective awareness and reduce chances of fatalities. Better policies - At work, home, schools, Panchayat offices! \textbf{\#TimesUp \#JusticeForRajalakshmi} \\
		\hline
		When Settlement amount is the motive by falsely charging a man' it's called \textbf{\#MeToo}  Pls tk action on ppl filing \textbf{\#FakeCases} \& bring {\bf \#GenderNeutralLaws \#MeToo4publicity \#MensCommission.} & A long term solution would be the exact opposite of the two suggested here - gender sensitisation, not segregation so that exchange between different genders is normalised instead of being stigmatised further. {\bf \#MeToo} \\
		\hline
	\end{tabular}
	\caption{{\bf Example tweets containing the hashtags that are most frequently posted by male and female users, in the MeToo dataset. Even though all tweets have high textual quality, the opinions expressed by the two groups of users are quite diverse.}}
	\label{tab:sample-metoo-tweets}
	\vspace*{-5mm}
\end{table*}

%\vspace{2mm}
%\noindent 
%{\bf Do tweets written by different user groups reflect different opinion?}
\subsection{Do tweets written by different user groups reflect different opinion?}
To answer this question, we asked our human annotators (those who prepared the gold standard summaries) to observe the tweets written by different user groups in the datasets. %three datasets.
For all three datasets, the annotators observed that the tweets posted by different social groups mostly contain very different information/opinion.

For instance, Table~\ref{tab:sample-metoo-tweets} shows some sample tweets written by male and female users in the MeToo dataset, along with some of the hashtags that are frequently posted by male and female users (highlighted). 
We observe that most tweets written by women support the \#MeToo movement, and give examples of relevant experiences of themselves or of other women. On the other hand, many of the tweets written by male users point out undesirable side-effects of the movement, and call for gender equality.

Similarly, in the US-Election dataset, the pro-republican tweets criticize Hillary Clinton and/or support the policies of Donald Trump (e.g., `{\it We must not let \#CrookedHillary take her criminal scheme into the Oval Office. \#DrainTheSwamp}'), while the pro-democratic tweets have the opposite opinion (e.g. `{\it Yes America. This is the election where Hillary's cough gets more furious coverage than Trump asking people to shoot her \#InterrogateTrump}'). 
The neutral tweets either give only information (and no opinion), or criticize both Clinton and Trump.
For the Claritin dataset as well, there is large difference in opinion among the tweets written by male and female users -- the female users criticize the drug much more than the male users (details omitted for brevity). 
Thus, it is clear that tweets posted by different social groups often reflect very different opinions.

\subsection{Need for fairness in summarization}
The fact that tweets written by different social groups are of very similar quality/merit implies that all groups should have `equality of opportunity'~\cite{roemer2009equality} for their opinions to be reflected in the summary. 
This fact, coupled with the diversity in opinion of the different groups, calls for a fair representation of the opinions of different groups in the summary.
%The need for fairness in summaries of crowdsourced data 
This is similar in spirit to the need for fairness in top crowdsourced recommendations~\cite{chakraborty2018equality} or top search results~\cite{Biega-fair-rank-sigir18}.
Since the tweets that get included in the summary are likely to get much more exposure than the rest of the information (just like how top search and recommendation results get much more exposure~\cite{chakraborty2018equality,Biega-fair-rank-sigir18}), under-representation of any of the social groups in the summary can severely suppress their opinion. 
These factors advocate the need for fair summaries when data %crowdsourced from
generated by various social groups is being summarized.

\section{Notions of Fair Summarization}\label{sec: notions}
%\vspace{-2mm}
%\section{Fairness notions for summarization} 
\label{sec:fairness-notions}
%\noindent \todo{This section needs to be improved. I feel the notions need to be explained first in general, and then it needs to be said what they mean in the context of summarization. -- Saptarshi}

Having established the need for fair summarization, we now define %some appropriate 
two fairness notions that are applicable in the context of summarization. Essentially, when the input data (e.g. tweets) are generated by users belonging to different social groups, we require the summaries to {\it fairly represent} these groups. Next, we consider two notions for fairness in representation.
%There are many notions of fairness that are prevalent in the fairness literature. Among those we have considered the following notions of fairness for fair summarization in this paper.

\subsection{Equal Representation}
%Let us consider the scenario of selecting a set of representatives for a population. The ideal set of representatives will be the one which has similar representation from all the socially salient groups that constitutes the population.
%To this end, this notion of fairness implies that the number of representatives from the different classes in the society has to be similar in the final selected set. 
%Equal representation is argued to be necessary to reverse the effect of historical biases~\cite{optimal-apportionment} that have been dominated by the majority class historically. 
The notion of equality finds its roots in the field of morality and justice, which advocates for the redress of undeserved inequalities (e.g. inequalities of birth %such as gender and race, and inequalities 
or due to natural endowment) 
%e.g. pro-republican, pro-democrat etc.)
~\cite{rawls2009theory}.
%Equality in its prescriptive usage is closely related with morality and justice. 
Formal equality suggests that when two people or two groups of people have equal status in at least one normatively relevant aspect, they must be treated equally~\cite{sep-equality}. 
%To this end, this notion of fairness i
In terms of selection, equal representation requires that the number of representatives from different classes in the society having comparable relevance has to be equal. 
%Equal representation is argued to be necessary to reverse the effect of historical biases~\cite{optimal-apportionment} that have been dominated by the majority class historically.

In the context of user-generated content, we observed that different sections of the society have different opinion on the same topic, either because of their gender or ideological leaning~\cite{babaei2018purple}. 
%For instance, as described in the previous section, females have significantly different opinion on the MeToo movement than men. Similarly, people having different political ideologies (e.g., democrats and republicans) hold very different opinion on many socio-political issues. 
However, if we consider the textual quality, i.e. their candidature for inclusion in the summary, then tweets from both the groups are comparable (as discussed in section~\ref{sec:background}). 
%To this end, an algorithm for summarizi user-generated content covering the entire ideology spectrum, needs to include opinions from different horizons in the final summary.
Thus, the notion of equal representation requires that 
%is the simplest notion of fairness. A
a summarization algorithm will be fair if different groups generating the input data are represented equally in the  output summary. Given the usefulness of summaries in many downstream applications, this notion of fairness ensures equal exposure to the opinions of different socially salient groups.

\subsection{Proportional Representation} 
Often it may not be possible to equally  represent different user groups in the summary, especially if the input data contains very different proportions from different groups. 
Hence, we consider another notion of fairness: {\it Proportional Representation} (also known as \textit{Statistical Parity}~\cite{luong2011k}). Proportional representation requires that the representation of different groups in the selected set should be proportional to their distribution in the input data. 

In certain scenarios such as hiring for jobs, relaxations of this notion are often used. For instance, the U.S. Equal Employment Opportunity Commission uses a variant of Proportional Representation to determine whether a company's hiring policy is biased against (has any adverse impact on) a demographic group~\cite{biddle-adverse-impact}.
According to this policy, a particular class $c$ is {\bf under-represented} in the selected set (or {\bf adversely impacted}), if the fraction of selected people belonging to class $c$ is less than $80\%$ of the fraction of selected people from the class %which has
having the highest selection rate. 

In the context of summarization, Proportional Representation requires that the proportion of content from different user groups in the summary should be  same as in the original input. A relaxed notion of proportional fairness is one which would ensure {\it no adverse impact} in the generated summary. 
In other words, `no adverse impact'  requires that the fraction of textual units from any class, that is selected for inclusion in the summary, {\it should not be} less than $80\%$ of the fraction of selected units from the class having the highest selection rate (in the summary).
These notions of fairness ensure that the probability of selecting an item is \textbf{independent} of which user group generated it.

%\vspace{2mm}
%\noindent {\bf Proportional Representation:} Often it may not be possible to satisfy equal representation of different classes in the summary, especially if the input data itself has very different proportions from the different classes. 
%The notion of {\it Proportional Representation} requires that the representation of different classes in the summary should be proportional to their distribution in the input data.
%\vspace{1mm}
%\noindent {\bf No Adverse Impact:}

\vspace{2mm}
\noindent 
%Note that the only prior work on fair summarization~\cite{fair-diverse-summarization} also considers the fairness notions of proportional representation and equal representation.
It should be noted that, we are {\it not} advocating for any particular notion of fairness to be better in the context of summarization. We also note that different applications may require different types of fairness. Hence, in this work, we propose mechanisms that can accommodate different notions of fairness, including the ones stated above, and produce fair summaries accordingly.

%\vspace{-2 mm}
\section{Do existing algorithms produce fair summaries?} \label{sec:existing}

Having discussed the need for fair summarization, we now check whether existing algorithms generate fair summaries. 
%Specifically, we consider extractive summarization algorithms, that select a subset of the textual units (tweets) for inclusion in the summary.

\subsection{Summarization algorithms}
%\vspace{2mm}
%\noindent {\bf Summarization algorithms:}

\noindent We consider a set of well-known extractive summarization algorithms, that select a subset of the textual units for inclusion in the summary. Some of the methods are unsupervised (the traditional methods) and some are recent supervised neural models.

\vspace{1mm}
\noindent \underline{\bf Unsupervised summarization algorithms:}
We consider six well-known summarization algorithms. 
These algorithms generally estimate an importance score for each textual unit (sentence / tweet) in the input,
and the $k$ textual units having the highest importance scores are selected to generate a summary of length $k$. \\
 %We consider the following algorithms: 
\noindent {\bf (1)~Cluster-rank}~\cite{GargFRH09}  %presents an extension of the TextRank algorithm~\cite{mihalcea-texrank} that 
which clusters the textual units to form a cluster-graph, and uses 
graph algorithms (e.g., PageRank) to compute the importance of each unit. \\
%which are then included in the summary in the decreasing order of this score.
%segments the input textual units into clusters and uses
%the clusters to construct a cluster-graph. 
%Then PageRank is used to compute an `importance score' for each cluster, and
%a centroid-based approach is used to score each unit within a cluster. 
%Text units are included in the summary in the decreasing order of this score. %, until the length constraint is satisfied. 
\noindent {\bf (2)~DSDR}~\cite{He-dsdr}  which
measures the relationship between the textual units using linear combinations and reconstructions,
and generates the summary by minimizing the reconstruction error. \\
%two objective functions --
%(i)~linear reconstruction, which approximates the document by linear
%combinations of selected units, and (ii)~non-negative linear reconstruction, which allows only additive linear combinations. 
%The summary is generated by minimizing the reconstruction error.
%\noindent {\bf (3)~Frequency Summarizer}~\cite{freqsum}, %which  assumes
%% This algorithm  attempts to extract those sentences which cover the main topics of a given document.
%which works on the simple idea 
%that if a textual unit contains the most frequent
%words, it is likely to be a good candidate for including in the summary.\\
%\footnote{Implementation obtained from \url{http://glowingpython.blogspot.in/2014/09/text-summarization-with-nltk.html}.} 
\noindent {\bf (3)~LexRank}~\cite{Erkan:2004}, which creates a graph representation based on similarity of the units,
where edges are placed depending on the intra-unit cosine similarity, and then computes the importance of textual units using eigenvector centrality on this graph. \\
%In this model, a connectivity matrix based on intra-sentence cosine similarity is used as the adjacency matrix of the graph representation of sentences.
%The units are then included in the summary in decreasing order of their importance.
\noindent {\bf (4)~LSA}~\cite{Gong:2001}, which  %decomposes the given document  into individual sentences.
%and uses these sentences to form a candidate set of sentences. 
%It then 
constructs a terms-by-units matrix, and estimates the importance of the textual units based on Singular Value Decomposition on the matrix. \\
%to obtain the singular value
%matrix (where each sentence is represented by the column vector). 
%The textual units  are included in the summary in decreasing order of importance that is estimated based on the SVD operation.
\noindent {\bf (5)~LUHN}~\cite{Luhn:1958}, which  
derives a `significance factor' for each textual unit based on 
occurrences and placements of frequent words within the unit. \\
%compiles a list of content words sorted by decreasing frequency (after stemming and stopword removal), the index providing a significance measure of
%the word. A `significance factor' is derived for each textual unit, that reflects the 
%number of occurrences and placement of significant words within  the unit. %, and the linear distance between them due to the intervention of non-significant words. 
%Units are included in the summary in decreasing order of this significance factor.
%Sentences are
%ranked in order of their significance factor, and the top ranking sentences are included in the summary.
\noindent {\bf (6)~SumBasic}~\cite{Nenkova05theimpact}, which uses frequency-based selection of textual units, 
and %re-weighting of the 
reweights word probabilities
to minimize redundancy. %\footnote{The implementation is available at \url{https://github.com/EthanMacdonald/SumBasic}.} 
%It starts with computing the probability distribution over
%the words appearing in the input. Then 
%Each textual unit is assigned a weight based on the average probability of occurrence of the words in the unit, and the
%units are then included in the summary based on this weight. 
%The best scoring sentence that contains
%the highest probability words is included in the summary. It then updates the probabilities. If the desired summary length has not been
%reached, the above steps are repeated.

%\noindent It is evident that the algorithms use a wide variety of approaches in judging the importance of textual units 
%(sentences in a document, or tweets in a given set of tweets) and thus selecting which units should be included in the summary.

\vspace{1mm}
\noindent \underline{\bf Supervised neural summarization algorithms:}
With the recent popularity of neural network based models, the state of the art techniques for summarization have shifted to data-driven supervised algorithms~\cite{dong2018survey}. 
We have considered two recently proposed extractive neural summarization models, proposed in~\cite{nallapati2017summarunner}:

\noindent {\bf (7)~SummaRuNNer-RNN}, a Recurrent Neural Network based sequence model that provides a binary label to each textual unit: -- a label of $1$ implies that the textual unit can be part of the summary, while $0$ implies otherwise. 
Each label has an associated confidence score. 
The summary is generated by picking textual units labeled $1$ in decreasing order of their confidence score.
%The proposed model is built around a two-layer bi-directional Gated Recurrent Unit based Recurrent Neural Network (GRU-RNN). 
%The first layer of RNN works at word-level while the second level of RNN accepts the average-pooled hidden states of the word-level RNN as input to encode the representation of sentences in a document.

\noindent {\bf (8)~SummaRuNNer-CNN} is a variant of the above model where the sentences are fed to a two layer Convolutional Neural Network (CNN) architecture before using GRU-RNN in the third layer. 

\noindent 
For both the SummaRuNNer models, the authors have made the pretrained models available\footnote{\url{https://github.com/hpzhao/SummaRuNNer}} which are trained on the CNN/Daily Mail news articles corpus\footnote{\url{https://github.com/deepmind/rc-data}}.
We directly used the pretrained models for the summarization.

%\noindent {\bf (1)~ESRL}~\cite{narayan2018ranking} \textcolor{blue}{conceptualizes extractive summarization as a sentence ranking task by globally optimizing the ROUGE evaluation metric through a reinforcement learning objective.}

\subsection{Verifying if the summaries are fair}

%\vspace{2mm}
%\noindent \underline{\bf Results of summarization:}

\noindent We applied the summarization algorithms stated above 
on the datasets described in Section~\ref{sub:datasets}, to
obtain summaries of length $50$ tweets each.
%Table~\ref{tab:claritin-data-results} shows the results of summarizing Dataset 1 (Claritin) --  
%shown are the percentages of tweets written by male and female users in the whole dataset (first row), and in the summaries
%obtained by the different methods (subsequent rows).
%Similarly,  Table~\ref{tab:election-data-results} shows the results of summarizing Dataset 2 (US election), i.e., 
%the percentages of Pro-Rep, Pro-Dem and Neutral tweets
%in the whole dataset and in the summaries obtained by the various summarization algorithms.
Table~\ref{tab:summ-results-claritin} shows the results of summarizing the Claritin dataset, while 
Table~\ref{tab:summ-results-uselection} and Table~\ref{tab:summ-results-metoo} show the results for the US-Election and MeToo datasets respectively.
In all cases, shown are the numbers of tweets of the different classes in the whole dataset (first row), and in the summaries
generated by the different summarization algorithms (subsequent rows), and the average ROUGE-1 and ROUGE-2 Recall and $F_1$ scores of the summaries.

We check whether the generated summaries are fair, according to the fairness notions of equal representation, proportional representation and the principle of {\it `no adverse impact'}~\cite{biddle-adverse-impact} (which were explained in Section~\ref{sec:fairness-notions}). 
%that is  used by the U.S. Equal Employment Opportunity Commission
%to determine whether a company's hiring policy is biased against has any adverse impact on 
%a demographic group~\cite{biddle-adverse-impact}.
%According to this policy, a particular class $c$ is {\bf under-represented} %(disadvantaged) 
%in the selected set, if the fraction of selected items belonging to class $c$ is less than $80\%$ of the fraction of selected items of the class %which has having the highest selection rate. 
We find under-representation of particular groups of users in the summaries generated by many of the algorithms; these cases are marked in
Table~\ref{tab:summ-results-claritin},  Table~\ref{tab:summ-results-uselection} and Table~\ref{tab:summ-results-metoo} with the symbols $\star$ (where equal representation is violated), $\dagger$ (where proportional representation is violated) and \# (cases where there is adverse impact). Especially, the minority groups are under-represented in most of the cases. 

%In the context of summarization, the textual units %that are included in the summary are `selected' by a summarization algorithm. Hence, a class $c$ of tweets is under-represented if the fraction of selected tweets of class $c$ %that is included in the summary, is less than $80\%$ of the fraction of tweets selected from %that class which has highest selection rate in the summary.the class with highest selection rate. Applying this rule, we find under-representation of particular classes of tweets in the summaries generated by many of the algorithms; these cases are marked with an asterisk (*) in Table~\ref{tab:summ-results-claritin} and Table~\ref{tab:summ-results-uselection}.

\begin{table}
%\small
\footnotesize
%\tiny
\centering
\begin{tabular}{|>{\centering}p{2.7 cm}||p{1.5 cm}|p{1.5 cm}||p{0.7 cm}|p{0.7 cm}|p{0.7 cm}|p{0.7 cm}|}
\hline
Method & \multicolumn{2}{c||}{Nos. of tweets}  & \multicolumn{2}{c|}{ROUGE-1} & \multicolumn{2}{c|}{ROUGE-2}
\\ \cline{2-7}
 & Female & Male & Recall & $F_1$ & Recall & $F_1$
\\ \hline
Whole data & 2,505 (62\%)    &   1,532 (38\%)  & NA  & NA & NA & NA
\\ \hline \hline
ClusterRank & 33 (66\%) & 17 $(34\%)^{\dagger\star}$  & 0.437  &  0.495 & 0.161 & 0.183
\\ \hline
DSDR & 31 (62\%)  & 19 $(38\%)^{\star}$  & 0.302  & 0.425 & 0.144 & 0.203 
\\ \hline
LexRank & 34 (68\%) & 16 $(32\%)^{\#\dagger\star}$  & 0.296 & 0.393 & 0.114  & 0.160
\\ \hline
LSA & 35 (70\%)  & 15 $(30\%)^{\#\dagger\star}$ & 0.515  & 0.504 & 0.151  & 0.147
\\ \hline
LUHN & 34 (68\%)  & 16 $(32\%)^{\#\dagger\star}$  & 0.380  & 0.405 & 0.128 & 0.136
\\ \hline
SumBasic & 27 $(54\%)^{\#\dagger}$ & 23 $(46\%)^{\star}$  & 0.314 & 0.434 & 0.108 & 0.149 
\\ \hline
SummaRNN & 33 (66\%) & 17 $(34\%)^{\dagger\star}$  & 0.342  & 0.375 & 0.126 & 0.147  
\\ \hline
SummaCNN & 30 $(60\%)^{\dagger}$ & 20 $(40\%)^{\star}$  & 0.377  & 0.409 & 0.126 & 0.146  
\\ \hline
% &   &   &   &   
%\\ \hline
\end{tabular}
\caption{\textbf{Results of summarizing the Claritin dataset: Number of tweets posted by the two user groups, in the whole dataset and in summaries of length $50$ tweets generated by different algorithms. 
Also given are ROUGE-1 and ROUGE-2 Recall and $F_1$ scores of each summary.
The symbols $\star$, $\dagger$ and \# respectively  indicate under-representation of a group according to the fairness notions of equal representation, proportional representation, and `no adverse impact'~\cite{biddle-adverse-impact}.}} 
\label{tab:summ-results-claritin}
\vspace*{-5mm}
\end{table}

\begin{table}
%\small
\footnotesize
%\tiny
\centering
\begin{tabular}{|>{\centering}p{2.5cm}||p{1.5cm}|p{1.5cm}|p{1.5cm}||p{0.7 cm}|p{0.6 cm}|p{0.7 cm}|p{0.6 cm}|}
\hline
Method & \multicolumn{3}{c||}{Nos. of tweets}  & \multicolumn{2}{c|}{ROUGE-1} & \multicolumn{2}{c|}{ROUGE-2}
\\ \cline{2-8}
 & Pro Rep & Pro Dem & Neutral & Recall & $F_1$ & Recall & $F_1$ 
\\ \hline
Whole data &  1,309 (62\%)  &   658 (31\%)  & 153 (7\%) & NA  & NA & NA & NA
\\ \hline \hline
ClusterRank &  32 (64\%) & 15 $(30\%)^{\star}$ & 3 $(6\%)^{\star}$ & 0.247 & 0.349 & 0.061 & 0.086  
\\ \hline
DSDR & 28 $(56\%)^{\#\dagger}$  & 19 (38\%)  & 3 $(6\%)^{\#\star}$  & 0.215 & 0.331  & 0.067 & 0.104
\\ \hline
LexRank & 27 $(54\%)^{\#\dagger}$  & 20 (40\%)  & 3 $(6\%)^{\#\star}$  & 0.252 & 0.367  & 0.078 & 0.114
\\ \hline
LSA & 24 $(48\%)^{\#\dagger}$  & 20 $(40\%)^{\#}$  & 6 $(12\%)^{\star}$  & 0.311 & 0.404  & 0.083 & 0.108
\\ \hline
LUHN & 34 (68\%)  & 13 $(26\%)^{\#\dagger\star}$  & 3 $(6\%)^{\#\star}$  & 0.281 & 0.375 & 0.085 & 0.113
\\ \hline
SumBasic &  27 $(54\%)^{\#\dagger}$ & 23 (46\%)  & 0 $(0\%)^{\#\dagger\star}$  & 0.200 & 0.311 & 0.051 & 0.080
\\ \hline
SummaRNN & 34 (68\%) & 15 $(30\%)^{\star}$  & 1 $(2\%)^{\#\dagger\star}$  & 0.347 & 0.436 & 0.120 & 0.160  
\\ \hline
SummaCNN & 32 (64\%) & 17 (34\%)  & 1 $(2\%)^{\#\dagger\star}$  & 0.337 & 0.423 & 0.108 & 0.145  
\\ \hline
% &   &   &   &   & 
%\\ \hline
\end{tabular}
\caption{\textbf{Results of summarizing the US-Election dataset: Number of tweets of the three groups in the whole data and summaries of length $50$ tweets generated by different algorithms. 
%Also given are ROUGE-1 and ROUGE-2 Recall scores of each summary.
The symbols $\star$, $\dagger$ and \# denote under-representation of the corresponding group, similar to Table~\ref{tab:summ-results-claritin}}.
}
\label{tab:summ-results-uselection}
\vspace*{-5mm}
\end{table}

\begin{table}
%\small
\footnotesize
%\tiny
\centering
\begin{tabular}{|>{\centering}p{2.7 cm}||p{1.5cm}|p{1.5cm}||p{0.7 cm}|p{0.7 cm}|p{0.7 cm}|p{0.7 cm}|}
\hline
Method & \multicolumn{2}{c||}{Nos. of tweets}  & \multicolumn{2}{c|}{ROUGE-1} & \multicolumn{2}{c|}{ROUGE-2}
\\ \cline{2-7}
 & Female & Male & Recall & $F_1$ & Recall & $F_1$
\\ \hline
Whole data &   275 (56.3\%)   &  213 (43.7\%)   &  NA & NA & NA & NA  
\\ \hline \hline
ClusterRank & 24 $(48\%)^{\#\dagger\star}$  & 26 (52\%)  & 0.550 & 0.560  & 0.216  & 0.223
\\ \hline
DSDR  &  32 (64\%) & 18 $(36\%)^{\#\dagger\star}$  & 0.233 & 0.358  & 0.092 & 0.141 
\\ \hline
LexRank  &  34 (68\%) & 16 $(32\%)^{\#\dagger\star}$  & 0.285 & 0.414  & 0.105 & 0.153
\\ \hline
LSA  &  20 $(40\%)^{\#\dagger\star}$ & 30 (60\%)  & 0.511 & 0.534  & 0.175 & 0.183
\\ \hline
LUHN  &  22 $(44\%)^{\#\dagger\star}$ & 28 (56\%)  & 0.520 & 0.522  & 0.219 & 0.184
\\ \hline
SumBasic  & 27 $(54\%)^{\dagger}$  & 23 $(46\%)^{\star}$  & 0.464 & 0.499  & 0.216 & 0.229
\\ \hline
SummaRNN  & 23 $(46\%)^{\#\dagger\star}$  &  27 (54\%) & 0.622 & 0.636  & 0.385 & 0.394
\\ \hline
SummaCNN  &  23 $(46\%)^{\#\dagger\star}$ & 27 (54\%) & 0.622 & 0.636  & 0.385 & 0.394
\\ \hline
\end{tabular}
\caption{\textbf{Results of summarizing the MeToo dataset: Number of tweets of the two classes, in the whole dataset and in summaries of length $50$ tweets generated by different algorithms. 
%Also given are ROUGE-1 and ROUGE-2 Recall and $F_1$ scores of each summary.
The symbols $\star$, $\dagger$ and \# denote under-representation of the corresponding group, similar to Table~\ref{tab:summ-results-claritin}.}
%\new{The symbols $\star$, $\dagger$ and \# respectively indicate under-representation of a class according to the fairness notions of equal representation, proportional representation, and `adverse impact'~\cite{biddle-adverse-impact} (details in text).} 
}
\label{tab:summ-results-metoo}
\vspace*{-5mm}
\end{table}

%It is clear that summaries produced by different summarization algorithms contain very different distributions of the gender / political leaning, as compared to the distribution in the whole input data. 
%For instance, for the Claritin dataset, 
%the LexRank, LSA and LUHN algorithms increases the proportion of tweets from female users in the summary,
%while SumBasic increase the proportion of tweets from male users in the summary.
%The observations are similar for the US Election dataset. 
%While the whole dataset is predominantly pro-Republican (61.74\%), all the algorithms reduce the proportion of 
%pro-Republican tweets in the summary. 

We repeated the experiments for summaries of lengths other than $50$ as well, such as for
$100, 200, \ldots, 500$ (details omitted due to lack of space).
We observed several cases where the same algorithm includes very 
different proportions of tweets of various groups, while generating summaries of different lengths. 
%Hence, whether summarization is fair depends on several factors, including the particular algorithm used and the length of summary.
%For instance, we observed cases where the summary distribution generated by a particular algorithm
%in the summary of length 100 matched closely with the input distribution, but different significantly when the 
%same algorithm was used to generate summary of length 200. 

Thus, there is no guarantee of fairness in the summaries generated by the existing summarization algorithms --
one or more groups are often under-represented in the summaries, even though the quality of the tweets written by different groups are quite similar (as was shown in Section~\ref{sec:background}).
%Having established the need for fairness-preserving summarization algorithms, we now proceed to propose three such algorithms in the subsequent sections.

%\vspace{2mm}
%\noindent %{\bf Section Summary:} 
%The experiments in this section establish that summaries generated by existing summarization algorithms are often not fair (according to standard notions of fairness), 
%and under-represent one or more classes, even though the textual quality of the textual units (tweets) of the different classes are quite similar (as was shown in Section~\ref{sec:background}).
%We now proceed to propose two algorithms for fairness-preserving summarization in the next two sections.

%\input{methodology}
%\section{Different approaches for achieving Fairness in Summarization}
\section{Achieving Fairness in Summarization}\label{sec: Framework}
Recently, there has been a flurry of research activities focusing on fairness issues in algorithmic decision making systems, with the main emphasis on classification algorithms~\cite{zafar2017fairness, zemel2013learning, dwork2012fairness}. %Such fairness aware algorithms 
Approaches proposed in these works can be broadly categorised into three types~\cite{friedler2019comparative}: {\it pre-processing, in-processing} and {\it post-processing}, based on the stage where the fairness intervention is applied. 
To achieve fairness,  pre-processing approaches attempt to change the input data/representation, in-processing approaches change the underlying algorithm itself, and post-processing methods change the outputs of the algorithm before they get used in downstream applications.

Following this line of work, in this paper, %To reduce the above mentioned unfairness, 
we develop three novel fairness-preserving summarization algorithms (adhering to the principles of pre-, in- and post-processing)  which select highly relevant textual units in the summary while maintaining fairness in the process. Next, we discuss the key ideas behind the proposed algorithms. Each of the algorithms will be explained in detail in subsequent sections.

%of the notions of fairness at different stages of the algorithm . In this section, we touch upon all the three approaches and describe how the different algorithms in our paper are fundamentally different from each other.

\vspace{1 mm}
\noindent
\underline{\bf (1) Pre-processing:} %The motivation behind
As mentioned above, pre-processing approaches attempt to change the input to the algorithms to make the outcome fair. The idea originated from classification algorithms where the biases in the training data may get translated into the learned model, and hence by making the training data or the input unbiased, the algorithm can be made non-discriminatory. 
%is the cause of the discrimination that a machine learning algorithm might learn, and so modifying it can keep a learning algorithm trained on it from discriminating. This could be because the training data itself captures historical discrimination or because there are more subtle patterns in the data, such as an under-representation of a ethnic-minority group, that makes the outcomes discriminate against the minority groups. In our context it can result in under representation of different socially salient groups (ethnic-minorities) in the final summary. 
In our context, to ensure fair representation, we propose a pre-processing technique {\it ClasswiseSumm} (described in Section~\ref{sub:classwise}),  where we first  group tweets on the basis of their association to different classes. Then, we propose to summarize each group separately using any state-of-the-art algorithm, and  generate $\{l_1, l_2, ...\}$ length summaries for different groups, where the lengths $\{l_1, l_2, ...\}$ would be determined based on the fairness objective. Finally, these individual summaries would be combined to generate the final fair summary. %summarise the input texts in different chunks as per their belongingness to different classes under consideration. Then depending on the underlying fairness constraints, the desired number of texts from different classes are selected for the final summary.}

\vspace{1 mm}
\noindent
\underline{\bf (2) In-processing: }
%Algorithm or Algorithmic Modifications:} Often different information filtering algorithms (e.g. search, recommendation, summarization) are keyed only towards optimizing for some measure of relevance, neglecting the unfair repercussions that it may bring. Hence,
In-processing methods work by changing the underlying learning  algorithms and making them adhere to the  fairness objectives (for instance, by putting  additional fairness constraints). 
%, have been by far the most common approach to affirm fairness. In this process, we make the entire decision making process i.e. the algorithm itself a fair process of generating the outcomes. 
Our proposed algorithm {\it FairSumm} (detailed in Section~\ref{sec:fairsumm}) is one such algorithm, where we summarize using a constrained sub-modular optimization, with the fairness criteria  applied as matroid constraints to an objective function ensuring goodness of the summary. 

\vspace{1 mm}
\noindent
\underline{\bf (3) Post-processing:} The third approach for bringing fairness into algorithmic systems is by modifying the outputs of an %previously designed 
algorithm to achieve the desired results for different groups. %We do not claim in anyway that ours is the best summarization algorithm that has ever been proposed.
Intervention at the output stage becomes necessary when the summarization algorithm is already decided, and there is no option to change its working. 
For example, in our context,  
if some organization intends to stick to its proprietary summarization algorithm, then post-processing on the generated summaries (or the ranked list of textual units)
 %produced by such an algorithm is another way 
becomes necessary to produce fair summaries. 
Hence, we propose {\it ReFaSumm} (\textbf{Re}ranking \textbf{Fa}irly the \textbf{Summ}arization outputs)    
where we attempt to fairly re-rank the outputs generated by existing summarization algorithms (detailed in Section~\ref{sub:fair-ranking}).

%\section{Proposed Algorithms for Fair Summarization}
%\label{sec:fairsumm}

%This section describes our proposed algorithms for fair summarization. We describe two algorithms - the first one based on constraint optimization, and the second one based on fair ranking.  

\section{FairSumm: In-processing algorithm for fair summarization}
\label{sec:fairsumm}
Our proposed in-processing algorithm, named {\tt FairSumm}, 
treats summarization as a constrained optimization problem 
%of a submodular, monotone objective function,
of an objective function.
The objective function is designed so that optimizing it is likely to result in a good quality summary, while the fairness requirements are applied as constraints which must be obeyed during the optimization process.
%We now describe the algorithm. 

\vspace{2 mm}
\noindent
\textbf{Some notations:}
Let $V$ denote the set of textual units (e.g., tweets) that is to be summarized.
Our goal is to find a subset $S$ ($\subseteq$ $V$) such that $|S| \le k$, where $k$ (an integer) is the desired length of the summary (specified as an input), 

\subsection{{\bf Formulating summarization as an optimization problem}} 
We need an {\it objective function} for extractive summarization, optimizing which is likely to lead to a good summary. 
Following the formulation by 
Lin {\it et al.}~\cite{Lin2011}, we consider two important aspects of an extractive text summarization algorithm, viz. {\it Coverage} and {\it Diversity reward}, described below.

\vspace{1mm}
\noindent \underline{\bf Coverage}: Coverage refers to amount of information covered in the summary $S$.
Clearly, the summary cannot contain the information in all the textual units. 
We consider the summary $S$ to cover the information contained in a particular textual unit $i \in V$ if either $S$ contains $i$, or if $S$ contains another textual unit $j \in V$ that is very similar to $i$. 
Here we assume a notion of similarity $sim(i,j)$ between two textual units $i \in V$ and $j \in V$, which can be measured in various ways.
Thus, the coverage will be measured by a function -- say, $\mathcal{L}$ -- whose 
generic form can be
\begin{equation}
\mathcal{L}(S) = \sum_{i \in S, j \in V} sim(i,j)
\label{eqn:coverage}
\end{equation}
Thus, $\mathcal{L}(S)$ measures the overall similarity of the textual units included in the summary $S$ with all the textual units in the input collection $V$.

%Note that, $\sum_{i \in V, j \in S} sim_{i,j}$ is monotone submodular. 
%Note that $\mathcal{L}$ is monotone submodular.
%$\mathcal{L}$ is monotonic since coverage increases by the addition of a new sentence in the summary. At the same time, $\mathcal{L}$ is submodular since the increase in $\mathcal{L}$ would be more when a sentence is added to a shorter summary, than when a sentence is added to a longer summary. 
%There can be several forms of $\mathcal{L}$ depending on how $sim_{i,j}$ is measured, which we will discuss later in this paper.

\vspace{1mm}
\noindent \underline{\bf Diversity reward}: The purpose of this aspect is to avoid redundancy and reward diverse information in the summary.
Usually, it is seen that the input set of textual units can be partitioned into groups, where each group contains textual units that are very similar to each other. 
A popular way of ensuring diversity in a summary is to partition the input set into such groups, and then select a representative element from each group~\cite{ensemble-summ}.

Specifically, let us consider that the set $V$ of textual units is partitioned into $K$ groups. 
Let $P_1, P_2, \ldots, P_K$ comprise a partition of $V$. That is, $\large\cup_i P_i = V$ ($V$ is formed by the {\it union} of all $P_i$) and $P_i \cap P_j$ = $\emptyset$ ($P_i$, $P_j$ have no element in common) for all $i$ $\neq$ $j$. 
For instance, the partitioning $P_1, P_2, \ldots, P_K$ can be achieved by clustering the set $V$ using any clustering algorithm (e.g., $K$-means), based on the similarity of items as measured by $sim(i,j)$.

Then, to reduce redundancy and increase diversity in the summary, including textual units from different partitions needs to be rewarded. 
Let the associated function for diversity reward be denoted as $\mathcal{R}$. A generic formulation of $\mathcal{R}$ is 
\begin{equation}
\mathcal{R}(S) = \sum_{i=1}^K \sqrt{\sum_{j \in P_i \cap S} r_j}
\label{eqn:diversity}
\end{equation}
where $r_j$ is a suitable function that estimates the importance of adding the textual unit $j \in V$ to the summary.
The function $r_j$ is called a `singleton reward function' since it estimates the reward of adding the singleton element $j \in V$ to the summary $S$. One possible way to define this function is by measuring the average similarity of $j$ to the other textual units in $V$. Mathematically,
\begin{equation}
r_j = \frac{1}{N} \sum_{i \in V} sim(i,j)
\label{eqn:singleton-reward}
\end{equation}

\vspace{1mm}
\noindent \underline{\bf Justifying the functional forms of Coverage and Diversity Reward}:
We now explain the significance of the functional form of $\mathcal{L}(S)$ in Equation~\ref{eqn:coverage} and $\mathcal{R}(S)$ in Equation~\ref{eqn:diversity}.
We give only an intuitive explanation here; more mathematical details are given in Appendix~\ref{App: FairSumm}.%\footnote{\url{http://cse.iitkgp.ac.in/~saptarshi/docs/DashEtAl-CSCW2019-fair-summarization-SuppleInfo.pdf}}.

The functions $\mathcal{L}(S)$ and $\mathcal{R}(S)$ are designed to be `monotonic non-decreasing submodular' functions (or  `monotone submodular' functions), since such functions are easier to optimize.  
A monotonic non-decreasing function is one that does not decrease (usually increases) as the set over which the function is employed grows.
A submodular function has the property of {\it diminishing returns} which intuitively means that as the set (over which the function is employed) grows, the increment of the function decreases.  

$\mathcal{L}$ is monotone submodular. $\mathcal{L}$ is monotonic since coverage increases by the addition of a new sentence in the summary.
At the same time, $\mathcal{L}$ is submodular since the increase in $\mathcal{L}$ would be
more when a sentence is added to a shorter summary, than when it is added to a longer summary.

Also $\mathcal{R}$ is a monotone submodular function. The diversity of a summary increases considerably only for the initial growth of the set (when new, `novel' elements are added to the summary) and stabilizes later on, and thus prevents the incorporation of similar elements (redundancy) in the summary.
$\mathcal{R}(S)$ rewards diversity since there is more benefit in selecting a textual unit from a partition (cluster) that does not yet have any of its elements included in the summary.
As soon as any one element from a cluster $P_i$ is included in the summary, the other elements in $P_i$ start having diminishing gains, due to the square root function in Equation~\ref{eqn:diversity}.

\vspace{1mm}
\noindent \underline{\bf Combining Coverage and Diversity reward}:
While constructing a summary, both coverage and diversity are important.
Only maximizing coverage may lead to lack of diversity in the resulting summary and vice versa. So, we define our objective function for summarization as follows:
\begin{equation}
\mathcal{F} = \lambda_1 \mathcal{L} + \lambda_2 \mathcal{R}
\label{eqn:summary-objective}
\end{equation}
where $\lambda_1$, $\lambda_1$ $\ge$ 0 are the weights given to coverage and diversity respectively.

Our proposed fairness-preserving summarization algorithm will maximize $\mathcal{F}$ in keeping with some fairness constraints.
Note that $\mathcal{F}$ is monotone submodular since it is a non-negative linear combination of two monotone submodular functions $\mathcal{L}$ and $\mathcal{R}$. We have chosen $\mathcal{F}$ such that it is monotone submodular, since there exist standard algorithms to efficiently optimize such functions (as explained later in the section).

%Note that, by Property (3), $\mathcal{F}$ is monotone submodular.
%We obtain L and R from equations (6) and (7) respectively (Section 5.1) of Lin et al.~\cite{Lin2011}. To compute $w_{i,j}$, apart from the formulation presented in this paper, we can use semantic similarity (e.g., cosine similarity between embedding vectors created from Word2Vec and GloVe). Finally, the {\it objective function} F is defined by equation (2) of this paper.

%\subsection{Proposed fair summarization scheme}
\vspace{2 mm}
\subsection{{\bf Proposed fair summarization scheme}} 
Our proposed scheme is based on the concept of {\it matriods} that are typically used to generalize the notion of liner independence in matrices~\cite{matroid-theory-book}.
Specifically, we utilize a special type of matroids, called {\it partition matroids}.
We give here a brief, intuitive description of our method. More details can be found in Appendix~\ref{App: FairSumm}.

\vspace{1mm}
\noindent \underline{\bf Brief background on matroids and related topics}:
In mathematical terms, a matroid is a pair $\mathcal{M}$ = ($\mathcal{Z}$, $\mathcal{I}$), defined over a finite set $\mathcal{Z}$ (called the ground set)  and a family of sets $\mathcal{I}$ (called the independent sets), that satisfies the three properties:
\begin{enumerate}
\item $\emptyset$ (empty set) $\in$ $\mathcal{I}$.
\item If $Y$ $\in$ $\mathcal{I}$ and $X$ $\subseteq$ $Y$, then $X$ $\in$ $\mathcal{I}$.
\item If $X$ $\in$ $\mathcal{I}$, $Y$ $\in$ $\mathcal{I}$ and $|Y|$ $>$ $|X|$, then there exists $e$ $\in$ $Y$ $\setminus$ $X$ such that $X$ $\cup$ $\{e\}$ $\in$ $\mathcal{I}$.
\end{enumerate}
Condition~(1) simply means that $\mathcal{I}$ can contain the empty set, i.e., the empty set is {\it independent}. 
Condition~(2) means that every subset of an independent set is also independent. 
Condition~(3) means that if $X$ is independent and there exists a larger independent set $Y$, $X$ can be extended to a larger independent set by adding an element in $Y$ but {\it not in} $X$.\footnote{For details, refer to \url{http://www-math.mit.edu/~goemans/18433S09/matroid-notes.pdf}}

\vspace{1mm}
{\it Partition matroids} refer to a special type of matroids where the ground set $\mathcal{Z}$ is partitioned into $s$ disjoint subsets $\mathcal{Z}_1$, $\mathcal{Z}_2$, ..., $\mathcal{Z}_s$ for some $s$, and $\mathcal{I}$ = \{$S$ | $S$ $\subseteq$ $\mathcal{Z}$ and $|S \cap \mathcal{Z}_i|$ $\le$ $c_i$, for all $i$ = 1, 2, ..., $s$\} 
for some given parameters $c_1$, $c_2$, ..., $c_s$.
Thus, $S$ is a subset of $Z$ that contains at least $c_i$ items from the partition $\mathcal{Z}_i$ (for all $i$), and $\mathcal{I}$ is the family of all such subsets.

%--------------------------------------------------

Consider that we have a set of control variables $z_j$ (e.g., `gender', `political leaning'). Each item in $\mathcal{Z}$ has a particular value for each $z_j$.
Also consider that $z_j$ takes $t_j$ distinct values, e.g., the control variable `gender' takes the two distinct values `male' and `female', while the control variable `political leaning' takes the values `Democrat', `Republican' and `Neutral'. 

For each control variable $z_j$,  we can partition $\mathcal{Z}$ into $t_j$ disjoint subsets $\mathcal{Z}_{j1}$, $\mathcal{Z}_{j2}$, ..., $\mathcal{Z}_{jt_j}$, each corresponding to a particular value of this control variable. 
We now define a partition matriod $\mathcal{M}_j$ = ($\mathcal{Z}$, $\mathcal{I}_j$) such that
\begin{center}
$\mathcal{I}_j$ = \{$S$ | $S$ $\subseteq$ $\mathcal{Z}$ and $|S \cap Z_{ji}|$ $\le$ $c_j$, for all $i = 1, 2, \ldots, t_j$\}
\end{center}
for some given parameters $c_1$, $c_2$, ..., $c_{t_j}$.

Now, for a given {\it submodular} objective function $f$, a submodular optimization under the partition matriod constraints with $P$ control variables can be designed as follows:
\begin{equation}
\label{lab:eq}
Maximize_{S \subseteq \mathcal{Z}} \;\; f(S)
\end{equation}
\begin{center}
subject to $S \in \bigcap_{j=1}^P \mathcal{I}$.
\end{center}
A prior work by Du {\it et al.}~\cite{Du2013} has established that this submodular optimization problem under the matroid constraints can be solved efficiently with provable guarantees (see~\cite{Du2013} for details).

%\subsection{Proposed summarization scheme}

\vspace{1mm}
\noindent \underline{\bf Formulating the fair summarization problem}:
In the context of the fair summarization problem, the ground set is $V$ (= $\mathcal{Z}$), the set of all textual units (sentences/tweets) which we look to summarize.
The control variables are analogous to the sensitive attributes with respect to which fairness is to be ensured, such as `gender' or `political leaning'.
In this work, we consider only one sensitive attribute for a particular dataset (the gender of a user for the Claritin and MeToo datasets, and political leaning for the US-Election dataset).
Let the corresponding control variable be $z$, and let $z$ take $t$ distinct values (e.g., $t = 2$ for the Claritin and MeToo datasets, and $t=3$ for the US-Election dataset).
Note that, the formulation can be extended to multiple sensitive attributes (control variables) as well.

Each textual unit in $V$ is associated with a class, i.e., a particular value of the control variable $z$ (e.g., is posted either by a male or a female). 
Let $Z_1$, $Z_2$, ..., $Z_t$ ($Z_i$ $\subseteq$ $V$, for all $i$) be disjoint subsets of the textual units from the $t$ classes, each associated with a distinct value of $z$.
%Let $Z_1$ ($\subseteq$ $V$) be the set of all the sentences posted by males and $Z_2$ ($\subseteq$ $V$) be the set of all sentences posted by females. 
We now define a partition matroid $\mathcal{M}$ = ($V$, $\mathcal{I}$) in which $V$ is partitioned into disjoint subsets $Z_1$, $Z_2$, ..., $Z_t$ and
\begin{center}
$\mathcal{I}$ = \{$S$ | $S$ $\subseteq$ $V$ and $|S \cap Z_i|$ $\le$ $c_i$, $i$ = 1, 2, ..., $t$\}
\end{center}
for some given parameters $c_1$, $c_2$, ..., $c_t$.
In other words, $I$ will contain all the sets $S$ containing at most $c_i$ textual units from $Z_i$, $i$ = 1, 2, ..., $t$.

Now we add the fairness constraints. Outside the purview of the matroid constraints, we maintain the restriction that $c_i$'s are chosen such that\\
(1) $\sum_{i=1}^t c_i = k$ (the desired length of the summary $S$), and\\ 
(2) a desired fairness criterion is maintained in $S$. For instance, if equal representation of all classes in the summary is desired, then $c_i = \frac{k}{t}$ for all $i$. 

%the 80\% {\it adverse impact fairness criterion}~\cite{biddle-adverse-impact} is maintained in $S$.

\noindent We now express our fairness-constrained summarization problem as follows:
\begin{equation}
\label{lab:eq2}
Maximize_{S \subseteq V} \;\; \mathcal{F}(S)
\end{equation}
\begin{center}
subject to $S$ $\in$ $\mathcal{I}$.
\end{center}
where the objective function $\mathcal{F}(S)$ is as stated in Equation~\ref{eqn:summary-objective}.
Given that $\mathcal{F}$ is a submodular function (as explained earlier in this section), the algorithm proposed by Du et al.~\cite{Du2013} is suitable to solve this constrained submodular optimization problem.

\vspace{1mm}
\noindent \underline{\bf An example}:
Let us illustrate the formulation of the fair summarization problem with an example. Assume that we are applying the equal representation fairness notion over the MeToo dataset, and we want a summary of length $k = 50$ tweets. 
Then, the control variable $z$ corresponds to the sensitive attribute `gender' which takes $t = 2$ values (`male' and `female') for this particular dataset.
The set of tweets $V$ will be partitioned into two disjoint subsets $Z_1$ and $Z_2$ which will comprise the tweets posted by male and female users respectively.
To enforce equal representation fairness constraint, we will set the parameters $c_1 = 25$ and $c_2 = 25$ (since we want equal number of tweets from $Z_1$ and $Z_2$ in the summary).
Thus, $I$ contains all the possible sets $S$ that contain at most $25$ tweets written by male users and $25$ tweets written by female users. Each such $S$ is a valid summary (that satisfies the fairness constraints).
Solving the optimization problem in Equation~\ref{lab:eq2} will give us that summary $S$ for which $\mathcal{F}(S)$ will be maximum, i.e. for which coverage and diversity reward will be the highest.

\begin{algorithm}[tb]
 \caption{: FairSumm (in-processing approach for fair summarization)}
 \label{algo:atga}
 \begin{algorithmic}[1]

  \State Set $d$ = $max_{z \in V}$$\mathcal{F}(\{z\})$. 
  \State Set $w_t$ = $\frac{d}{(1+\delta)^t}$ for $t$ = 0, $\ldots$, $l$ where $l$ = $argmin_i$ [$w_i$ $\le$ $\frac{\delta d}{N}$], and $w_{l+1}$ = 0.
 % /* Here $N$ is the size of $V$, $\delta$ is chosen as 0.01 by Du et al.~\cite{Du2013} */
  \State Set $G$ = $\emptyset$
  \For{$t$ = 0, 1, $\ldots$, $l$, $l+1$}
  	\For{each $z \in V$ and $G\cup \{z\}$ $\in$ $I$}
    	\If{$\mathcal{F}$($G\cup \{z\}$) - $\mathcal{F}$($G$) $\ge$ $w_t$}
        	\State Set $G$ $\leftarrow$ $G\cup \{z\}$
        \EndIf
    
    \EndFor

  \EndFor
  \State Output $G$ as the summary
  
 \end{algorithmic}
\end{algorithm}

\vspace{1mm}
\noindent \underline{\bf Algorithm for fair summarization}:
Algorithm~\ref{algo:atga} presents the algorithm to solve this constrained submodular optimization problem, based on the algorithm developed by Du et al.~\cite{Du2013}. 
The $G$ output by Algorithm~\ref{algo:atga} is the solution of Equation~\ref{lab:eq2}. 
We now briefly describe the steps of Algorithm~\ref{algo:atga}.

In Step 1, the maximum value of the objective function $\mathcal{F}$ that can be achieved for a text unit $z$ ($\in$ $V$) is calculated and stored in $d$. The purpose of this step is to compute the maximum value of $\mathcal{F}$ for a single text unit $z$ and set a selection threshold (to be described shortly) with respect to this value. 
This step will help in the subsequent selection of textual units for the creation of the summary to be stored in $G$. 
$w_t$ (defined in Step 2) is such a threshold at the $t^{th}$ time step. $w_t$ is updated (decreased by division with a factor $1 + \delta$) for $t$ = 0, 1, $\ldots$, $l$. 
$l$ is the minimum value of $i$ for which $w_i$ $\le$ $\frac{\delta d}{N}$ holds (see Du et al.~\cite{Du2013} for details) and $w_{l+1}$ is set to zero. 
In Step 3, $G$ (the set that will contain the summary) is initialized as an empty set. Note that $G$ is supposed to be an independent set according to the definition of matroid given earlier in this section. 
By condition (1) in the definition of matroids (stated earlier in this section), an empty set is independent. 
Step (4) iterates through the different values of $t$. Step (5) tests, for each $z$ (text element) $\in$ $V$, if $G$ remains an independent set by the inclusion of $z$. Only those $z$'s are chosen in this step whose inclusion expands $G$ (already an independent set) to another independent set. 
Step (6) selects a $z$ (permitted by Step (5)) for inclusion in $G$ if $\mathcal{F}$($G\cup \{z\}$) - $\mathcal{F}$($G$) $\ge$ $w_t$. This $z$ is added to $G$ in Step (7). That is, $z$ is added to $G$ if the increment of $\mathcal{F}$ by the addition of $z$ is not less than the threshold $w_t$. For $t$ = 0, $w_t$ = $d$, that is, the maximum value of $\mathcal{F}$ for any $z$ ($\in$ $V$). 
This means, the $z$ which maximizes $\mathcal{F}$ is added to $G$. Note that, there can be multiple $z$'s for which $\mathcal{F}$ is maximized. In that case, the tie is broken arbitrarily. The remaining $z$'s may or may not be added to $G$ based on the threshold value.

Another important point to note is that, our chosen $\mathcal{F}$ (see equation (\ref{eqn:summary-objective})) is designed to maximize both coverage and diversity. So, even if multiple $z$'s satisfy Step (5), they may not be added to $G$ in Step (7) if they contain redundant information. The value of $w_t$ is relaxed for the subsequent values of $t$ to allow text elements $z$ producing relatively lower increments of $\mathcal{F}$ to be considered for possible inclusion in $G$. $w_{l+1} = 0$ indicates that for the final value of $t$, at least one text unit $z$ which does not decrement $\mathcal{F}$ is added to $G$. This ensures that the coverage of the summary produced is not compromised while preserving diversity. This process (Steps (5) to (7)) is repeated for $t$ = 0, 1, $\ldots$, $l$, $l+1$ resulting in the final output $G$.

The reason for the efficiency of Algorithm~\ref{algo:atga} is the fact that this algorithm does {\it not} perform exhaustive evaluation of all the possible submodular functions evolving in the intermediate steps of the algorithm. The reduction in the number of steps in the algorithm is achieved mainly by decreasing $w_t$ geometrically by a factor of $1 + \delta$. In addition, multiple elements $z$ can be added to $G$ for a single threshold which also expedites the culmination of the algorithm.

\section{Pre and Post-processing mechanisms for fair summarization}\label{sec:otherMeth}

In this section, we discuss our proposed pre-processing and post-processing summarization algorithms to produce fair summaries. 

\subsection{ClasswiseSumm: Pre-processing algorithm for fair summarization} \label{sub:classwise}

We now describe a simple pre-procesing algorithm for  fair summarization.
Suppose that the textual units in the input belong to $t$ classes $Z_1, Z_2, \ldots, Z_t$, and to conform to a desired fairness notion, the summary should have $c_i$ units from class $Z_i$, $i = 1, 2, \ldots, t$ 
%where $\sum_{i=1}^{t} c_i = k$ 
(using the same notations as in Section~\ref{sec:fairsumm}). 
The simplest way to generate a fair summary is to {\it separately summarize the textual units belonging to each class} $Z_i$, to produce a summary of length $c_i$, and finally to combine all the $t$ summaries to obtain the final summary of length $k$.
We refer to this method as the {\bf ClasswiseSumm} method. 
Specifically, in this work, we use our proposed algorithm FairSumm, without any fairness constraints, to summarize each class separately. However, any other summarization algorithm can be used to summarize each class separately.
%While this is the easiest way to generate a fair summary, it is not clear how good the resultant summaries will be. 

%\subsection{Algorithm II for fair summarizaton}
\subsection{ReFaSumm: Post-processing algorithm for fair summarization}
\label{sub:fair-ranking}

In this section, we discuss our proposed post-processing mechanism for generating fair summaries, which can be used along with any existing summarization algorithm.
Many summarization algorithms (including the ones stated in Section~\ref{sec:existing}) 
generate an importance score of each textual unit in the input.
The textual units are then ranked in decreasing order of this importance score, and the top-ranked $k$ units are selected to form the summary.
Hence, if the ranked list of the textual units can be made fair (according to some desired fairness notion), then selecting the top $k$ from this fair ranked list can generate a fair summary. We refer to this algorithm as ReFaSumm ({\bf Re}-ranking {\bf Fa}irly the {\bf Summ}arization output).

Fairness in ranking systems is an important problem that has been addressed recently by some works~\cite{Zehlike2017,Biega-fair-rank-sigir18}. 
We adopt the fair-ranking methodology developed by Zehlike {\it et al.}~\cite{Zehlike2017} to generate fair summaries. 
%since some summarization algorithms which produce summaries as a ranked list of textual units (sentences or tweets).
The fair-ranking scheme in~\cite{Zehlike2017} considers  a {\bf two-class} setting, with a `majority class' and a `minority class' for which fairness has to be ensured adhering to a {\it ranked group fairness criterion} .
Their proposed ranking algorithm (named {\it FA*IR}~\cite{Zehlike2017}) ensures that the proportion of the candidates/items from the minority class in a ranked list never falls below a certain specified threshold. 
Specifically, two fairness criteria are ensured -- {\it selection utility} which means every selected item is more qualified than those not selected, and {\it ordering utility} which means for every pair of selected candidates, either the more qualified is ranked above or the difference in their qualifications is small~\cite{Zehlike2017}. 

\if{0}
\vspace{2mm}
\noindent \underline{{\bf Definitions:}} We state some definitions that are necessary to explain the methodology. The reader is referred to~\cite{Zehlike2017}for details.

\vspace{1mm}
\noindent \textit{Fair representation condition:} Let $F(x; n, p)$ be the cumulative distribution function for a binomial distribution $B$($n$, $p$).
Let $\mathcal{S}_{k,n}$ be the set of all permutations of size $k$ of the set $[n] = \{1, 2, ..., n\}$ of items to be ranked.
A set $\mathcal{s}$ $\subseteq$ $\mathcal{S}_{k,n}$  having $\mathcal{s}_p$ protected candidates fairly represents a protected class with minimum proportion $p$ at significance level $\alpha$ if $F(\mathcal{s}_p; k, p)$ $>$ $\alpha$.

\vspace{1mm}
\noindent \textit{Ranked group fairness condition:} A ranking $\mathcal{s}$ $\in$ $\mathcal{S}_{k,n}$ is said to satisfy this condition if for every prefix $\mathcal{s}|_{i}$ = $<$ $\mathcal{s}_{1}$, $\mathcal{s}_{2}$, ..., $\mathcal{s}_{i}$$>$, 1 $\le$ $i$ $\le$ $k$ and for $p$ and $\alpha$, if $\mathcal{s}|_{i}$ satisfies the \textit{fair representation condition} with proportion $p$ and significance $\alpha_c$ = c($\alpha$, $k$, $p$) which is corrected using~\cite{YangCorr2016}.
\fi
%%%%%%%%%%%%%%%%% COMMENTED OUT TILL HERE    %%%%%%%%%%%%%%%%%%%%%

We propose to use the algorithm in~\cite{Zehlike2017} for fair extractive text summarization as follows. Note that this scheme is only applicable to cases where there are two groups (e.g., the Claritin and MeToo datasets).
We consider that group to be the majority class which has the higher number of textual units (tweets) in the input data, while the group having lesser textual units in the input is considered the minority class.

\vspace{1mm}
\noindent \underline{\bf Input and Parameter settings:} The algorithm takes as input a set of textual units (to be summarized); the other input parameters ($k$, $q_i$, $g_i$ and $p$) taken by the algorithm in~\cite{Zehlike2017} are set as follows.\\
$\bullet$  \textit{Qualification ($q_i$) of a candidate:} In our summarization setting, this is the goodness value of a textual unit in the data to be summarized.
We set this value to the importance score computed by some standard summarization algorithm (e.g., the ones discussed in Section~\ref{sec:existing}) that ranks the text units by their importance scores.\\
$\bullet$  \textit{Expected size ($k$) of the ranking:} The expected number of textual units in the summary ($k$).\\
$\bullet$ \textit{Indicator variable ($g_i$) indicating if the candidate is protected:} 
%In our summarization setting, we do not have a pre-defined protected class. 
%However, to be consistent with the premise of {\it FA*IR} algorithm, 
We consider that group to be the minority class which has the lesser number of textual units in the input data. All tweets posted by the minority group are marked as `protected'.\\
$\bullet$  \textit{Minimum proportion ($p$) of protected candidates:} We will set this value in the open interval $]$0, 1$[$ (0 and 1 excluded) so that a particular notion of fairness is ensured in the summary. For instance, if we want equal representation of both classes in the summary, we will set $p = 0.5$.\\
$\bullet$  \textit{Adjusted significance level ($\alpha_c$):} We regulate this parameter in the open interval $]$0, 1$[$.

\vspace{1mm}
\noindent \underline{\bf Working of the algorithm:} Two priority queues $P_0$ (for the textual units of the majority class) and $P_1$ (for the textual units of the minority class), each with capacity $k$, are set to empty. 
$P_0$ and $P_1$ are initialized by the goodness values ($q_i$) of the majority and minority textual units respectively.
Then a ranked group fairness table is created which calculates the minimum number of minority textual units at each rank, given the parameter setting. 
If this table determines that a textual unit from the minority class needs to added to the summary (being generated), the algorithm adds the best element from $P_1$ to the summary $S$; otherwise it adds the overall best textual unit (from $P_0 \cup P_1$) to $S$. 
%Finally, the algorithm produces a fair summary $S$ of the input set $V$.
Thus a fair summary ($S$) of desired length $k$ is generated, adhering to a particular notion of fairness (decided by the parameter setting).

%\vspace{1mm}
%\noindent \underline{Output:} A fair summary ($S$) of desired length $k$, adhering to a particular notion of fairness.

\vspace{1mm}
\noindent 
Note that since the {\it FA*IR} algorithm provides fair ranking for two classes only~\cite{Zehlike2017}, we look to apply this algorithm for summarization of data containing tweets from exactly two social groups (i.e., the Claritin and MeToo datasets only).
%So, we report the summarization results using this methodology only on the Claritin dataset that has two classes -- Male and Female.
It is an interesting future work to design a fair ranking algorithm for more than two classes, and then to use the algorithm for summarizing data from more than two social groups.

\section{Experiments and Evaluation} \label{sec:expts}

We now experiment with different %the various 
methodologies of generating fair summaries, over the three datasets described in Section~\ref{sub:datasets}.

\vspace{2mm}
\noindent {\bf Pre-processing the datasets:} We performed standard pre-processing on the datasets including stopword removal, and stemming using Porter Stemmer. All summarization algorithms were executed on these pre-processed datasets.

\vspace{-2mm}
\subsection{Parameter settings of algorithms}

\noindent The following parameter settings are used.

\noindent $\bullet$ For all datasets, we generate all summaries of $k = 50$ tweets.

\noindent $\bullet$ The proposed algorithm (FairSumm) uses a similarity function $sim(i,j)$ to measure the similarity between two tweets $i$ and $j$.
We experimented with the following two similarity functions:\\
\underline{TFIDFsim} -- we compute TF-IDF scores for each word (unigram) in a dataset, and hence obtain a TF-IDF vector for each textual unit. 
The similarity $sim(i,j)$ is computed as the cosine similarity between the TF-IDF vectors of $i$ and $j$.\\
\underline{Embedsim} -- we obtain an embedding (a vector of dimension $300$) for each distinct word in a dataset, either by training Word2vec~\cite{Mikolov2013} on the dataset, or by considering pre-trained GloVe embeddings~\cite{pennington2014glove}.
%In either way, we get an embedding (a vector of dimension $300$) for each distinct word, which is expected to capture the semantics of the word.
We obtain an embedding for a tweet by taking the mean embedding of all words contained in the tweet. 
%(note that Word2vec vectors are additive in nature~\cite{Mikolov2013}). 
$sim(i,j)$ is computed as the cosine similarity between the embeddings of tweets $i$ and $j$.\\
%We experimented with these two similarity measures and 
We found that the performance of the FairSumm algorithm is very similar for both the similarity measures. %using either one. 
Hence, we report results for the {\it TFIDFsim} similarity measure.

\noindent $\bullet$ For our post-processing algorithm (the one based on fair ranking), the value of the parameter $\alpha_c$ needs to be decided (see Section~\ref{sub:fair-ranking}). 
We try different values of $\alpha_c$ in the interval $]0, 1[$ using grid search, and finally use $\alpha_c = 0.5$ since this value obtained the best ROUGE scores on the Claritin and MeToo datasets.

\if 0
%\vspace{-2mm}
%\subsection{Baseline: Summarizing classes separately} \label{sub:classwise}

%We consider a baseline fair summarization algorithm.
%Suppose that the textual units in the input belong to $t$ classes $Z_1, Z_2, \ldots, Z_t$, and to conform to a desired fairness notion, the summary should have $c_i$ units from class $Z_i$, $i = 1, 2, \ldots, t$ 
%where $\sum_{i=1}^{t} c_i = k$ 
%(using the same notations as in Section~\ref{sec:fairsumm}). 
%The simplest way to generate a fair summary is to {\it separately summarize the textual units belonging to each class} $Z_i$, to produce a summary of length $c_i$, and finally to combine all the $t$ summaries to obtain the final summary of length $k$.
%We refer to this method as the {\bf ClasswiseSumm} method -- specifically, we use our proposed algorithm FairSumm, without any fairness constraints, to summarize each class separately. 
%While this is the easiest way to generate a fair summary, it is not clear how good the resultant summaries will be. 
%We compare the performance of this simple baseline method with that of the two proposed fair summarization algorithms in the next section.
\fi

\subsection{Baselines}
To our knowledge, there is no existing fair text summarization algorithm. Hence we consider the standard text summmarization algorithms stated in Section~\ref{sec:existing} as baselines. Additionally, we have used our proposed FairSumm algorithm (described in Section~\ref{sec:fairsumm}) without considering any fairness constraints as a separate baseline. We call this summarization algorithm \textit{DiCoSumm}, a summarization algorithm that is optimized for both diversity and coverage, but not considering any fairness constraints.
Next, we compare the performance of the different fair summarization algorithms with all the  baselines.

\begin{table}[tb]
%\small
\footnotesize
%\tiny
\centering
\begin{tabular}{|>{\centering}p{2 cm}|>{\centering}p{2.7 cm}||p{1.0 cm}|p{1.0 cm}||p{0.7 cm}|p{0.7 cm}|p{0.7 cm}|p{0.7 cm}|}
\hline
Approach & Algorithm & \multicolumn{2}{c||}{Nos. of tweets}  & \multicolumn{2}{c|}{ROUGE-1} & \multicolumn{2}{c|}{ROUGE-2}
\\ \cline{2-8}
 & & Female & Male & Recall & $F_1$ & Recall & $F_1$
\\ \hline
 & Whole data & 2,505 & 1,532   &   &  & &  \\
   &   & (62\%) & (38\%) &   &  & & 
\\ \hline \hline
\multicolumn{8}{|c|}{{\bf Baselines (which do not consider fairness)}} \\
\hline
& ClusterRank & 33  & $17$  & 0.437  &  0.495 & 0.161 & 0.183
\\ \cline{2-8}
& DSDR & 31  & $19$  & 0.302  & 0.425 & 0.144 & 0.203 
\\ \cline{2-8}
&  LexRank & 34  & $16$  & 0.296 & 0.393 & 0.114  & 0.160
\\ \cline{2-8}
& LSA & 35  & $15$ & 0.515  & 0.504 & 0.151  & 0.147
\\ \cline{2-8}
& LUHN & 34  & $16$  & 0.380  & 0.405 & 0.128 & 0.136
\\ \cline{2-8}
& SumBasic & $27$ & $23$  & 0.314 & 0.434 & 0.108 & 0.149 
\\ \cline{2-8}
& SummaRNN & 33 & $17$  & 0.342  & 0.375 & 0.126 & 0.147  
\\ \cline{2-8}
& SummaCNN & $30$ & $20$  & 0.377  & 0.409 & 0.126 & 0.146  
%\\ \hline \hline
%\multicolumn{7}{|c|}{{\bf Without any fairness constraint}} 
\\ \cline{2-8}
& DiCoSumm & 37  & 13  & 0.548  &  0.545    &   0.172 & 0.171
\\ \hline \hline

\multicolumn{8}{|c|}{{\bf Fairness: Equal representation}} 
\\ \hline
In-processing & FairSumm & 25  & 25  & {\bf 0.560}  & \textbf{0.552} & \bf {0.188} & \textbf{0.185}
\\ \hline
Pre-processing & ClasswiseSumm & 25  & 25  & 0.545 & 0.538  &   0.172 & 0.170
\\ \hline
& Fair-ClusRank & 25 & 25 & 0.433 & 0.481 & 0.135 & 0.162
\\ \cline{2-8}
 & Fair-DSDR & 25 & 25 & 0.285 & 0.400 &  0.139 & 0.206 
\\ \cline{2-8}
Post-processing & Fair-LexRank & 25 & 25 & 0.290 & 0.370  & 0.110 & 0.153
\\ \cline{2-8}
(ReFaSumm used & Fair-LSA & 25 & 25 & 0.513 & 0.493 & 0.114 & 0.109
\\ \cline{2-8}
with existing & Fair-LUHN & 25 & 25 & 0.415 & 0.429 & 0.114 & 0.118
\\ \cline{2-8}
summarization & Fair-SumBasic & 25 & 25 & 0.314 & 0.436 & 0.111 & 0.154
\\ \cline{2-8}
algorithms) & Fair-SummaRNN & 25 & 25 & 0.356 & 0.410 & 0.126 & 0.154
\\ \cline{2-8}
& Fair-SummaCNN & 25 & 25  & 0.356 & 0.410 & 0.126 & 0.154
\\ \hline \hline

\multicolumn{8}{|c|}{{\bf Fairness: Proportional representation}} 
\\ \hline
In-processing & FairSumm & 31  & 19  & {\bf 0.572}  & \textbf{0.568} &  {\bf 0.206} & \textbf{0.202}
\\ \hline
Pre-processing & ClasswiseSumm & 31 & 19  & 0.550  &  0.541 &  0.180 & 0.173
\\ \hline 
& Fair-ClusRank & 31 & 19 & 0.439 & 0.483 & 0.133 & 0.159
\\ \cline{2-8}
& Fair-DSDR & 31 & 19 & 0.302 & 0.425 & 0.145 & 0.204
\\ \cline{2-8}
& Fair-LexRank & 31 & 19 & 0.312 & 0.406 & 0.115 & 0.160
\\ \cline{2-8}
Post-processing & Fair-LSA & 31 & 19 & 0.502 & 0.487 & 0.118 & 0.115
\\ \cline{2-8}
(ReFaSumm) & Fair-LUHN & 31 & 19 & 0.426 & 0.435 & 0.119 & 0.121
\\ \cline{2-8}
& Fair-SumBasic & 31 & 19 & 0.318 & 0.435 &0.116 & 0.159
\\ \cline{2-8}
& Fair-SummaRNN & 31 & 19  & 0.340 & 0.394 & 0.120 & 0.147
\\ \cline{2-8}
& Fair-SummaCNN & 31 & 19  & 0.340 & 0.394 & 0.120 & 0.147
\\ \hline
%\multicolumn{5}{|c|}{{\bf Fairness: Adverse Impact}} 
%\\ \hline
%FairSumm & 29  & 31  & 0.5616  & 0.1863
%\\ \hline
%FairSumm & 30 & 20  & {\bf 0.5665}  & {\bf 0.1941}
%\\ \hline
%FairSumm & 32 & 18 & 0.5662 & 0.1960
%\\ \hline
%FairSumm & 33 & 17 & 0.5640 & 0.1940
%\\ \hline

% &   &   &   &   
%\\ \hline
\end{tabular}
\caption{\textbf{Summarizing the Claritin dataset: Number of tweets written by the two user groups, in the whole dataset and the summaries of length $50$ tweets generated by different algorithms. 
Also given are the ROUGE-1 and ROUGE-2 Recall and $F_1$ scores of each summary.}}
\label{tab:fair-summ-results-claritin}
\vspace*{-10mm}
\end{table}

\begin{table}[tb]
%\small
\footnotesize
%\tiny
\centering
\begin{tabular}{|>{\centering}p{2 cm}|>{\centering}p{2.7 cm}||p{1.0 cm}|p{1.0 cm}||p{0.7 cm}|p{0.7 cm}|p{0.7 cm}|p{0.7 cm}|}
\hline
Approach & Algorithm & \multicolumn{2}{c||}{Nos. of tweets}  & \multicolumn{2}{c|}{ROUGE-1} & \multicolumn{2}{c|}{ROUGE-2}
\\ \cline{2-8}
&  & Female & Male & Recall & $F_1$ & Recall & $F_1$
\\ \hline
 & Whole data & 275 & 213   &   &  & &  \\
    &  & (56.3\%) & (43.7\%) &   &  & & 
\\ \hline \hline
\multicolumn{8}{|c|}{{\bf Baselines (which do not consider fairness)}} \\
\hline
& ClusterRank & $24$  & 26  & 0.550 & 0.560  & 0.216  & 0.223
\\ \cline{2-8}
& DSDR  &  32 & $18$  & 0.233 & 0.358  & 0.092 & 0.141 
\\ \cline{2-8}
& LexRank  &  34 & $16$  & 0.285 & 0.414  & 0.105 & 0.153
\\ \cline{2-8}
& LSA  &  $20$ & 30  & 0.511 & 0.534  & 0.175 & 0.183
\\ \cline{2-8}
& LUHN  &  $22$ & 28  & 0.520 & 0.522  & 0.219 & 0.184
\\ \cline{2-8}
& SumBasic  & $27$  & $23$  & 0.464 & 0.499  & 0.216 & 0.229
\\ \cline{2-8}
& SummaRNN  & $23$  &  27 & 0.622 & 0.636  & 0.385 & 0.394
\\ \cline{2-8}
& SummaCNN  &  $23$ & 27 & 0.622 & 0.636  & 0.385 & 0.394
\\ \cline{2-8}
%\multicolumn{7}{|c|}{{\bf Without any fairness constraint}} 
%\\ \hline
& DiCoSumm & 30  & 20  &  0.563 & 0.569  & 0.229   & 0.249
\\ \hline \hline

\multicolumn{8}{|c|}{{\bf Fairness: Equal representation}} 
\\ \hline
In-processing& FairSumm & 25  & 25  & 0.616  & 0.613 & 0.285 & 0.296
\\ \hline
Pre-processing& ClasswiseSumm & 25  & 25  & 0.587 & 0.569  & 0.189 & 0.196
\\ \hline 
& Fair-ClusRank & 25 & 25 & 0.499 & 0.532 & 0.186 & 0.198
\\ \cline{2-8}
& Fair-DSDR & 25 & 25 & 0.558 & 0.574 & 0.157  & 0.162
\\ \cline{2-8}
& Fair-LexRank & 25 & 25 & 0.511 & 0.564  & 0.209 & 0.230
\\ \cline{2-8}
Post-processing & Fair-LSA & 25 & 25 & 0.556 & 0.541 & 0.196 & 0.191
\\ \cline{2-8}
(ReFaSumm)& Fair-LUHN & 25 & 25 & 0.527 & 0.537 & 0.207 & 0.211
\\ \cline{2-8}
& Fair-SumBasic & 25 & 25 & 0.541 & 0.567 & 0.180 & 0.189
\\ \cline{2-8}
& Fair-SummaRNN & 25 & 25 & 0.623 & 0.629 & 0.371 & 0.375
\\ \cline{2-8}
& Fair-SummaCNN & 25 & 25 & \textbf{0.623} & \textbf{0.629} & \textbf{0.371} & \textbf{0.375}
\\ \hline \hline

\multicolumn{8}{|c|}{{\bf Fairness: Proportional representation}} 
\\ \hline
In-processing & FairSumm & 28  & 22  & \textbf{0.631}  &\textbf{0.648}& 0.311  & 0.338
\\ \hline
Pre-processing & ClasswiseSumm & 28 & 22  &  0.605 & 0.622  & 0.279  & 0.298
\\ \hline
& Fair-ClusRank  & 28 & 22  & 0.499  & 0.528  & 0.174  & 0.184
\\ \cline{2-8}
& Fair-DSDR  & 28 & 22  &  0.565 & 0.577  & 0.168  & 0.172
\\ \cline{2-8}
& Fair-LexRank  & 28 & 22  & 0.518  & 0.564  & 0.210  & 0.228
\\ \cline{2-8}
Post-processing & Fair-LSA  & 28 & 22  & 0.560  & 0.544  & 0.197  & 0.191
\\ \cline{2-8}
(ReFaSumm)& Fair-LUHN  & 28 & 22  & 0.533  & 0.541  & 0.213  & 0.216
\\ \cline{2-8}
& Fair-SumBasic & 28 & 22  & 0.546  & 0.569  & 0.190  & 0.198
\\ \cline{2-8}
& Fair-SummaRNN  & 28 & 22  & 0.622  & 0.636  & 0.385  & 0.394
\\ \cline{2-8}
& Fair-SummaCNN  & 28 & 22 & 0.621  & 0.636  & \textbf{0.385}  & \textbf{0.394}
\\ \hline
%\multicolumn{5}{|c|}{{\bf Fairness: Adverse Impact}} 
%\\ \hline
%FairSumm & 29  & 31  & 0.5616  & 0.1863
%\\ \hline
%FairSumm & 30 & 20  & {\bf 0.5665}  & {\bf 0.1941}
%\\ \hline
%FairSumm & 32 & 18 & 0.5662 & 0.1960
%\\ \hline
%FairSumm & 33 & 17 & 0.5640 & 0.1940
%\\ \hline

% &   &   &   &   
%\\ \hline
\end{tabular}
\caption{\textbf{Summarizing the MeToo dataset: Number of tweets written by the two user groups, in the whole dataset and the summaries of length $50$ tweets generated by different algorithms. 
Also given are the ROUGE-1 and ROUGE-2 Recall and $F_1$ scores of each summary.} 
}
\label{tab:fair-summ-results-metoo}
\vspace*{-8 mm}
\end{table}

\subsection{Results and Insights}

%\vspace{2mm}
%\noindent 
%{\bf Results:}

We now describe the results of applying various fair summarization algorithms over the three datasets. Some sample summaries obtained by using various algorithms are given in Appendix~\ref{App: Summaries}. %Supplementary Information. %(\url{https://tinyurl.com/y8gkc4he} -- anonymized URL).

To evaluate the quality of summaries, we compute ROUGE-1 and ROUGE-2 Recall and $F_1$ scores by matching the algorithmically generated summaries with the gold standard summaries (described in  Section~\ref{sub:datasets}).
Table~\ref{tab:fair-summ-results-claritin} reports the results of summarizing the Claritin dataset. 
We compute summaries without any fairness constraint, and considering the two fairness notions of {\it equal representation} and {\it proportional representation} (explained in Section~\ref{sec:fairness-notions}). 
In each case, we state the number of tweets in the summary from the two user groups, and the ROUGE scores of the summary. 
Similarly, Table~\ref{tab:fair-summ-results-metoo} and Table~\ref{tab:fair-summ-results-uselection} report the results for the MeToo dataset and the US-Election dataset respectively.

\begin{table}[tb]
%\small
\footnotesize
%\tiny
\centering
\begin{tabular}{|>{\centering}p{2cm}|>{\centering}p{2cm}||p{1.2 cm}|p{1.2 cm}|p{1.2 cm}||p{0.7 cm}|p{0.7 cm}|p{0.7 cm}|p{0.7 cm}|}
\hline
Approach & Algorithm & \multicolumn{3}{c||}{Nos. of tweets}  & \multicolumn{2}{c|}{ROUGE-1} & \multicolumn{2}{c|}{ROUGE-2}
\\ \cline{3-9}
 & & Pro-Rep & Pro-Dem & Neutral & Recall & $F_1$ & Recall & $F_1$
\\ \hline
& Whole data &  1,309    &   658   & 153  &   &  & & \\
  &         & (62\%) & (31\%) &  (7\%) &    &  & &
\\ \hline \hline
\multicolumn{9}{|c|}{{\bf Baselines (which do not consider fairness)}} \\
\hline 
& ClusterRank &  32 & $15$ & $3$ & 0.247 & 0.349 & 0.061 & 0.086  
\\ \cline{2-9}
& DSDR & $28$  & 19  & $3$  & 0.215 & 0.331  & 0.067 & 0.104
\\ \cline{2-9}
& LexRank & $27$  & 20  & $3$  & 0.252 & 0.367  & 0.078 & 0.114
\\ \cline{2-9}
& LSA & $24$  & 20 & $6$  & 0.311 & 0.404  & 0.083 & 0.108
\\ \cline{2-9}
& LUHN & 34  & $13$  & $3$  & 0.281 & 0.375 & 0.085 & 0.113
\\ \cline{2-9}
& SumBasic &  $27$ & 23  & $0$  & 0.200 & 0.311 & 0.051 & 0.080
\\ \cline{2-9}
& SummaRNN & 34 & $15$  & $1$  & 0.347 & 0.436 & 0.120 & 0.160  
\\ \cline{2-9}
& SummaCNN & 32 & 17  & $1$  & 0.337 & 0.423 & 0.108 & 0.145  
%\\ \hline \hline
%\multicolumn{8}{|c|}{{\bf Without any fairness constraint}} 
\\ \cline{2-9}
& DiCoSumm & 34  & 12  & 4  & 0.359  &  0.460 & 0.074 & 0.091
\\ \hline \hline

\multicolumn{9}{|c|}{{\bf Fairness: Equal representation}} 
\\ \hline
In-processing & FairSumm &  17 & 17  & 16  & {\bf 0.368} & \textbf{0.467}  & {\bf 0.078} & \textbf{0.096}
\\ \hline
Pre-processing & ClasswiseSumm &  16 & 16  & 18  &  0.363 & 0.467 & 0.071 & 0.088
\\ \hline \hline

 \multicolumn{9}{|c|}{{\bf Fairness: Proportional representation}} 
\\ \hline
In-processing & FairSumm & 31  &  15 &  4 &  {\bf 0.376} & \textbf{0.490}& {\bf 0.094} & \textbf{0.116}
\\ \hline
Pre-processing & ClasswiseSumm & 30 & 15  & 5  &  0.367 & 0.454 & 0.081 & 0.100
%\\ \hline
% &   &   &   &   & 
\\ \hline \hline 

\multicolumn{9}{|c|}{{\bf Fairness: No Adverse Impact}} 
\\ \hline
& FairSumm & 29  &  17 & 4 &  0.371 & 0.484  & 0.086 & 0.102
\\ \cline{2-9}
& FairSumm & 30 & 16  & 4  &  0.372 &  0.489 & 0.087 & 0.109
\\ \cline{2-9}
In-processing & FairSumm & 31  &  15 &  4 &  {\bf 0.376} & \textbf{0.490} & {\bf 0.094} & \textbf{0.116}
\\ \cline{2-9}
& FairSumm & 31  & 16 &  3 &  0.371 & 0.477 &  0.085 & 0.096
\\ \cline{2-9}
& FairSumm & 32  & 15 &  3 &  0.371 & 0.473 & 0.085 & 0.093
\\ \hline

\end{tabular}
\caption{\textbf{Summarizing the US-Election dataset: Number of tweets of the three classes, in the whole dataset and the summaries of length $50$ tweets generated by different algorithms. 
Also given are the ROUGE-1 and ROUGE-2 Recall and $F_1$ scores of each summary.}
}
\label{tab:fair-summ-results-uselection}
\vspace*{-6mm}
\end{table}

The FairSumm algorithm (in-processing algorithm) and the ClasswiseSumm algorithm (pre-processing algorithm) are executed over all three datasets. 
For the two-class Claritin and MeToo datasets, we also apply our post-processing methodology (stated in Section~\ref{sub:fair-ranking}) where a fair ranking scheme is used for fair summarization (results in Table~\ref{tab:fair-summ-results-claritin} and Table~\ref{tab:fair-summ-results-metoo}). 
Specifically, we use our post-processing methodology over the existing summarization algorithms described in Section~\ref{sec:existing} such as ClusterRank, LexRank, SummaRNN, SummaCNN, etc. 
The resulting fair summarization algorithms are denoted as Fair-ClusRank, Fair-LexRank, Fair-SummaRNN, Fair-SummaCNN, and so on. 
Note that, for generating a fixed length summary, the neural models use only the textual units labeled with $1$, ranked as per their confidence scores. %unlike the  traditional summarization algorithms that rank all the textual units. 
Hence, in Fair-SummaRNN and Fair-SummaCNN methods, we have considered the ranked list of only those textual units that are labeled with $1$.
%\footnote{For the SummaRNN and SummaCNN methods, theoretically, it may so happen that there is not sufficient representation (required for a certain fairness notion) of a certain class in the set of textual units that are labeled with $1$. In such cases, a fair ranking algorithm can not make the final summaries fair. However, we did not encounter such cases for the datasets we used.}

\vspace{2mm}
\noindent \underline{\bf Insights from the results:}
We make the following observations from the results shown in Table~\ref{tab:fair-summ-results-claritin}, Table~\ref{tab:fair-summ-results-metoo} and Table~\ref{tab:fair-summ-results-uselection}.

\vspace{2mm}
%\noindent $\bullet$ {\bf Summarizing different classes separately does not yield good summarization:}
\noindent $\bullet$ {\bf In-processing and post-processing methods perform better than pre-processing:}
Across all datasets, the in-processing FairSumm algorithm achieves higher ROUGE scores than 
ClasswiseSumm, considering the same fairness notion. 
Note that in the ClasswiseSumm approach, the same FairSumm algorithm is used on each class separately.
Hence, the pre-processing approach of separately summarizing each class leads to relatively poor summaries, as compared to the in-processing FairSumm methodology.
This difference in performance is probably because, if similar tweets / opinions are posted by different social groups, ClasswiseSumm can include multiple similar posts in the summary, thereby leading to redundancy in the summary. On the other hand, FairSumm optimizes coverage and diversity across all textual units taken together, thereby avoiding redundancy in the summary.

The in-processing FairSumm algorithm and some of the post-processing approaches achieve comparable performances. For instance, while FairSumm performs decidedly better than all other algorithms for the Claritin dataset, the post-processing approaches Fair-SummaCNN and Fair-SummaRNN perform better in most cases over the MeToo dataset.

Note that the performances of the pre-processing and post-processing algorithms depend on that of the original summarization algorithm that is used. 
As such, the pre-processing and post-processing algorithms can be useful in situations where an existing summarization algorithm is preferred, e.g., when a firm has a proprietary summarization algorithm.

\vspace{2mm}
\noindent $\bullet$ {\bf Proposed algorithms are generalizable to different fairness notions:}
Table~\ref{tab:fair-summ-results-claritin}, Table~\ref{tab:fair-summ-results-metoo} and Table~\ref{tab:fair-summ-results-uselection} demonstrate that the proposed algorithms are generalizable to various fairness notions. We demonstrate summaries conforming to equal representation and proportional representation for all three datasets.
Additionally, Table~\ref{tab:fair-summ-results-uselection} shows different summaries that can be generated using FairSumm considering the `no adverse impact' fairness notion (such rows are omitted from other tables for brevity).

In general, summaries conforming to proportional representation achieve higher ROUGE scores than summaries conforming to other fairness notions, 
probably because the human assessors intuitively attempt to represent different opinions (coming from different social groups) in a similar proportion in the gold standard summaries as what occurs in the input data (even though they were not told anything about ensuring fairness while writing the gold standard summaries).

\vspace{2mm}
\noindent $\bullet$ {\bf Ensuring fairness does not lead to much degradation in summary quality:}
For all three datasets, we observe that FairSumm with fairness constraints always achieves higher ROUGE scores than DiCoSumm (without fairness constraints). 
%In some cases, ClasswiseSumm with fairness constraints also achieve higher ROUGE scores than FairSumm without fairness constraints.
Also, we can compare %Table~\ref{tab:summ-results-claritin} with Table~\ref{tab:fair-summ-results-claritin} (both on Claritin dataset), and 
%Table~\ref{tab:summ-results-metoo} with Table~\ref{tab:fair-summ-results-metoo} (both on MeToo dataset)to compare 
the performances of the existing summarization algorithms (e.g., DSDR, LexRank, SummaRNN) without any fairness constraint, and after their outputs are made fair using the methodology in Section~\ref{sub:fair-ranking}.
We find that the performances in the two scenarios are comparable to each other. 
%-- while most of the ROUGE-1 Recall scores are higher in Table~\ref{tab:fair-summ-results-claritin} (when the summary is made fair),
%most ROUGE-2 Recall scores are higher in Table~\ref{tab:summ-results-claritin} (in the original summaries).
%The trends are similar in case of $F_1$ Scores.
In fact, for a few cases, the ROUGE scores marginally improve after the summaries generated by an algorithm are made fair, over those of the original summary generated by the same algorithm.
%If better summarization algorithms are used (which better judge the importance of textual units), this mechanism can also generate good quality fair summaries.
%However, it is uncertain whether the fair summaries will achieve similar quality as the original summaries output by the summarization algorithm.
%Note that the input to the FA*IR ranking-based summarization algorithm is the ranked textual units from an existing summarization algorithm (say $Algo_1$). 
%Let the summary produced after FA*IR ranking be called $S_{rank}$ and that originally produced by $Algo_1$ be $S_1$. 
%The quality of $S_{rank}$ (as measured by ROUGE scores) depends upon the quality of ranking produced by $Algo_1$, and the quality of $S_{rank}$ may improve if better ranking of textual units can be provided. 
%However, as depicted by Table~\ref{tab:fair-summ-results-claritin}, we can {\it not} confidently conclude that using a fairness-preserving ranking algorithm like FA*IR will produce better quality summaries or summaries at least as good as $S_1$ (as measured by ROUGE scores).
Thus, {\it making summaries fair does not lead to much degradation in summary quality} (as measured by ROUGE scores).
%, and can improve quality of summaries in some cases (as measured by ROUGE scores).

\if 0

\vspace{2mm}
\noindent {\bf Comparing FairSumm with other algorithms:}
As demonstrated in Table~\ref{tab:summ-results-claritin} and  Table~\ref{tab:summ-results-uselection}, none of the existing summarization algorithms generate fair summaries for both datasets. 
For the US-Election dataset, the recently proposed supervised neural algorithms achieve higher ROUGE scores than the unsupervised algorithms; however, the summaries produced by the neural methods under-represent the minority neutral group in the US-Election dataset. 

The summaries generated by FairSumm, apart from ensuring fairness, achieve very comparable ROUGE scores as the best-performing algorithms.
For both the Claritin and US-Election datasets, FairSumm with `proportional representation' achieves higher ROUGE scores than all other methods. 

\fi

\vspace{1mm}
\noindent 
Overall, the results signify that, the proposed fair summarization algorithms can not only ensure various fairness notions in the summaries, but also can generate summaries that achieve comparable (or better) ROUGE scores than many well-known summarization algorithms (which often do not generate fair summaries, as demonstrated in Section~\ref{sec:existing}).
%Additionally, ensuring that summaries are fair according to standard fairness notions can actually ensure good quality of the summaries.\red{the last line seems repeatation}

%\vspace{-2 mm}
\subsection{Effects of degraded information of demographic details}
\label{sec:DegradedInfo}

Our proposed algorithms assume the availability of class information of the textual units under consideration e.g., gender information for the tweets in Claritin and MeToo  datasets, and %sentiment of the tweets
ideological leaning for tweets in the US-Election dataset. 
%Note that, such information is not impossible to get.
In cases where such class information is not available, we need to resort to inference mechanisms. For instance, there are multiple methodologies to infer demographic details of %the writer of the tweets 
people from their writing styles with a high level of accuracy~\cite{shariff2016correlation, olteanu2016characterizing, sloan2013knowing}. Moreover, in social media, additional information can be utilized to infer demographic details of their users, such as user names and profile pictures~\cite{chakraborty2017makes}. 
Similarly, sentiment analysis can be deployed to infer the opinions.
%However, we can not always divide the tweets or opinions based on the demographics of the writers of the tweets. Sometimes the classifications can also be done based on the sentiments of the content of the tweets i.e. the opinion that they (tweets) represent e.g. for and against a particular issue. Sentiment analysis has been one of the well studied domain in computational linguistics. Many prior works~\cite{pak2010twitter, agarwal2011sentiment, kouloumpis2011twitter} have proposed methodologies that infer the sentiment of tweets, which can be used to classify tweets based on the sentiment of the opinions of the tweeters. 
However, such inferences may not be absolutely perfect, and the accuracy of these inference methodologies will eventually decide how well the predicted labels replicate the true class labels. %gold standard labels. Since our proposed algorithm also depend on these information, 
%In this section,
In this section, we check how the inaccuracy in the inference mechanism may impact the performance 
%robustness check of the in-processing fair summarization algorithm 
of the proposed FairSumm algorithm.

\vspace{1 mm}
\noindent
\textbf{Experimental Setup:} 
We performed these experiments for all the three datasets, and obtained qualitatively similar results. Hence, for brevity, we are reporting the experimental results only on the MeToo dataset. 
We assume the existence of a classifier which can infer the gender information of the authors of textual posts (required for our FairSumm algorithm), with certain level of error, for the MeToo dataset. % and Claritin datasets. 
To simulate the effect of a classifier with $x\%$ error rate, we change the class labels (i.e., consider the inferred label to be `male' if the true label was `female', and vice versa) of {\it randomly selected} $x\%$ of the tweets in the dataset. 
Now this degraded information of gender labels is given as input to the FairSumm algorithm to create the fair summaries. 
We experiment with error rates $x = 10\%, 20\%, 30\%$. 
For every error rate, we repeat the experiment $100$ times, and then report the average results over all experiments. 
Note that, for checking the fairness property (i.e., the proportion of tweets from various groups in the summaries), we use the true labels (and not the inferred labels).

\begin{table}[tb]
	%\small
	\footnotesize
	%\tiny
	\centering
	\begin{tabular}{|>{\centering}p{1.5 cm}||p{1.0 cm}|p{1.0 cm}||p{0.7 cm}|p{0.7 cm}|p{0.7 cm}|p{0.7 cm}|}
		\hline
		Error Rate & \multicolumn{2}{c||}{Nos. of tweets}  & \multicolumn{2}{c|}{ROUGE-1} & \multicolumn{2}{c|}{ROUGE-2}
		\\ \cline{2-7}
		& Female & Male & Recall & $F_1$ & Recall & $F_1$
		\\ \hline
		Whole data & 275 & 213   &   &  & &  \\
		& (56.3\%) & (43.7\%) &   &  & & 
		\\ \hline \hline
		\multicolumn{7}{|c|}{{\bf Fairness: Equal representation}} 
		\\ \hline
		0 \% & 25  & 25  & \textbf{0.616}  & \textbf{0.613} & \textbf{0.285} & \textbf{0.296}
		\\ \hline
		10 \% & 27  & 23  & 0.612  & 0.604 & 0.285 & 0.286
		\\ \hline
		20 \% & 29  & 21  & 0.596 & 0.582  & 0.282 & 0.287
		\\ \hline
		30 \% & 29  & 21  & 0.591 & 0.590  & 0.282 & 0.287
		\\ \hline \hline
		
		\multicolumn{7}{|c|}{{\bf Fairness: Proportional representation}} 
		\\ \hline
		0 \% & 28  & 22  & \textbf{0.631}  &\textbf{0.648}& \textbf{0.311}  & \textbf{0.338}
		\\ \hline
		10 \% & 29  &  21  & 0.624  & 0.633 & 0.304  & 0.317
		\\ \hline
		20 \% & 30  &  20  &  0.615 & 0.610  & 0.297  & 0.302
		\\ \hline 
		30 \% & 31  &  19  &  0.614 & 0.610  & 0.293  & 0.300
		\\ \hline 
	\end{tabular}
	\caption{\textbf{Effect of degraded information of gender in MeToo dataset: Number of tweets written by the two user groups, in the whole dataset and the summaries of length $50$ tweets generated by FairSumm algorithm. 
	Also given are the ROUGE-1 and ROUGE-2 Recall and $F_1$ scores of each summary. Each result is averaged over 100 random experiments for each of the error rates.} 
	}
	\label{tab: degraded-fair-summ-results-metoo}
	\vspace*{-5mm}
\end{table}

\vspace{1 mm}
\noindent
\textbf{Observations:} 
%The results of the experiments for Claritin and MeToo datasets are reported in Tables~\ref{tab: degraded-fair-summ-results-claritin} and~\ref{tab: degraded-fair-summ-results-metoo} respectively. 
Table~\ref{tab: degraded-fair-summ-results-metoo} reports the results obtained over the MeToo dataset, with increasing amount of noise/error in the demographic labels. 
As expected, the error/noise in the  demographic inference has some effects on the fairness property that the summaries are meant to satisfy.
%Even though the summaries satisfy the fairness constraints based on the degraded information, the same does not hold true when we consider the %gold standard
%true gender information. 
%Another point to observe is, 
With increasing error rate in the prediction of the class labels, %more is the chance of compromising on fairness.
the summaries deviate further from the desired fairness criterion.
%\red{Especially, the majority group (Female in case of MeToo dataset) gets favored by such random noise, the reason being as follows. 
%Due to randomly flipping the labels of $x\%$ of the tweets, a larger number of tweets posted actually by Female users were mislabeled as Male, as opposed to tweets actually posted by male users being mislabeled as Female.
%If any of these mislabled tweets that were actually posted by females get included in the summary, it disturbs the fairness goal in favor of the majority female group.}
However, it is interesting to note that even with degraded demographic information, the summaries generated by FairSumm have better fairness property than the summaries generated by many of the baseline algorithms. 
%It is evident from Tables~\ref{tab: degraded-fair-summ-results-claritin} and~\ref{tab: degraded-fair-summ-results-metoo} 
Also it is evident from Table~\ref{tab: degraded-fair-summ-results-metoo}
that the degradation in the availability of demographic details does {\it not} affect the quality of the summaries much (as measured by ROUGE scores). 
In fact, even with the degraded gender information, the FairSumm algorithm outperforms many of the existing summarization algorithms in terms of ROUGE scores.
%, as is seen from the fact that ROUGE scores diminish only slightly when 10\% labels are wrongly inferred. 
%As more and more noise is added, the ROUGE scores diminish gracefully.

Hence, we can conclude that, though  
the accuracy of the inference methodology %plays a vital part in the improvement of fairness in the end summary 
is an important factor when the actual class information is not present (especially in order to achieve the fairness goals), the quality of summaries produced by FairSumm is generally robust to such noise in the inferred labels.

%But the quality and fairness of the summary output by FairSumm fall gracefully as the amount of noise in the inferred demographic labels increases.

\section{Concluding Discussion} \label{sec:conclu}

\noindent 
To our knowledge, this work is the first attempt to consider  
fairness in textual  summarization.
Through experiments on several user-generated microblog datasets, we show that existing algorithms often produce summaries that are not fair, even though the text written by different social groups are of comparable quality. 
Note that, we do {\it not} claim %that any of 
the existing algorithms to be intentionally biased towards/against any social group. Since these algorithms attempt to optimize %some other 
only one metric (e.g., textual quality of the summary), the unfairness comes as a side-effect.
We further propose algorithms to generate high-quality summaries that conform to various standard notions of fairness (implementations available at \url{https://github.com/ad93/FairSumm.git}). 
%In fact, ensuring the fairness of summaries often leads to enhancing the quality of the summary as well.
%Recently, multiple research works have investigated the unfairness of automated decision making systems, where an algorithm is labelled unfair if it denies `a desired outcome' to an individual or a group of individuals on grounds that are inappropriate, e.g., based on a protected attribute according to laws of a country~\cite{zafar2017fairness}.
%In the context of summarization, the desired outcome (for a textual unit) is to be included in the summary, assuming that only the textual units included in the summary will get visibility or be seen by humans (and not the entire input data). 
%The question of fairness is especially significant 
%In scenarios where the textual units in the input data (sentences, tweets, etc.) are associated with a class %sensitive or protected 
%attribute (such as the political leaning of the text, gender/race/religion of the author, or the news media source of the text), 
%Additionally, it is also important to check the fairness of a summary when the textual units in the input data come from multiple sources (e.g., multiple news media, or multiple socially salient groups) -- in such cases, 
%it needs to be investigated whether all such classes are being represented fairly in the summary.
These algorithms will help in addressing the concern that using a (inadvertently) `biased' summarization algorithm can reduce the visibility of the voice/opinion of certain social groups in the summary. 
Moreover, downstream applications that use the summaries (e.g., for opinion classification and rating inference~\cite{lloret2010experiments}) would benefit from a fair summary.

\vspace{1 mm}
\noindent
\textbf{Limitations and future directions: }
There are potential limitations of the fair summarization algorithms presented in this paper.
All three algorithms need as input the class (e.g., socially salient group) information to which each textual unit belongs. Where such class information is not readily available, we need to infer these information, which may impact the fairness objectives to some extent, as discussed in Section~\ref{sec:DegradedInfo}. 
A major limitation of ReFaSumm comes from the fair ranking algorithm being  applied in the post-processing phase. The FA*IR framework is designed for scenarios where only two socially salient groups are defined (e.g., male and female). Hence our post-processing algorithm is presently applicable only in such cases. We plan to extend ReFaSumm to more than two classes in future work.

We also note that, in certain special cases, the fair summarization algorithms developed in this work may lead to degradation in the summary quality. For example, let us assume a scenario where we are summarizing tweets posted by two equally-sized groups of users, e.g., group $A$ and group $B$, and the fairness objective is to achieve {\it Equal Representation}, i.e., both groups should have the same number of tweets in the summary. 
Now, if the variability of opinions within the groups are different -- e.g., everyone from group $A$ has the same opinion on an issue, while people in group $B$ have many varied opinions on the same issue -- then the proposed method will not generate a good summary because there will be redundant tweets posted by users of group $A$, while some of the diverse opinions posted by users of group $B$ may not be included in the summary. 
Hence, if the distribution of opinions is very different from the distribution of people belonging to different social groups, then the summaries may not be of good quality. 
This situation leads us to an interesting question of whether to look for {\it fair representation across demographic groups}, or for {\it fair representation across the different opinions} -- a question that we would like to investigate in future work.

Going beyond summarization, deciding the social grouping is often a normative question in many socio-technical applications, which requires decisions at the policy level. In some applications, legal doctrines may suggest what should be the social grouping (which often emerges from long deliberations and historical contexts); whereas, in other cases, the corresponding online platforms (such as social media sites like Facebook or Twitter) may have their own %internal 
guidelines to decide groups they want to be considerate about. 
Similar to most of the recent algorithms trying to incorporate group fairness (including fair classification algorithms~\cite{zafar2017fairness, zemel2013learning, dwork2012fairness}), the algorithms proposed in this paper also consider the grouping to be given apriori. 
However, questions on what constitute the right grouping are important, and should be more widely discussed in the  research community.

%\vspace{1 mm}
%\noindent
%\new{The proposed algorithm is color sensitive, in the sense it requires to know the classes to which the different tweets belong. To this end, it can be argued that probably introducing the sensitive attributes might not be the best way to create fair summaries. However, it is an important first step towards solving the bigger problem. These limitations in the proposed methodologies lay proper foundation for a more generic color-blind approach to bring in fairness in summarization. In many practical social issues, usually the division in opinions does not always align with different socially salient groups. In such scenarios, generating summaries that provide maximum satisfaction to the authors of the textual units is the way ahead. This will not only take care of the author satisfaction but also cover all the different perspectives in the given issues, irrespective of the spread of opinions across groups.}

Finally, looking at a higher level, fairness-preserving information filtering algorithms like the ones proposed in this paper are of significant societal importance. 
Today social media sites are the gateway of information for a large number of people worldwide; and the algorithms (search, recommendation, sampling, summarization etc.) deployed in these sites act as the gatekeepers. 
%People get to see what these different algorithms tailor for them to see. However,
If these algorithms %being keyed to \textit{relevance} or {\it quality}, 
lack the sense of embedded ethics or civic responsibilities, %and hence 
they may not be fully suitable to curate information for the heterogeneous society. 
%As a result, the flow of information in the society is no longer balanced. Lack of perceptions in issues involving different social sub-groups can eventually make the society disintegrated. Hence, a sense of
Thus, incorporating fairness in algorithms is the need of the hour. As discussed in Section~\ref{sec:related}, %recently algorithms are undergoing a fair-revolution. This revolution must bring the journalistic ethics to the Web; barring which a smooth functioning of the society seems improbable. 
recent research works are taking correct steps in that direction. Likewise, we believe that our work will open up multiple  interesting research problems on fair summarization, such as extending the concept of fairness to abstractive summaries, or estimating user preferences for fair summaries in various applications, and will be an important addition to the emerging literature on fairness in algorithmic decision making.

\bibliographystyle{ACM-Reference-Format}
\bibliography{fairness}

\newpage
\noindent {\bf {\Huge APPENDICES}}

\appendix
%\section{FairSumm: Proposed In-processing Algorithm for Fair Summarization}
\section{FairSumm: Mathematical details}
\label{App: FairSumm}

This section describes FairSumm, our proposed in-processing algorithm for fair summarization, in more detail. 
This algorithm
treats summarization as an optimization problem of a submodular, monotone objective function, where the fairness requirements are applied as constraints.
%Our proposed fairness-preserving summarization algorithm treats summarization as an optimization problem of a submodular, monotone objective function, where the fairness requirements are applied as constraints. 
In this section, we first establish the theoretical premise of our algorithm, and then describe our algorithm. 
The symbols used in this section are given in Table~\ref{tab:symbols}.

\subsection{Submodularity and Monotonicity}

%In this subsection, we will briefly some of important concepts underlying our proposed algorithm.

%\subsubsection*{Definitions}
\noindent \underline{\bf Definitions:}
Let $V = \{v_1, v_2, ..., v_n\}$ be the set of elements to be summarized, where each $v_i$ represents a textual unit (e.g., a tweet). 
We define a function $\mathcal{F}$ : $2^V$ $\rightarrow$ $\mathbb{R}$ (the set of real numbers)
%(where $\mathbb{R}$ is the set of real numbers) 
that assigns a real value to a subset (say, $S$) of $V$.
%, i.e. $S$ $\subseteq$ $V$. 
Our goal is to find $S$ ($\subseteq$ $V$) such that $|S| \le k$, where $k \in \mathbb{R}$ is the desired length of the summary (specified as an input), 
and for which $\mathcal{F}$ is maximized (i.e. $S$ = $argmax_{B\subseteq V} \mathcal{F}(B)$). 
So, from a set of textual units $V$, we look to find a summary $S$ that maximizes an objective function $\mathcal{F}$.

\vspace{2mm}
\noindent \textbf{Definition 1} (Discrete derivative)~\cite{Krause2014}: For $e$ $\in$ $V$ and a function  $f: 2^V \rightarrow \mathbb{R}$, let $\Delta(e|S)$ = $f(S \cup \{e\})$ -- $f(S)$ be the \textit{discrete derivative} of $f$ at $S$ with respect to $e$.

\vspace{2mm}
\noindent \textbf{Definition 2} (Submodularity): a function $f$ is \textit{submodular} if for every $A$ $\subseteq$ $B$ $\subseteq$ $V$ and $e$ $\in$ $V$ $\setminus$ $B$ (i.e. $e$ $\in$ $V$ and $e$ $\notin$ $B$), %the following holds
\begin{equation}
\Delta(e|A) \ge \Delta(e|B)
\end{equation}
This property is popularly called the {\it property of diminishing returns}.

\vspace{2mm}
\noindent \textbf{Definition 3} (Monotonicity): The function $f$ is \textit{monotone} (or \textit{monotone nondecreasing}) if for every $A$ $\subseteq$ $B$ $\subseteq$ $V$, $f(A)$ $\le$ $f(B)$

\begin{table} [h]
\centering
%\small
\footnotesize
\renewcommand{\arraystretch}{1.2}
\begin{tabular}{c|p{0.7\columnwidth}}
\hline
\textbf{Symbols} & \textbf{Meanings}\\
\hline
$V$ & Set of textual units to be summarized %(sentences in a document or tweets in a tweet collection) 
\\ \hline
$N$ & $|V|$, number of textual units to be summarized \\ \hline
$t$ & Number of classes to which the textual units in $V$ belong (e.g., $t=2$ for MeToo dataset, $t=3$ for US-Election) \\ \hline
$Z_1, Z_2, \ldots, Z_t$ & The $t$ classes of the textual units \\ \hline
$S$ & Summary conforming to fairness notions ($S \subseteq V$) \\ \hline
$k$ & Desired length of summary, $|S|$ \\ \hline
$c_i$, $i = 1 \ldots t$ & Minimum number of textual units from class $Z_i$ to be included in $S$ (to satisfy a given fairness notion) \\ \hline

$\mathcal{F}$ & Objective function (overall goodness measure of $S$)\\ \hline
$\mathcal{L}$ & Coverage function 
(goodness measure of $S$)\\ \hline
$\mathcal{R}$ & Diversity reward function (goodness measure of $S$)\\ \hline
$sim(i,j)$ & Similarity score between two textual units $i, j \in V$ %and $j \in V$ 
\\ \hline
$\mathcal{M}$ & A (partition) matriod\\ \hline
$\mathcal{I}$ & Set of partitions of matriod $\mathcal{M}$\\ \hline
%$P$ & Number of partition matriods\\ \hline
$G$ & The optimized fair summary produced by Algorithm~\ref{algo:atga}\\
\hline
\end{tabular}
\caption{\textbf{The main symbols used in Appendix~\ref{App: FairSumm}.}}
\label{tab:symbols}
\end{table}

\vspace{2mm}
%\subsubsection*{Properties} 
\noindent \underline{\bf Properties:}
We now discuss some of the important properties of monotone submodular functions which we will exploit in our problem formulation.

\vspace{2mm}
\noindent \textbf{Property 1} (The class of submodular functions is closed under non-negative linear combinations): Let $f_1$, $f_2$, ..., $f_n$ defined by \\$f_i$ : $2^V$ $\rightarrow$ $\mathbb{R}$ ($i$ = 1, 2, ..., $n$) be submodular functions and $\lambda_1$, $\lambda_2$, ..., $\lambda_n$ be non-negative real numbers. Then $\sum_{i=1}^{n} \lambda_i f_i$ is also submodular.\\
\textbf{Proof}: This is a well-known property of submodular functions (e.g., see \cite{Lin2011}),
hence the proof is omitted.
% \footnote{See \url{https://en.wikipedia.org/wiki/Submodular_set_function} as seen on 19th May, 2018.} 

\vspace{2mm}
\noindent \textbf{Property 2} (The class of monotone functions is closed under non-negative linear combinations): Let $f_1$, $f_2$, ..., $f_n$ defined by \\ $f_i$ : $2^V$ $\rightarrow$ $\mathbb{R}$ ($i$ = 1, 2, ..., $n$) be monotone functions and $\lambda_1$, $\lambda_2$, ..., $\lambda_n$ be non-negative real numbers. Then $\sum_{i=1}^{n} \lambda_i f_i$ is also monotone.\\
\textbf{Proof}: Let $A_1$ $\subseteq$ $A_2$ $\subseteq$ $V$. Then, since each $f_i$ is a monotone, $f_i(A_1)$ $\le$ $f_i(A_2)$. Let $F$ = $\sum_{i=1}^{n} \lambda_i f_i$.\\
\noindent \textit{Case I}: Let $\lambda_i$ $=$ 0, for all $i = 1, 2, \ldots, n$. 
Then, $F$ = 0 and is a constant (non-decreasing) function.\\
\textit{Case II}: Let $\lambda_i$ $>$ 0, for all $i = 1, 2, \ldots, n$. Then we can write
\begin{align*}
&A_1 \subseteq A_2
\Rightarrow f_i(A_1) \le f_i(A_2), \; \forall i \;\; \Rightarrow \sum_{i}^{n} f_i(A_1) \le \sum_{i}^{n}  f_i(A_2) \\
&\Rightarrow \sum_{i}^{n} \lambda_i f_i(A_1) \le \sum_{i}^{n} \lambda_i f_i(A_2) \;\;
[\because \; \lambda_i > 0, \forall i] \\
%[\mbox{since} \; \lambda_i > 0, \forall i] \\
&\Rightarrow F(A_1) \le F(A_2).
\end{align*}
\textit{Case III}: Let only some $m$ ($m$ $>$ 0 and $m$ $<$ $n$) $\lambda_i$s $>$ 0. Let, without any loss of generality,  such $\lambda_i$s be $\lambda_1$, $\lambda_2$, ..., $\lambda_m$. Then we have
\begin{align*}
&A_1 \subseteq A_2 
\Rightarrow f_i(A_1) \le f_i(A_2), \forall i \; \Rightarrow \sum_{i}^{m} f_i(A_1) \le \sum_{i}^{m}  f_i(A_2) \\
&\Rightarrow \sum_{i}^{m} \lambda_i f_i(A_1) \le \sum_{i}^{m} \lambda_i f_i(A_2) \hspace{1mm}%\\
[\because \lambda_i > 0, \mbox{for} \; i = 1, 2, \ldots, m)] \\
%[\mbox{since} \lambda_i > 0, \mbox{for} \; i = 1, 2, \ldots, m)] \\
&\Rightarrow \sum_{i}^{n} \lambda_i f_i(A_1) \le \sum_{i}^{n} \lambda_i f_i(A_2) \hspace{1mm} %\\
[\because \lambda_i = 0, \mbox{for} \; i = m+1, m+2, \ldots, n)] \\
%[\mbox{since} \lambda_i = 0, \mbox{for} \; i = m+1, m+2, \ldots, n)] \\
&\Rightarrow F(A_1) \le F(A_2)  
\end{align*}
This completes the proof.

\vspace{2mm}
\noindent \textbf{Property 3} (The class of monotone submodular functions is closed under non-negative linear combinations): Let $f_1$, $f_2$, ..., $f_n$ defined by $f_i$ : $2^V$ $\rightarrow$ $\mathbb{R}$ ($i$ = 1, 2, ..., $n$) be monotone submodular functions and $\lambda_1$, $\lambda_2$, ..., $\lambda_n$ be non-negative real numbers. Then $\sum_i \lambda_i f_i$ is also monotone submodular.\\
\textbf{Proof}: This follows trivially from Properties 1 and 2.

\vspace{2mm}
\noindent \textbf{Property 4}: Given functions $F$: $2^V$ $\rightarrow$ $\mathbb{R}$ and $f$: $\mathbb{R}$ $\rightarrow$ $\mathbb{R}$, the composition $F'$ = $f$ $\circ$ $F$ : $2^V$ $\rightarrow$ $\mathbb{R}$ (i.e., $F'(S)$ = $f(F(S))$) is nondecreasing sub-modular, if $f$ is non-decreasing concave and $F$ is nondecreasing submodular.\\ 
%~\cite{Lin2011}\\
%Sum of monotones is monotones, weighted sum of submodular is submodular
\textbf{Proof}: This is also a well-known property of submodular functions (e.g., see~\cite{Lin2011}), hence the proof is omitted.
% \footnote{See \url{https://en.wikipedia.org/wiki/Submodular_set_function} as seen on 19th May, 2018.} 

\vspace{-2mm}
\subsection{Formulation of the objective function for summarization}

We now look for an objective function for the task of extractive summarization. 
Following the formulation by 
Lin {\it et al.}~\cite{Lin2011}, we use monotone submodular functions to construct the objective function. We consider the following two aspects of an extractive text summarization algorithm:

\vspace{1mm}
\noindent \underline{\bf Coverage}: Coverage refers to amount of information covered in the summary $S$, measured by a function, say, $\mathcal{L}$. The generic form of $\mathcal{L}$ can be 
\begin{equation}
\mathcal{L}(S) = \sum_{i \in S, j \in V} sim_{i,j}
\label{eqn:coverage}
\end{equation}
where $sim_{i,j}$ denotes the similarity between two textual units (tweets, sentences, etc.) $i \in V$ and $j \in V$. 
Thus, $\mathcal{L}(S)$ measures the overall similarity of the textual units included in the summary $S$ with all the textual units in the input collection $V$.

%Note that, $\sum_{i \in V, j \in S} sim_{i,j}$ is monotone submodular. 
Note that $\mathcal{L}$ is monotone submodular.
$\mathcal{L}$ is monotonic since coverage increases by the addition of a new sentence in the summary. At the same time, $\mathcal{L}$ is submodular since the increase in $\mathcal{L}$ would be more when a sentence is added to a shorter summary, than when a sentence is added to a longer summary. 
There can be several forms of $\mathcal{L}$ depending on how $sim_{i,j}$ is measured, which we will discuss later in this paper.

\vspace{1mm}
\noindent \underline{\bf Diversity reward}: The purpose of this aspect is to avoid redundancy and reward diverse information in the summary. 
Let the associated function be denoted as $\mathcal{R}$. A generic formulation of $\mathcal{R}$ is 
\begin{equation}
\mathcal{R}(S) = \sum_{i=1}^K \sqrt{\sum_{j \in P_i \cap S} r_j}
\label{eqn:diversity}
\end{equation}
where $P_1, P_2, \ldots, P_K$ comprise a partition of $V$ such that $\large\cup_i P_i = V$ and $P_i \cap P_j$ = $\emptyset$ for all $i$ $\neq$ $j$; 
$r_j$ is a suitable monotone submodular function that estimates the importance of adding the textual unit $j$ to the summary.
The partitioning $P_1, P_2, \ldots, P_K$ can be achieved by clustering the set $V$ using any clustering algorithm (e.g., $K$-means), based on the similarity of items as measured by $sim(i,j)$.

$\mathcal{R}(S)$ rewards diversity since there is more benefit in selecting a textual unit from a partition (cluster) that does not yet have any of its elements included in the summary.
As soon as any one element from a cluster $P_i$ is included in the summary, the other elements in $P_i$ start having diminishing gains, due to the square root function.

The function $r_j$ is a `singleton reward function' since it estimates the reward of adding the singleton element $j \in V$ to the summary $S$. One possible way to define this function is:
\begin{equation}
r_j = \frac{1}{N} \sum_i sim(i,j)
\label{eqn:singleton-reward}
\end{equation}
which measures the average similarity of $j$ to the other textual units in $V$. 
Note that $\mathcal{R}(S)$ is monotone submodular by Property (4), since square root is a non-decreasing concave function. This formulation will remain monotone submodular if any other non-decreasing concave function is used instead of square root.

While constructing a summary, both coverage and diversity are important.
Only maximizing coverage may lead to lack of diversity in the resulting summary and vice versa. So, we define our objective function for summarization as follows:
\begin{equation}
\mathcal{F} = \lambda_1 \mathcal{L} + \lambda_2 \mathcal{R}
\label{eqn:summary-objective}
\end{equation}
where $\lambda_1$, $\lambda_1$ $\ge$ 0 are the weights given to coverage and diversity respectively. Note that, by Property (3), $\mathcal{F}$ is monotone submodular.
%We obtain L and R from equations (6) and (7) respectively (Section 5.1) of Lin et al.~\cite{Lin2011}. To compute $w_{i,j}$, apart from the formulation presented in this paper, we can use semantic similarity (e.g., cosine similarity between embedding vectors created from Word2Vec and GloVe). Finally, the {\it objective function} F is defined by equation (2) of this paper.

Our proposed fairness-preserving summarization algorithm will maximize $\mathcal{F}$ in keeping with some fairness constraints. We now discuss this step.
\vspace*{-4mm}
\subsection{Submodular Maximization using Partition Matroids} \label{sub:matroid}

For fairness-preserving summarization, we essentially need to optimize the submodular function $\mathcal{F}$ while ensuring that the summary includes at least a certain number of textual units from each class present in the input data.
%of the textual units (e.g., textual units written by male authors, and those written by female authors).
This problem can be formulated using the concept of partition matroids, as described below.

\vspace{2mm}
%\subsubsection*{Definitions}
\noindent \underline{\bf Definitions:} We start with some definitions that are necessary to formulate the constrained optimization problem.

\vspace{1mm}
\noindent \textbf{Definition 4} (Matroid): A matroid is a pair $\mathcal{M}$ = ($\mathcal{Z}$, $\mathcal{I}$), defined over a finite set $\mathcal{Z}$ (called the ground set)  and a family of sets $\mathcal{I}$ (called the independent sets), that satisfies the three properties:
\begin{enumerate}
\item $\emptyset$ (empty set) $\in$ $\mathcal{I}$.
\item If $Y$ $\in$ $\mathcal{I}$ and $X$ $\subseteq$ $Y$, then $X$ $\in$ $\mathcal{I}$.
\item If $X$ $\in$ $\mathcal{I}$, $Y$ $\in$ $\mathcal{I}$ and $|Y|$ $>$ $|X|$, then there exists $e$ $\in$ $Y$ $\setminus$ $X$ such that $X$ $\cup$ $\{e\}$ $\in$ $\mathcal{I}$.
\end{enumerate}

\vspace{1mm}
\noindent \textbf{Definition 5} (Partition Matroids): Partition matroids refer to a special type of matroids where the ground set $\mathcal{Z}$ is partitioned into disjoint subsets $\mathcal{Z}_1$, $\mathcal{Z}_2$, ..., $\mathcal{Z}_s$ for some $s$ and \\
$\mathcal{I}$ = \{$S$ | $S$ $\subseteq$ $\mathcal{Z}$ and $|S \cap \mathcal{Z}_i|$ $\le$ $c_i$, for all $i$ = 1, 2, ..., $s$\} 
for some given parameters $c_1$, $c_2$, ..., $c_s$.
Thus, $S$ is a subset of $Z$ that contains at least $c_i$ items from the partition $\mathcal{Z}_i$ (for all $i$), and $\mathcal{I}$ is the family of all such subsets.

\vspace{2mm}
%\subsubsection*{Formulation}
\noindent \underline{\bf Formulation of the constrained maximization problem:}
Consider that we have a set of control variables $z_j$ (e.g., gender, political leaning), each of which takes $t_j$ distinct values (e.g., male and female, democrat and republican). Each item in $\mathcal{Z}$ has a particular value for each $z_j$.

For each control variable $z_j$,  we can partition $\mathcal{Z}$ into $t_j$ disjoint subsets $\mathcal{Z}_{j1}$, $\mathcal{Z}_{j2}$, ..., $\mathcal{Z}_{jt_j}$, each corresponding to a particular value of this control variable. 
We now define a partition matriod $\mathcal{M}_j$ = ($\mathcal{Z}$, $\mathcal{I}_j$) such that
\begin{center}
$\mathcal{I}_j$ = \{$S$ | $S$ $\subseteq$ $\mathcal{Z}$ and $|S \cap Z_{ji}|$ $\le$ $c_j$, for all $i = 1, 2, \ldots, t_j$\}
\end{center}
for some given parameters $c_1$, $c_2$, ..., $c_{t_j}$.

Now, for a given submodular objective function $f$, a submodular optimization under the partition matriod constraints with $P$ control variables can be designed as follows:
\begin{equation}
\label{lab:eq}
Maximize_{S \subseteq \mathcal{Z}} \;\; f(S)
\end{equation}
\begin{center}
subject to $S \in \bigcap_{j=1}^P \mathcal{I}$.
\end{center}
A prior work by Du {\it et al.}~\cite{Du2013} has established that this submodular optimization problem under the matroid constraints can be solved efficiently with provable guarantees (see~\cite{Du2013} for details).

\subsection{Proposed summarization scheme}

In the context of the summarization problem, the ground set is $V$ (= $\mathcal{Z}$), the set of all textual units (sentences/tweets) which we look to summarize.
The control variables (stated in Section~\ref{sub:matroid}) are analogous to the sensitive attributes with respect to which fairness is to be ensured.
In this work, we consider only one sensitive attribute for a particular dataset (the gender of a user for the Claritin dataset, political leaning of a tweet for the US-Election dataset, and the media source for the DUC06 dataset).
Let the corresponding control variable be $z$, and let $z$ take $t$ distinct values (e.g., $t = 2$ for the Claritin dataset, and $t=3$ for the US-Election dataset).
Note that, as described in Section~\ref{sub:matroid}, the formulation can be extended to multiple sensitive attributes (control variables) as well.

Each textual unit in $V$ is associated with a class, i.e., a particular value of the control variable $z$ (e.g., is posted either my a male or a female). 
Let $Z_1$, $Z_2$, ..., $Z_t$ ($Z_i$ $\subseteq$ $V$, for all $i$) be the disjoint subsets of the textual units from the $t$ classes, each associated with a distinct value of $z$.
%Let $Z_1$ ($\subseteq$ $V$) be the set of all the sentences posted by males and $Z_2$ ($\subseteq$ $V$) be the set of all sentences posted by females. 
We now define a partition matroid $\mathcal{M}$ = ($V$, $\mathcal{I}$) in which $V$ is partitioned into disjoint subsets $Z_1$, $Z_2$, ..., $Z_t$ and
\begin{center}
$\mathcal{I}$ = \{$S$ | $S$ $\subseteq$ $V$ and $|S \cap Z_i|$ $\le$ $c_i$, $i$ = 1, 2, ..., $t$\}
\end{center}
for some given parameters $c_1$, $c_2$, ..., $c_t$.
In other words, $I$ will contain all the sets $S$ containing at most $c_i$ sentences from $Z_i$, $i$ = 1, 2, ..., $t$.

Outside the purview of the matroid constraints, we maintain the restriction that $c_i$'s are chosen such that\\
(1) $\sum_{i=1}^t c_i = k$ (the desired length of the summary $S$), and\\ 
(2) a desired fairness criterion is maintained in $S$. For instance, if equal representation of all classes in the summary is desired, then $c_i = \frac{k}{t}$ for all $i$. 

%the 80\% {\it adverse impact fairness criterion}~\cite{biddle-adverse-impact} is maintained in $S$.

\noindent We now express our fairness-constrained summarization problem as follows:
\begin{equation}
\label{lab:eq2}
Maximize_{S \subseteq V} \mathcal{F}(S)
\end{equation}
\begin{center}
subject to $S$ $\in$ $\mathcal{I}$.
\end{center}
where the objective function $\mathcal{F}(S)$ is as stated in Equation~\ref{eqn:summary-objective}.

\begin{algorithm}[tb]
 \caption{: FairSumm (for fair summarization)}
 \label{algo:atga}
 \begin{algorithmic}[1]

  \State Set $d$ = $max_{z \in V}$$\mathcal{F}(\{z\})$. 
  \State Set $w_t$ = $\frac{d}{(i+\delta)^t}$ for $t$ = 0, $\ldots$, $l$ where $l$ = $argmin_i$ [$w_i$ $\le$ $\frac{\delta d}{N}$], and $w_{l+1}$ = 0.
 % /* Here $N$ is the size of $V$, $\delta$ is chosen as 0.01 by Du et al.~\cite{Du2013} */
  \State Set $G$ = $\emptyset$
  \For{$t$ = 0, 1, $\ldots$, $l$, $l+1$}
  	\For{each $z \in $ and $G\cup \{z\}$ $\in$ $I$}
    	\If{$\mathcal{F}$($G\cup \{z\}$) - $\mathcal{F}$($G$) $\ge$ $w_t$}
        	\State Set $G$ $\leftarrow$ $G\cup \{z\}$
        \EndIf
    
    \EndFor

  \EndFor
  \State Output $G$
  
 \end{algorithmic}
\end{algorithm}

The suitable algorithm to solve this constrained submodular optimization problem, proposed by Du et al.~\cite{Du2013}, is presented as Algorithm~\ref{algo:atga}. The $G$ produced by Algorithm~\ref{algo:atga} is the solution of Equation~\ref{lab:eq2}. 
This algorithm is an efficient alternative to the greedy solution which has a time complexity of $\mathcal{O}(kN)$ where $N$ = $|V|$ and $k$ = $|S|$. On the other hand, it can be shown that the time complexity of Du et al.~\cite{Du2013} is $\mathcal{O}(min\{|G|N, \frac{N}{\delta}log\frac{N}{\delta}\})$, where $\delta$ is a factor (to be explained shortly). 
The reason for this efficiency is the fact that this algorithm does {\it not} perform exhaustive evaluation of all the possible submodular functions evolving in the intermediate steps of the algorithm. 
Instead, it keeps on decreasing the threshold $w_t$ by a dividing factor (1 + $\delta)$, which skips the evaluation of many submodular functions and sets the threshold to zero when it is small enough. 
It selects elements $z$ from the ground set (in our case, $V$) only at each threshold value to evaluate the marginal gain ($\mathcal{F}$($G\cup \{z\}$) - $\mathcal{F}$($G$)) without violating any constraints.

%The $G$ produced by Algorithm~\ref{algo:atga} is the solution of Equation~\ref{lab:eq2}. 
Note that, the theoretical guarantee of Algorithm~\ref{algo:atga} depends upon the number of partition matroids ($P$), i.e., the number of control variables, and the curvature ($c_f$) of $\mathcal{F}$ given by
\begin{center}
$c_f = max_{j \in V, \mathcal{F}(j)>0} \; \frac{\mathcal{F}(j) - \mathcal{F}(j|V \setminus \{j\})}{\mathcal{F}(j)}$
\end{center}
The approximation ratio is $\frac{1}{P+c_f}$ (see~\cite{Du2013} for details); in our setting, the number of partition matroids $P$ is $1$.

\vspace{1mm}
\noindent \underline{\bf Algorithm for fair summarization}:
Algorithm~\ref{algo:atga} presents the algorithm to solve this constrained submodular optimization problem, based on the algorithm developed by Du et al.~\cite{Du2013}. 
The $G$ produced by Algorithm~\ref{algo:atga} is the solution of Equation~\ref{lab:eq2}. 
We now briefly describe the steps of Algorithm~\ref{algo:atga}.

In Step 1, the maximum value of the objective function $\mathcal{F}$ that can be achieved for a text unit $z$ ($\in$ $V$) is calculated and stored in $d$. The purpose of this step is to compute the maximum value of $\mathcal{F}$ for a single text unit $z$ and set a selection threshold (to be described shortly) with respect to this value. 
This step will help in the subsequent selection of textual units for the creation of the summary to be stored in $G$. 
$w_t$ (defined in Step 2) is such a threshold at the $t^{th}$ time step. $w_t$ is updated (decreased by division with a factor $1 + \delta$) for $t$ = 0, 1, $\ldots$, $l$. 
$l$ is the minimum value of $i$ for which $w_i$ $\le$ $\frac{\delta d}{N}$ holds (see Du et al.~\cite{Du2013} for details) and $w_{l+1}$ is set to zero. 
In Step 3, $G$ (the set that will contain the summary) is initialized as an empty set. Note that $G$ is supposed to be an independent set according to the definition of matroid given earlier in this section. 
By the condition (1) in the definition of matroids (stated earlier in this section), an empty set is independent. 
Step (4) iterates through the different values of $t$. Step (5) tests, for each $z$ (text element) $\in$ $V$, if $G$ remains an independent set by the inclusion of $z$. Only those $z$'s are chosen in this step whose inclusion expands $G$ (already an independent set) to another independent set. 
Step (6) selects a $z$ (permitted by Step (5)) for inclusion in $G$ if $\mathcal{F}(G\cup \{z\}) - \mathcal{F}(G) \ge w_t$. This $z$ is added to $G$ in Step (7). That is, $z$ is added to $G$ if the increment of $\mathcal{F}$ by the addition of $z$ is not less than the threshold $w_t$. For $t$ = 0, $w_t$ = $d$, that is, the maximum value of $\mathcal{F}$ for any $z$ ($\in$ $V$). 
This means, the $z$ which maximizes $\mathcal{F}$ is added to $G$. Note that, there can be multiple $z$'s for which $\mathcal{F}$ is maximized. In that case, the tie is broken arbitrarily. The remaining $z$'s may or may not be added to $G$ based on the threshold value.

Another important point to note is that, our chosen $\mathcal{F}$ (see equation (\ref{eqn:summary-objective})) is designed to maximize both coverage and diversity. So, even if multiple $z$'s satisfy Step (5), they may not be added to $G$ in Step (7) if they contain redundant information. The value of $w_t$ is relaxed for the subsequent values of $t$ to allow text elements $z$ producing relatively lower increments of $\mathcal{F}$ to be considered for possible inclusion in $G$. $w_{l+1} = 0$ indicates that for the final value of $t$ at least one text unit $z$ which does not decrement $\mathcal{F}$ is added to $G$. This ensures that the coverage of the summary produced is not compromised while preserving diversity. This process (Steps (5) to (7)) is repeated for $t$ = 0, 1, $\ldots$, $l$, $l+1$ resulting in the final output $G$.

The reason for the efficiency of Algorithm~\ref{algo:atga} is the fact that this algorithm does {\it not} perform exhaustive evaluation of all the possible submodular functions evolving in the intermediate steps of the algorithm. The reduction in the number of steps in the algorithm is achieved mainly by decreasing $w_t$ geometrically by a factor of $1 + \delta$. In addition, multiple elements $z$ can be added to $G$ for a single threshold which also expedites the culmination of the algorithm.

\newpage
\section{Sample Summaries of the Datasets}\label{App: Summaries}

We give some examples of summaries generated by the different summarization algorithms discussed in the paper, on two datasets -- (i)~MeToo dataset, and (ii)~US Election dataset. The datasets are described in detail in the original paper. All summaries showen here are  of length $k = 20$ tweets.

\subsection{Summaries of MeToo Dataset}

Here are few summaries produced with different fairness constraints of the MeToo Dataset.

\vspace{2 mm}
\noindent
\textbf{Summary produced by Sumbasic algorithm: (14 tweets by Females  \& 6 tweets by Males)}
\vspace{2 mm}
\begin{compactitem}
    \item Abusers aren't Victims, Opportunists aren't Crusaders. Slut shamers shudn't talk about Slut Shaming \& cowards shudn't pretend to be warriors. This is what Sonam Mahajan who I once considered a friend did to me. Telling this now seeing her blatant bigotry \& hypocrisy around \#MeToo
    \item 1. All the rapes and \#MeToo \#MeTooIndia allegations give raise to one question. 'when will it stop?" But the questions must be "when did it start?" And "How long has it taken for us to standup against it?"
    \item I need influencers to help me get the word out. I need 1000s if people to follow me \& RT. I need a Dateline long format interview \& to talk to Oprah.  I need funding and support. I am Alex Jones ex-wife \& I am NOT the underdog. I am you. I am America, \#metoo \& can change this.
    \item Apparently I'm only scared because of \#metoo even though the movement didn't exist at the time
    \item Good to see India's top uni take a no-nonsense approach to sexual harassment in \#academia \#Metoo: IISc shows door to top professor on charges of sexual harassment 
    \item The first story on Talk, Dammit was mine. The first person I opened up to about having been molested as a child was Faizan.   Please dont believe this attention seeking bully of a woman, Faizan has been there for me and countless others even before there was a \#metoo movement. 
    \item Also - the jokes. The jokes. If I now shake hands with you, will you yell \#MeToo?  It is that funny, is it?  Women and so many men finally found the strength to share the stories of sexual misconduct and assault and some men need to joke about it.
    \item And it would be because of the actions of a few girls like \#MeghaSharma who abuse \#metoo , one day it will probably have its own \#metoo movement as a mark of protest.
    \item @SmartKehan @MIKENY78 @MailOnline \#metoo ignores evidence and thats not right because a Man is Human and he has the rights as Woman but they doesnt understand that . \#Feministsaremonster
    \item Allowing a clear serial sexual predator to run one of your labels is a bad look, @AtlanticRecords. You need to drop @adam22 now. \#MeToo \#BelieveWomen
    \item My model friend @kawaljitanand has taken the initiative to come forward and speak about his  \#Metoo experience in the fashion industry. LEAD THE WAY  KAWAL. Proud of you.  Hope more men come forward to expose their experience with the predators around. \#KawaljitAnand \#MeTooIndia 
    \item It is becoming more commonplace for schools to offer support groups for boys. Here they feel safe to discuss their emotions and learn to navigate their relationship with girls in the \#metoo movement \#UBSW510 Boy Talk: Breaking Masculine Stereotypes
    \item \#Metoo dies a little more everyday. Bryan Singer is back at work. Woody Allen is getting support. James Gunn is back at work. Harvey Weinstein is pretty much on the verge of being cleared of any charges. CeeLo Green is back on the voice. America really doesn't care about rape.
    \item For ladies in \#Gambia.   Im working on a series of survivor stories within the context of \#MeToo, told exactly as the survivors would choose. If youre a survivor or know someone who is comfortable with it, and would want to share their story, please DM me.
    \item @danprimack @UCLA ask: in the light of \#metoo movements and Harvey Weinsteins, how do they feel about using money and lawyers to silence victims under NDAs instead of letting justice run its course, how did they address the problem in their organizations?
    \item @HillaryClinton Trump is right; a vote for Republicans on November 6th is a vote for him. Its a vote against women, a vote against freedom of speech, a vote for racism, a vote for hate crimes, a vote for acceptance of his lies. \#MeToo \#JamalKhashoggi \#MAGABomber \#PittsburghSynagogueShooting
    \item If the fear about \#MeToo is false accusations against men, then the GOP SHOULDN'T BE OFFERING \$\$\$ TO WOMEN TO MAKE FALSE ALLEGATIONS AGAINST ROBERT MUELLER!!! \#AreWeGreatEnoughYet
    \item Still, better than the sports masseur who said I lit him up like a Christmas tree, stroked the side of my breast, put his penis in my upturned palm \& said hed love to put cameras all over my house so he could watch me all day ...I just cant even...  \#metoo \#sexualharassment
    \item @annavetticad @TheQuint Salmans is an issue of domestic violence. Not to be conflated with \#MeToo. Diluting the def of \#MeToo will dampen its impact. Its def should be restricted to people in power sexually exploiting the weak. Not domestic violence or failed relationships
    \item It takes immense courage to this young woman ananya to express and write her story! Really the time's up!! This young actress faced lots of mental, emotional \& physical abuse from another woman!! \#MeToo please read her story in the FB link with this tweet.
\end{compactitem}

\vspace{2 mm}
\noindent
\textbf{Summary produced by applying Proportional Representation on the Sumbasic algorithm (11 tweets by Females \& 9 tweets by Males) [Algorithm II]}
\vspace{2 mm}

\begin{compactitem}
    \item I need influencers to help me get the word out I need 1000s if people to follow me \& RT I need a Dateline long format interview \& to talk to Oprah  I need funding and support I am Alex Jones ex-wife \& I am NOT the underdog I am you I am America, \#metoo \& can change this.
    \item @WomenInFilm \#womeninfilm \#BelieveSurvivors \#timesup \#metoo \#ChristineBlaseyFord \#MKUTRA \#Hypocrites \#Hypocrisy \#BoycottHollywood    The \#Truth is, you \#Women 'Walked Out' of having any \#CommonSense You're all 'Puppets of \#PedoWood' \#WalkAway.
    \item Apparently I'm only scared because of \#metoo even though the movement didn't exist at the time.
    \item 1 All the rapes and \#MeToo \#MeTooIndia allegations give raise to one question 'when will it stop" But the questions must be "when did it start" And "How long has it taken for us to standup against it".
    \item \#METOO MEN WHO SEXUALLY HARASS WOMEN DONT SEE IT AS SEXUAL HARASSMENT \#WOMENTOO DOES THE SAME.
    \item Today I spoke to a victim of \#FalseAllegations we stay in contact and touch base regularly He's still angry at he's treatment by the so called justice system He spent just over 3 months in prison till he's acquittal As a pensioner he spent over \$100k on he's defense \#MeToo.
    \item In the current climate of \#MeToo if someone did that to you at your place of work would you just let it go Or should it be considered sexual assault.
    \item @skhndh Another thing you are not alone in Its been mentally consuming for me It was was one of your tweets shared but @ghaatidancer that pushed me through my silence and made me share my story We are all in this together and its high time So be proud \#MeToo.
    \item And it would be because of the actions of a few girls like \#MeghaSharma who abuse \#metoo , one day it will probably have its own \#metoo movement as a mark of protest.
    \item The left has hyjacked the \#metoo movement The left does not believe in innocent until proven guilty The left does not believe in America first instead they would rather put every country ahead of the United States.
    \item In many organizations, content is also raising its \#MeToo voice for the way it is treated there \#contentstrategy.
    \item All the gentlemen who have been urging me via direct messages to talk about my personal \#MeToo experience(s) - please look for this soft porn gratification from the low cost data services now available Or you could make a quick trip to Bangkok (without bhabhiji) \#NoJudgement.
    \item @SmartKehan @MIKENY78 @MailOnline \#metoo ignores evidence and thats not right because a Man is Human and he has the rights as Woman but they doesnt understand that  \#Feministsaremonster.
    \item Any feminist lawyers out there want to join forces and learn cutting edge defamation law so we can protect people who speak out in the \#metoo movement Im in Vancouver and want to join a collective to level the playing field.
    \item @Chinmayi  Can you take back the shame and humiliation this guy faced bcoz of what you have done to him He should seriously file defamation case against you \#MeTooControversy  \#metoo.
    \item I hope the \#metoo sexual assault allegations in India tear down every perpetrator of sexual violence in Bollywood and beyond I hope the industry crumbles from within and rebuilds something new.
    \item Party coming to SM disclosing his proof of innocense He is having access to those ministers who r appointed as GoM in \#MeToo campaign but he preferred to come to SM bcoz he knows \#Vote matters even they would have not listened to \#GenderAppeasement Now pl think about \#NOTA4Men.
    \item @rosemcgowan You are the founder of \#metoo and as such I feel that you should stand for anyone being abused Johnny Depp made the mistake of marrying Amber Heard but shes trashed his name when shes really the abuser  She should be accountable and is not Here are the court.
    \item A year ago today I thought my world was falling apart I woke up to find out that the hashtag \#metoo had gone viral and I didn't see any of the work I laid out over the previous decade attached to it I thought for sure I would be erased from a thing I worked so hard to build.
    \item Excellent news Should have happened right away but good that it happened now Thanks to @TataCompanies for doing the right thing Meanwhile Suhel Seth continues to hide in whatever hole he's crawled into \#MeToo.
\end{compactitem}

\vspace{2 mm}
\noindent
\textbf{Summary produced by applying Equal Representation on the Sumbasic algorithm (10 tweets by Females \& 10 tweets by Males )  [Algorithm II]}
\vspace{2 mm}

\begin{compactitem}
    \item I need influencers to help me get the word out I need 1000s if people to follow me \& RT I need a Dateline long format interview \& to talk to Oprah  I need funding and support I am Alex Jones ex-wife \& I am NOT the underdog I am you I am America, \#metoo \& can change this.
    \item @WomenInFilm \#womeninfilm \#BelieveSurvivors \#timesup \#metoo \#ChristineBlaseyFord \#MKUTRA \#Hypocrites \#Hypocrisy \#BoycottHollywood    The \#Truth is, you \#Women 'Walked Out' of having any \#CommonSense You're all 'Puppets of \#PedoWood' \#WalkAway  
    \item Apparently I'm only scared because of \#metoo even though the movement didn't exist at the time.
    \item 1 All the rapes and \#MeToo \#MeTooIndia allegations give raise to one question 'when will it stop" But the questions must be "when did it start" And "How long has it taken for us to standup against it".
    \item \#METOO MEN WHO SEXUALLY HARASS WOMEN DONT SEE IT AS SEXUAL HARASSMENT \#WOMENTOO DOES THE SAME.
    \item Today I spoke to a victim of \#FalseAllegations we stay in contact and touch base regularly He's still angry at he's treatment by the so called justice system He spent just over 3 months in prison till he's acquittal As a pensioner he spent over \$100k on he's defense \#MeToo.
    \item @skhndh Another thing you are not alone in Its been mentally consuming for me It was was one of your tweets shared but @ghaatidancer that pushed me through my silence and made me share my story We are all in this together and its high time So be proud \#MeToo.
    \item The \#MeToo movement is officially dead.   The left only cares about women's sexual assault allegations when it is against a rival political party.  They NEVER cared about the claims of the victims!  \#ThesePeopleAreSick \#HypocriticalLeft \#VoteDemsOut \#PedoWood \#RapeIsntAJoke
    \item In the current climate of \#MeToo if someone did that to you at your place of work would you just let it go Or should it be considered sexual assault 
    \item All the gentlemen who have been urging me via direct messages to talk about my personal \#MeToo experience(s) - please look for this soft porn gratification from the low cost data services now available Or you could make a quick trip to Bangkok (without bhabhiji) \#NoJudgement 
    \item The left has hyjacked the \#metoo movement The left does not believe in innocent until proven guilty The left does not believe in America first instead they would rather put every country ahead of the United States.
    \item Any feminist lawyers out there want to join forces and learn cutting edge defamation law so we can protect people who speak out in the \#metoo movement Im in Vancouver and want to join a collective to level the playing field.
    \item In many organizations, content is also raising its \#MeToo voice for the way it is treated there \#contentstrategy.
    \item I hope the \#metoo sexual assault allegations in India tear down every perpetrator of sexual violence in Bollywood and beyond I hope the industry crumbles from within and rebuilds something new.
    \item @SmartKehan @MIKENY78 @MailOnline \#metoo ignores evidence and thats not right because a Man is Human and he has the rights as Woman but they doesnt understand that  \#Feministsaremonster.
    \item @rosemcgowan You are the founder of \#metoo and as such I feel that you should stand for anyone being abused Johnny Depp made the mistake of marrying Amber Heard but shes trashed his name when shes really the abuser  She should be accountable and is not  Here are the court .
    \item A year ago today I thought my world was falling apart I woke up to find out that the hashtag \#metoo had gone viral and I didn't see any of the work I laid out over the previous decade attached to it I thought for sure I would be erased from a thing I worked so hard to build .
    \item @Chinmayi Can you take back the shame and humiliation this guy faced bcoz of what you have done to him He should seriously file defamation case against you \#MeTooControversy  \#metoo 
     \item Excellent news Should have happened right away but good that it happened now Thanks to @TataCompanies for doing the right thing Meanwhile Suhel Seth continues to hide in whatever hole he's crawled into \#MeToo.
    \item Party coming to SM disclosing his proof of innocense He is having access to those ministers who r appointed as GoM in \#MeToo campaign but he preferred to come to SM bcoz he knows \#Vote matters even they would have not listened to \#GenderAppeasement Now pl think about \#NOTA4Men.
\end{compactitem}

\vspace{2 mm}
\noindent
\textbf{Summary produced by applying Proportional Representation  on proposed FairSumm Algorithm (11 tweets by Females \& 9 tweets by Males) [Algorithm I]}
\vspace{2 mm}

\begin{compactitem}
    \item I need influencers to help me get the word out. I need 1000s if people to follow me \& RT. I need a Dateline long format interview \& to talk to Oprah.  I need funding and support. I am Alex Jones ex-wife \& I am NOT the underdog. I am you. I am America, \#metoo \& can change this.
    \item IMPORTANT : My DMs are open. Let's hear \#HimToo \#MenToo and the \#MeToo stories of Men. Tell me about cases where you as a man or a man you know has been abused, sexually assaulted, harassed, intimidated, falsely accused and not even heard let alone believed. REACH OUT \& RT.
    \item It is not just Suhel Seth's silence. It's also the unacceptable silence of many in media who knew him professionally and socially. I did too like many others. I spoke up but I can count on one hand those of us  who have demanded accountability from him \#MeToo : Where are others?
    \item @SenatorCollins Too bad you don't really believe in the power of truth. How can you live with yourself, knowingly voting for an alcoholic abuser of women. Next time I am sexually assaulted I will first ask if anyone would like to witness for prosperity's sake \#metoo screw you.
    \item I for one believe all women which is why i will believe this Mueller accuser until it is proven she is or isn't lying. Sure is odd All the media is trying to get out in front of this before we even have the accuser's name though for those noting.. \#Metoo
    \item A year ago today I thought my world was falling apart. I woke up to find out that the hashtag \#metoo had gone viral and I didn't see any of the work I laid out over the previous decade attached to it. I thought for sure I would be erased from a thing I worked so hard to build. 
    \item I admire the courage of @priyaramani \& all the victims who spoke out against sexual harassment \& abuse. @mjakbar s resignation is a moral victory for everyone. The defamation case cant change that. The \#MeToo movement needs support across party lines.
    \item @MIKENY78 @SmartKehan @MailOnline That to they lied and now they want to lynch every Man on Earth . The Whole \#metoo is Stupid . Its Dangerous to believe every woman . No Man is safe anymore when someone doesnt like you they cry \#metoo . It can happen to everyone
    \item @varshapillai @womensweb Are women that weak that they need a hashtag to speak out ? I'm sure women are strong enough to fight without social media. Atleast women around me are. \#metoo needs to be stopped because it does really promote more fake cases. Instead fight the real one bravely and we support u.
    \item As I cry and succumb to mental exhaustion because of \#MeToo060 , I remember the friendships women forged and solidarity we showered upon each other. Each one of you who has helped, supported, fought the war, been there relentlessly - I will never forget.  Thank you. I love you.
    \item She has no cellphone and doesn't use hashtags. "What is misleading is that people say \#MeToo is an urban elite thing that must reach the poor. It's actually the other way around." 'They Deserve Justice': Mother Of India's \#MeToo Speaks Out
    \item Excellent news Should have happened right away but good that it happened now Thanks to @TataCompanies for doing the right thing Meanwhile Suhel Seth continues to hide in whatever hole he's crawled into \#MeToo.
    \item Im so fucking stupid everyone always says dont get in a strangers car. Im fucking smarter than this. I hate this world you can never trust anyoneud83eudd2eIm fine, he just tried to hold my hand, put his hand on my leg. I got home safely but still... he was 60 and Im 17 \#metoo
    \item These wounds come from a culture infection; we know where it began. It's effects has lasted for generations and the time to cleanse it all out is now.   This isn't our shame to hold any longer. This was never ours to begin with.  \#IndigenousMeToo \#Indigenous \#MeToo
    \item If a guy says he's nervous about \#MeToo, just remind him that we come down pretty hard on murderers too, and ask him why that doesn't make him nervous. If he says, "Because I haven't murdered anyone," then you've learned something new about your friend.
    \item I had to cancel a follow up gynecologist appointment today for my iud. My anxiety is through the roof. I can't even deal with the travel right now... any thought of going to that appointment brings panic through my body. I'll try again, but not today. \#survivorculture \#MeToo
    \item I have been asking media houses to debate this but these spineless guys don't ever discuss such incidents. Biased media could only talk  \#MeToo at prime time to demean the men of this country. Poor guard with no mistake is behind bars .   Jago  Jago @Dev\_Fadnavis @CMOMaharashtra
    \item If these same signs had pictures of transgender African Americans holding \#metoo signs, there would be rioting, looting, and protests. The right would be called white supremacists and nazis for it. The double standard is real and easy to see.
    \item @Annamahof1 @Gatonyenn @Darius\_M4 The question wasn't about if women are dangerous or not, it was if women are more agreable then men, don't change the topic.  The answer is NO -Men give more present than Women -Men help more others on their works -Men don't do \#metoo -Men don't ask for divorce at 75% and so on..
    \item If a woman shares a \#metoo without evidence, it's taken to be true coz it's a women's testimony, a man coming out with \#HeToo story, people would be doubtful,  \& question the evidences, the intent \& will never except the man as victim. \#misandry must be understood. \#SpeakUpMan
\end{compactitem}

\vspace{2 mm}
\noindent
\textbf{Summary produced by applying Equal Representation on proposed FairSumm Algorithm  (10 tweets by Females \& 10 tweets by Males) [Algorithm I]}
\vspace{2 mm}
\begin{compactitem}
    \item I need influencers to help me get the word out. I need 1000s if people to follow me \& RT. I need a Dateline long format interview \& to talk to Oprah.  I need funding and support. I am Alex Jones ex-wife \& I am NOT the underdog. I am you. I am America, \#metoo \& can change this.
    \item IMPORTANT : My DMs are open. Let's hear \#HimToo \#MenToo and the \#MeToo stories of Men. Tell me about cases where you as a man or a man you know has been abused, sexually assaulted, harassed, intimidated, falsely accused and not even heard let alone believed. REACH OUT \& RT.
    \item It is not just Suhel Seth's silence. It's also the unacceptable silence of many in media who knew him professionally and socially. I did too like many others. I spoke up but I can count on one hand those of us  who have demanded accountability from him \#MeToo : Where are others? 
    \item @SenatorCollins Too bad you don't really believe in the power of truth. How can you live with yourself, knowingly voting for an alcoholic abuser of women. Next time I am sexually assaulted I will first ask if anyone would like to witness for prosperity's sake \#metoo screw you.
    \item I for one believe all women which is why i will believe this Mueller accuser until it is proven she is or isn't lying. Sure is odd All the media is trying to get out in front of this before we even have the accuser's name though for those noting.. \#Metoo
    \item A year ago today I thought my world was falling apart. I woke up to find out that the hashtag \#metoo had gone viral and I didn't see any of the work I laid out over the previous decade attached to it. I thought for sure I would be erased from a thing I worked so hard to build. 
    \item I admire the courage of @priyaramani \& all the victims who spoke out against sexual harassment \& abuse. @mjakbar s resignation is a moral victory for everyone. The defamation case cant change that. The \#MeToo movement needs support across party lines. 
    \item @MIKENY78 @SmartKehan @MailOnline That to they lied and now they want to lynch every Man on Earth . The Whole \#metoo is Stupid . Its Dangerous to believe every woman . No Man is safe anymore when someone doesnt like you they cry \#metoo . It can happen to everyone
    \item @varshapillai @womensweb Are women that weak that they need a hashtag to speak out ? I'm sure women are strong enough to fight without social media. Atleast women around me are. \#metoo needs to be stopped because it does really promote more fake cases. Instead fight the real one bravely and we support u.
    \item As I cry and succumb to mental exhaustion because of \#MeToo , I remember the friendships women forged and solidarity we showered upon each other. Each one of you who has helped, supported, fought the war, been there relentlessly - I will never forget.  Thank you. I love you.
    \item She has no cellphone and doesn't use hashtags. "What is misleading is that people say \#MeToo is an urban elite thing that must reach the poor. It's actually the other way around." 'They Deserve Justice': Mother Of India's \#MeToo Speaks Out 
    \item Excellent news Should have happened right away but good that it happened now Thanks to @TataCompanies for doing the right thing Meanwhile Suhel Seth continues to hide in whatever hole he's crawled into \#MeToo.
    \item Im so fucking stupid everyone always says dont get in a strangers car. Im fucking smarter than this. I hate this world you can never trust anyone Im fine, he just tried to hold my hand, put his hand on my leg. I got home safely but still... he was 60 and Im 17 \#metoo
    \item These wounds come from a culture infection; we know where it began. It's effects has lasted for generations and the time to cleanse it all out is now.   This isn't our shame to hold any longer. This was never ours to begin with.  \#IndigenousMeToo \#Indigenous \#MeToo
    \item If a guy says he's nervous about \#MeToo, just remind him that we come down pretty hard on murderers too, and ask him why that doesn't make him nervous. If he says, "Because I haven't murdered anyone," then you've learned something new about your friend.
    \item I had to cancel a follow up gynecologist appointment today for my iud. My anxiety is through the roof. I can't even deal with the travel right now... any thought of going to that appointment brings panic through my body. I'll try again, but not today. \#survivorculture \#MeToo
    \item We need to send out the message This not how we behave in our sector. Brave words from @shyhousinggirl. @LuchiaFitz100 and I have been calling for such action daily @24housing  Opinion \#MeToo movement in housing: Are we standing up for change? 
    \item If these same signs had pictures of transgender African Americans holding \#metoo signs, there would be rioting, looting, and protests. The right would be called white supremacists and nazis for it. The double standard is real and easy to see. 
    \item @Annamahof1 @Gatonyenn @Darius\_M4 The question wasn't about if women are dangerous or not, it was if women are more agreable then men, don't change the topic.  The answer is NO -Men give more present than Women -Men help more others on their works -Men don't do \#metoo -Men don't ask for divorce at 75% and so on..
    \item If a woman shares a \#metoo without evidence, it's taken to be true coz it's a women's testimony, a man coming out with \#HeToo story, people would be doubtful,  \& question the evidences, the intent \& will never except the man as victim. \#misandry must be understood. \#SpeakUpMan 
\end{compactitem}

\subsection{Summaries of US-Election Dataset}

Here are few summaries (of length $k = 20$ tweets) produced with different fairness constraints of the US-Election Dataset.

\vspace{2 mm}
\noindent
\textbf{Summaries produced by Sumbasic Algorithm: (3 pro-democratic tweets \& 15 pro-republican tweets \& 2 neutral tweets)}
\vspace{2 mm}

\begin{compactitem}
    \item There is no other Donald Trump. This is it.
    \item REPEAL AND REPLACE!!!  \#ObamaCareInThreeWords
    \item 13 states have voter registration deadlines TODAY: FL, OH, PA, MI, GA, TX, NM, IN, LA, TN, AR, KY, SC.  
    \item COMING UP @GenFlynn @newtgingrich on @foxandfriends
    \item "A rough night for Hillary Clinton"  ABC News.
    \item There's only one candidate in this election who's ready to be America's Commander-in-Chief.
    \item Barack\& Michelle\& Joe\& Bernie\& Elizabeth\& You?  
    \item Crooked Hillary Clinton likes to talk about the things she will do but she has been there for 30 years - why didn't she do them?
    \item That was me! I was there... 
    \item MAKE AMERICA GREAT AGAIN!
    \item 'Donald Trump: A President for All Americans' 
    \item Do you think Hillary has all the questions that she is going to be asked debate Retweet before you vote to give your prople a chance to vote
    \item The media and establishment want me out of the race so badly -  I WILL NEVER DROP OUT OF THE RACE, WILL NEVER LET MY SUPPORTERS DOWN! \#MAGA
    \item We call on the FBI to immediately release all emails pertinent to their investigation. Americans have the right to know before Election Day.
    \item \#NEW Reuters/Ipsos Polls  \#Iowa: Trump 49\% (+8) Clinton 41\%  \#Florida: Trump 50\% (+4) Clinton 46\%  \#Colorado Trump 43\% (+3) Clinton 40\%
    \item Thank you to @foxandfriends for the great review of the speech on immigration last night. Thank you also to the great people of Arizona!
    \item Russia took Crimea during the so-called Obama years. Who wouldn't know this and why does Obama get a free pass?
    \item Praying for the families of the two Iowa police who were ambushed this morning. An attack on those who keep us safe is an attack on us all.
    \item @realDonaldTrump campaign raised a record \$100 million in September. Number of donors up 25\%.
    \item my son is voting for the first time i'm so proud 
\end{compactitem}

\vspace{2 mm}
\noindent
\textbf{Summary produced by applying Proportional Representation on FairSumm Algorithm (6 pro-democratic tweets \& 12 pro-republican tweets\& 2 neutral tweets) [Algorithm I]}
\vspace{2 mm}

\begin{compactitem}
    \item You can change your vote in six states. So, now that you see that Hillary was a big mistake, change your vote to MAKE AMERICA GREAT AGAIN!
    \item Wow, Hillary Clinton was SO INSULTING to my supporters, millions of amazing, hard working people. I think it will cost her at the Polls!
    \item WHY ARE PEOPLE VOTING TRUMP BECAUSE "HILLARY IS A CRIMINAL" WHEN THE FBI ALREADY CLEARED HER AND TRUMP IS GOING TO COURT FOR A RAPE CLAIM
    \item Donald Trump still refuses to say @POTUS was born in America. And this man wants to be our president?  When will he stop this ugliness?
    \item Just heard from NC Campaign. Hillary Campaign is calling voters saying they have a chance to win \$5k if they vote Hillary! Paying 4 votes!  
    \item Voter fraud! Crooked Hillary Clinton even got the questions to a debate, and nobody says a word. Can you imagine if I got the questions?
    \item Trump said his campaign could be his "single greatest waste of time, of energy \& money." Congrats, Donald! Today you are a Trump U graduate.
    \item In order to \#DrainTheSwamp \& create a new GOVERNMENT of, by, \& for the PEOPLE, I need your VOTE! Go to - LET'S \#MAGA!
    \item Imagine if Hillary Clinton said on her email server - "I don't talk about it anymore." And we all just moved on. That's not how it works.
    \item As @realDonaldTrump just showed the American people, no matter what happens he will not be deterred \& he will not give up fighting for you!
    \item Wow, did you just hear Bill Clinton's statement on how bad ObamaCare is. Hillary not happy. As I have been saying, REPEAL AND REPLACE!
    \item Hearing privately from many reporters that Hillary had a "terrible night" \& Trump had best debate. Hope this will make it into print, on air
    \item Hillary Clinton didn't go to Louisiana, and now she didn't go to Mexico. She doesn't have the drive or stamina to MAKE AMERICA GREAT AGAIN!
    \item The people are really smart in cancelling subscriptions to the Dallas \& Arizona papers \& now USA Today will lose readers! The people get it!
    \item Hey everyone, the Obama speechwriters who came up with "If you like your plan you can keep it" are saying Hillary's health is just fine.
    \item We hear Hillary will try to get under Trump's skin tomorrow night. How does this differ with what she does to all of us daily? \#Debates
    \item The fact that Trump's answer to a question about racial injustice is just "we need law and order" pretty much says it all. \#debatenight
    \item Win or lose, Trump made our world uglier. I don't want any reality shows from him after this, no cutesy late night sketches. I want nothing.
    \item How are so many people JUST NOW offended by Trump? It's like getting to the 7th Harry Potter book \& realizing Voldemort might be a bad guy.
    \item "I call on all Republicans to put people before politics and finally vote on a clean funding bill to fight Zika right here in Florida."
\end{compactitem}

\vspace{2 mm}
\noindent
\textbf{Summary produced by applying Equal Representation on  FairSumm Algorithm  (7 pro-democratic tweets \& 7 pro-republican tweets \& 6 neutral tweets) [Algorithm I]}
\vspace{2 mm}

\begin{compactitem}
    \item You can change your vote in six states. So, now that you see that Hillary was a big mistake, change your vote to MAKE AMERICA GREAT AGAIN!
    \item Wow, Hillary Clinton was SO INSULTING to my supporters, millions of amazing, hard working people. I think it will cost her at the Polls!
    \item WHY ARE PEOPLE VOTING TRUMP BECAUSE "HILLARY IS A CRIMINAL" WHEN THE FBI ALREADY CLEARED HER AND TRUMP IS GOING TO COURT FOR A RAPE CLAIM
    \item Donald Trump still refuses to say @POTUS was born in America. And this man wants to be our president?  When will he stop this ugliness?
    \item Voter fraud! Crooked Hillary Clinton even got the questions to a debate, and nobody says a word. Can you imagine if I got the questions?
    \item In order to \#DrainTheSwamp \& create a new GOVERNMENT of, by, \& for the PEOPLE, I need your VOTE! Go to - LET'S \#MAGA!
    \item Imagine if Hillary Clinton said on her email server - "I don't talk about it anymore." And we all just moved on. That's not how it works.
    \item As @realDonaldTrump just showed the American people, no matter what happens he will not be deterred \& he will not give up fighting for you!
    \item Wow, did you just hear Bill Clinton's statement on how bad ObamaCare is. Hillary not happy. As I have been saying, REPEAL AND REPLACE!
    \item The people are really smart in cancelling subscriptions to the Dallas \& Arizona papers \& now USA Today will lose readers! The people get it!
    \item We have to stand up to this hate.  We have to send a clear message:  America is better than this. America is better than Donald Trump.
    \item Trump surrogates now attacking former Miss Universe 4 doing a nude photo spread. You know, just like Melania Trump did. Do they ever think?
    \item Trump said his campaign could be his "single greatest waste of time, of energy \& money." Congrats, Donald! Today you are a Trump U graduate.
    \item Win or lose, Trump made our world uglier. I don't want any reality shows from him after this, no cutesy late night sketches. I want nothing.
    \item How are so many people JUST NOW offended by Trump? It's like getting to the 7th Harry Potter book \& realizing Voldemort might be a bad guy.
    \item "I call on all Republicans to put people before politics and finally vote on a clean funding bill to fight Zika right here in Florida."
    \item It happened: @Cubs win World Series. That's change even this South Sider can believe in. Want to come to the White House before I leave?
    \item Trump: I alone can fix! Hillary: Together we can fix it! Stein: They can't fix it and neither can I!  Johnson: [Googles definition of fix]
    \item If Hillary wins- God is still God If Trump wins- God is still God I may have lost hope in America, but never in God.
    \item Thank you to every nation that moved to bring the Paris Agreement into force. History will judge today as a turning point for our planet.
\end{compactitem}

\vspace{2 mm}
\noindent
NOTE: The difference in length of tweets in the MeToo dataset and that of US-Election dataset is due to the change in the maximum number of characters allowed in a tweet from $140$ to $280$ in the year 2017.

\end{document}